\documentclass{aa}  

\usepackage{graphicx}
\usepackage{array}
\usepackage{txfonts}
\usepackage{caption}
\usepackage{lipsum}
\usepackage{hyperref}
\usepackage{url}
\usepackage{natbib}
\usepackage{breqn}
\usepackage{diagbox}
\usepackage{cuted} 
\usepackage[mathscr]{euscript}

\newcommand{\orcid}[1]{\href{https://orcid.org/#1}{\includegraphics[width=10pt]{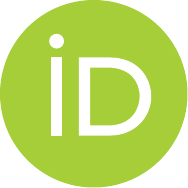}}}

\newcolumntype{C}{>{\centering\arraybackslash}p{0.2\textwidth}}
\usepackage{enumitem}
\usepackage{booktabs}
\bibpunct{(}{)}{;}{a}{}{,} % to follow the AA style
\begin{document} 
\title{The rotational disruption of porous dust aggregates from ab-initio kinematic calculations}
\titlerunning{The rotational disruption of porous dust}
\author{Stefan Reissl \orcid{0000-0001-5222-9139} \inst{\ref{inst1}} \and Philipp Nguyen \inst{\ref{inst1}} \and Lucas M. Jordan\inst{\ref{inst2}} \and Ralf S. Klessen  \orcid{0000-0002-0560-3172} \inst{ \inst{\ref{inst1},\ref{inst3}}  } }

\institute{
\centering Universit\"{a}t Heidelberg, Zentrum f\"{u}r Astronomie, Institut f\"{u}r Theoretische Astrophysik, Albert-Ueberle-Straße 2, \\  D-69120 Heidelberg, Germany \label{inst1} 
\and
\centering Institut f\"{u}r Astronomie und Astrophysik, Universit\"{a}t T\"{u}bingen, Auf der Morgenstelle 10, 72076 T\"{u}bingen, Germany \label{inst2}
\and
\centering Universit\"{a}t Heidelberg, Interdisziplin\"{a}res Zentrum f\"{u}r Wissenschaftliches Rechnen, Im Neuenheimer Feld 205, \\ D-69120 Heidelberg, Germany  \label{inst3}
}
						
%\date{Received  XXXX; accepted  XXXX}
\abstract
   {The sizes of dust grains in the interstellar medium follows a  distribution where most of the dust mass is in smaller grains. However, the re-distribution from larger grains towards smaller sizes especially by means of rotational disruption is poorly understood.}
{We aim to study the dynamics of porous grain aggregates under accelerated ration. Especially, we determine the deformation of the grains and the maximal angular velocity up to the rotational disruption event by caused by centrifugal forces.}
   {We pre-calculate porous grain aggregate my means of ballistic aggregation analogous to the interstellar dust as input for subsequent numerical simulations. In detail, we perform three-dimensional N-body simulations mimicking the radiative torque spin-up process up to the point where the grain aggregates become rotationally disrupted.}
   {Our simulations results are in agreement with theoretical models predicting a characteristic angular velocity $\omega_{\mathrm{disr}}$ of the order of ${ 10^8 - 10^9\ \mathrm{rad\ s^{-1}} }$, where grains become rotationally disrupted. In contrast to the theoretical predictions, we show that for large porous grain aggregates ($\gtrapprox 300\ \mathrm{nm}$) $\omega_{\mathrm{disr}}$ does not strictly decline but reaches a lower asymptotic value. Hence, such grains can withstand an accelerated ration more efficiently up to a factor of 10 because the displacement of mass by centrifugal forces and the subsequent mechanical deformation supports the build up of new connections within the aggregate. Furthermore, we report that the rapid rotation of grains deforms an ensemble with initially 50:50 prolate and oblate shapes, respectively,  preferentially into oblate shapes. Finally, we present a best fit formula to predict the average rotational disruption of an ensemble of porous dust aggregates dependent on internal grain structure, total number of monomers, and applied material properties.}
 {}
  \maketitle
%
%________________________________
\section{Introduction}
Dust is a key component of the interstellar medium (ISM). It is important for regulating the properties of astrophysical objects across a wide range of scales: from the cooling of collapsing  molecular clouds and subsequent star-formation down to the formation of planetary systems \citep{Spitzer1978,Dorschner1995}. However, the origin of dust, its initial physical properties and the redistribution of grain sizes is still a field of ongoing research \citep{ODonnell1997,Ormel2009,Birnstiel2010,Guillet2018,Draine2021}.\\
Dust composition and grain size distribution may be derived from the observed interstellar extinction curve and starlight polarization \citep{Mathis1977,Draine1984,Guillet2018,Draine2021}. Initially, the ISM is enriched 
by  intermediate-mass stars at the asymptotic giant branch (AGB) supernova ejecta coming with a certain grain size distribution \citep{Nozawa2007,Gail2009,Barlow2010,Matsuura2011,Karovicova2013,Zhukovska2015,Bevan2016}. Later, the grains grow in dense molecular clouds by accretion of abundant elements and coagulation  
\citep{Spitzer1978,Chokshi1993,ODonnell1997}. Naturally, this process results in porous dust aggregates rather than solid bodies \citep{Ossenkopf1993,Dominik1997,Wada2007}.
Dust destruction processes such as gas-grain sputtering or grain-grain collision (shattering) may redistribute the grown grains towards smaller sizes \citep{Dwek1980,Tielens1994,Hirashita2009A}.\\
More recently, the pioneering work presented by \cite{Hoang2019}  describes a new dust destruction mechanism. Here, a radiation field causes radiative torques (RAT) acting on dust grains which lead to an angular acceleration\citep[see e.g.][]{LazarianHoang2007,Hoang2014}. Given a sufficiently luminous environment the grains would inevitably be disrupted by the emerging centrifugal force \citep{Silsbee2016,Hoang2019,Hoang2020Galaxy}. The disruption process is usually quantified by the maximal tensile strength $\mathcal{S}_{\mathrm{max}}$ which is a measure how materials respond to stretching \citep{Silsbee2016,Tatsuuma2021}. For porous materials and dust grain analogs the tensile strengths may be determined by numerical N-body simulations  \citep[][]{Kataoka2013X,Seizinger2013Tensile,TatsuumaKataoka2019}. However,  what simulations of $\mathcal{S}_{\mathrm{max}}$ are missing is that individual building blocks (so called monomers) do not just feel the stretching force between its neighbours, but additionally the global centrifugal force acting on the entire aggregate. Hence, up to this point, it remains unclear if the parameter $\mathcal{S}_{\mathrm{max}}$ describes accurately the internal processes of material displacement within the rotating aggregate.\\
The rotational disruption of fractal grain aggregates was already studied indirectly in \citep[see][RMK22 hereafter]{Reissl2022} by evaluating $\mathcal{S}_{\mathrm{max}}$ in the context of rapid grain ration caused by a differential gas-dust velocity. However, in this paper we aim to simulate the dynamics of rapidly rotating interstellar grains directly by taking the time evolution of the internal aggregate structure into account. The aim is to develop a model of the average disruption process of large ensembles of porous grains. This paper is structured as follows: In Sect.~\ref{sect:GrainComposition} we discuss the most likely composition of elements and minerals to be present in dust aggregates. An algorithm to mimic the growth of porous dust aggregates is outlined in Sect.~\ref{sect:GrainGrowth} in detail. Here, we also introduce the methods to quantify the grain shape and porosity. In Sect.~\ref{sect:MonomerContact} we discuss the processes and forces acting between the monomers connected within the aggregates. The  spin-up process of grains by means of RATs is outlined in Sect.~\ref{sect:RATD}. In Sect.~\ref{sect:NumericalSetup} we discus the numerical implementation of the set of equations that governs the internal aggregate dynamics under rapid rotation. In Sect.~\ref{sect:ResultsDiscussion} we present and discuss our N-body simulation results. Finally, in Sect.~\ref{sect:Summary} we summarize our findings.
%\cite{DraineWeingartner1996}
%The  excitation  of  interstellar is  a  long-standing  problem  in  astrophysics,  which remains to be solved today.
\begin{table*}[ht!]
\centering
 \begin{tabular}{|ccccc|c|} 
 \hline
   $\mathrm{Mg/O}$  &  $\mathrm{Si/O}$ & $\mathrm{Fe/O}$ & $\mathrm{Mg/Fe}$ &  $\mathrm{(Mg+Fe)/Si}$ & References \\
   \hline
  0.31     & 0.26    & 0.15    & 2.03 & 1.82 & \cite{Min2007} \\
     X     & X    & X    & 1.09 & 2.25 & \cite{Voshchinnikov2010} \\
   0.23    & 0.19 & 0.19 & 1.25 & 2.25 & \cite{Compiegne2011} \\
   0.19    & 0.15 & 0.17 & 1.07 & 2.34 & \cite{Hensley2021} \\
   0.23  & 0.27 & 0.21 & 1.08 & 1.60 & this work\\
 \hline
\end{tabular}
\caption{ Ratio of different element abundances as observed in the ISM in comparison with the composition of the co-S material applied in this work. }
\label{table:abundance}
\end{table*}
\begin{figure*}[ht!]
     \begin{center}
     \begin{tabular}{ | c | c | c | c | }
     \hline % \diagbox{$a_{\mathrm{eff}}$}{$D_{\mathrm{f}}$}
       & $a_{\mathrm{eff}}=200\ \mathrm{nm}$ & $a_{\mathrm{eff}}=350\ \mathrm{nm}$ & $a_{\mathrm{eff}}=500\ \mathrm{nm}$ \\

			\hline
			\raisebox{-\totalheight}{$\mathrm{BA}$}

			&
      \raisebox{-\totalheight}{\vspace{-10mm}\includegraphics[width=0.28\textwidth]{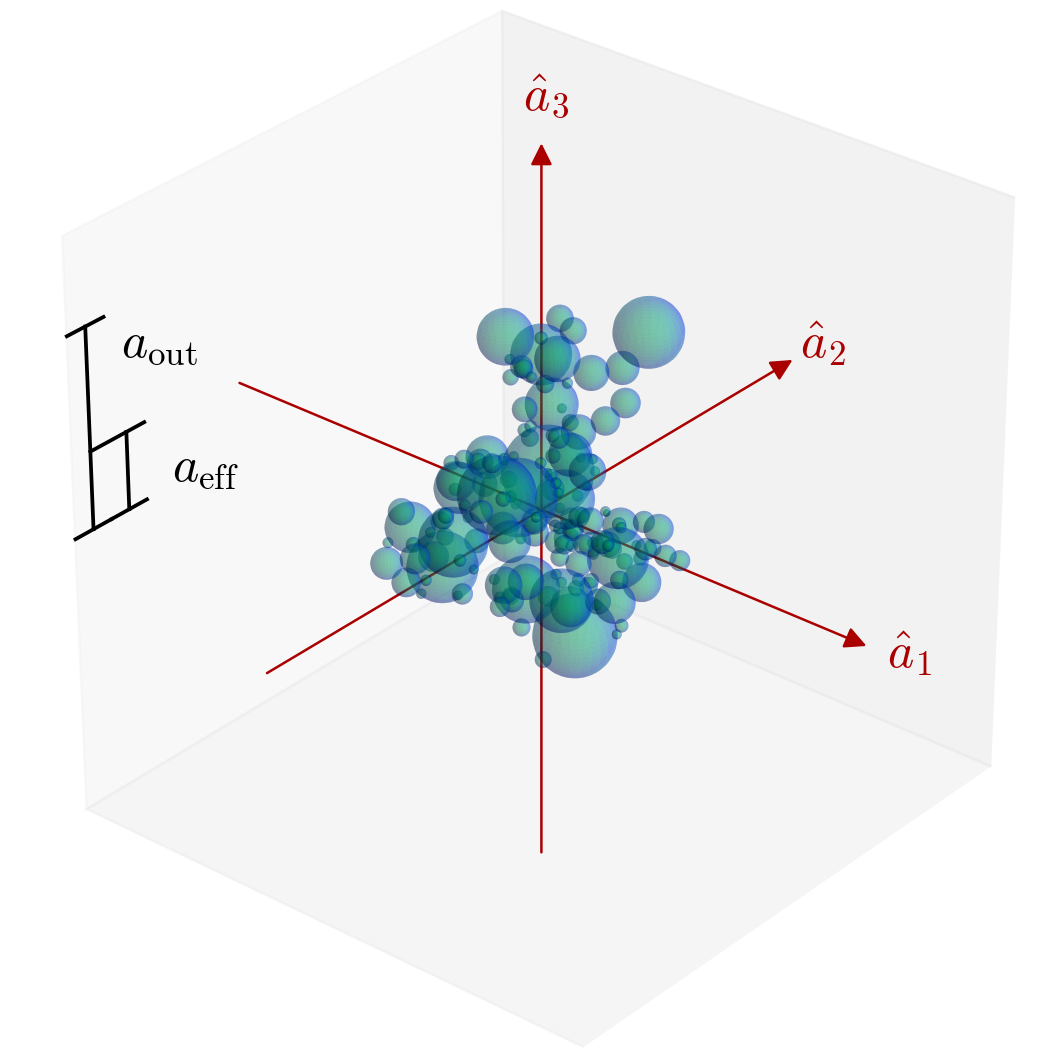}}

      & 
      \raisebox{-\totalheight}{\vspace{-10mm}\includegraphics[width=0.28\textwidth]{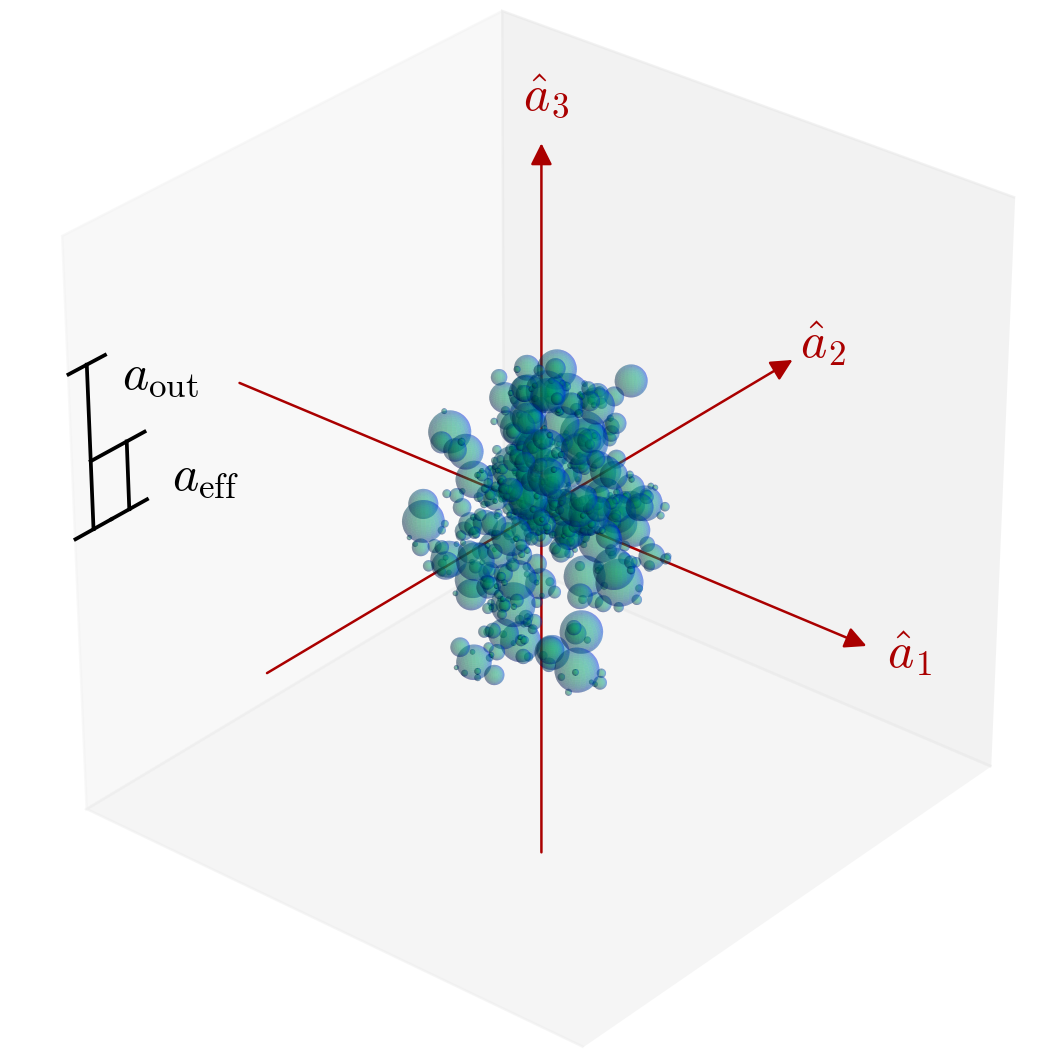}}
      & 
      \raisebox{-\totalheight}{\vspace{-10mm}\includegraphics[width=0.28\textwidth]{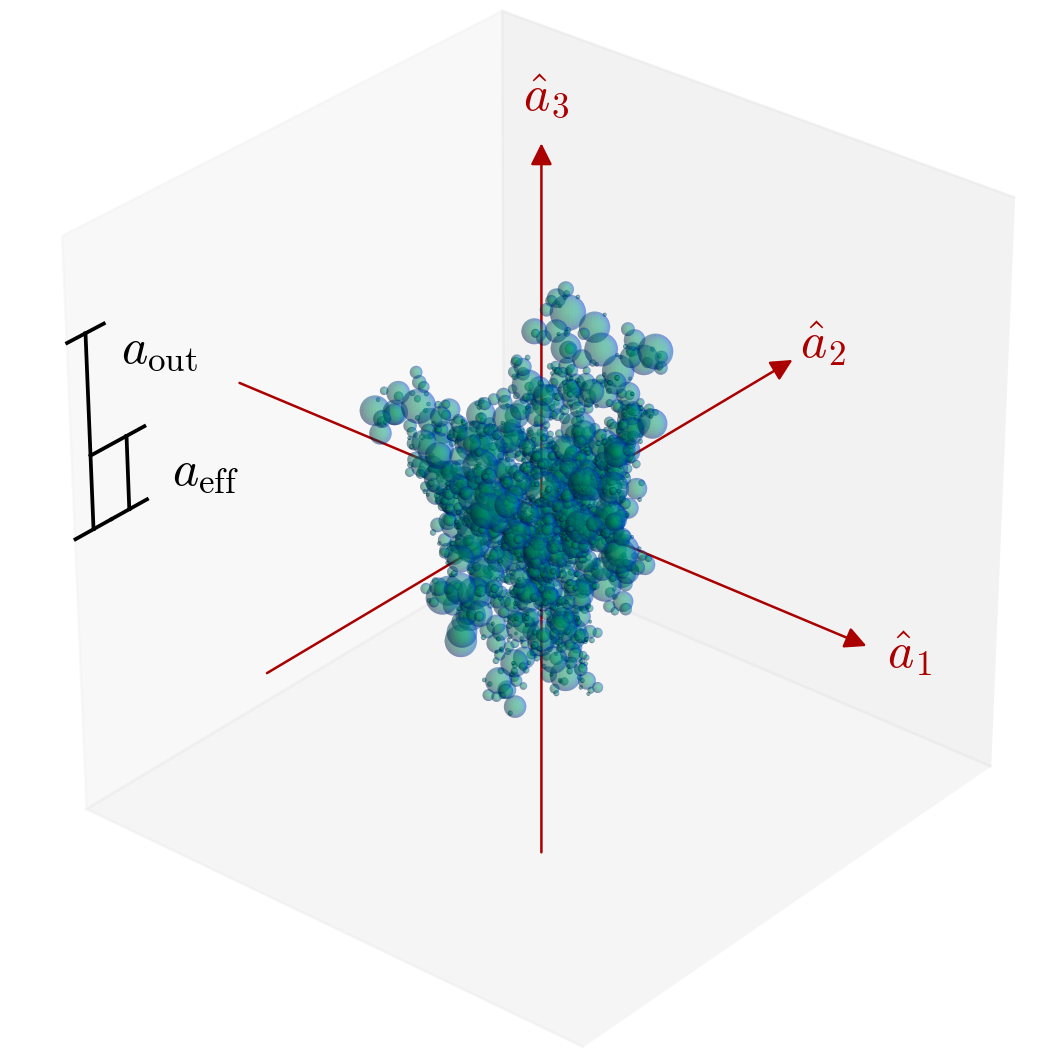}} \vspace{-1mm}
      \\ 
			& 
			$a_{\mathrm{out}}=543.1\ \mathrm{nm}$  & 
			$a_{\mathrm{out}}=882.6\ \mathrm{nm}$  & 
			$a_{\mathrm{out}}= 1385.7\ \mathrm{nm}$   \\
			
		    & 
			$N_{\mathrm{mon}}=160$, $N_{\mathrm{con}}=162$ & 
            $N_{\mathrm{mon}}=521$, $N_{\mathrm{con}}=521$  & 
		    $N_{\mathrm{mon}}=1773$, $N_{\mathrm{con}}=1772$\\
		    
		    %& 
			%$N_{\mathrm{con}}=162$ & 
            %$N_{\mathrm{con}}=521$  & 
		    %$N_{\mathrm{con}}=1772$\\
			
	\hline
\raisebox{-\totalheight}{$\mathrm{BAM1}$}
			&
      \raisebox{-\totalheight}{\vspace{-10mm}\includegraphics[width=0.28\textwidth]{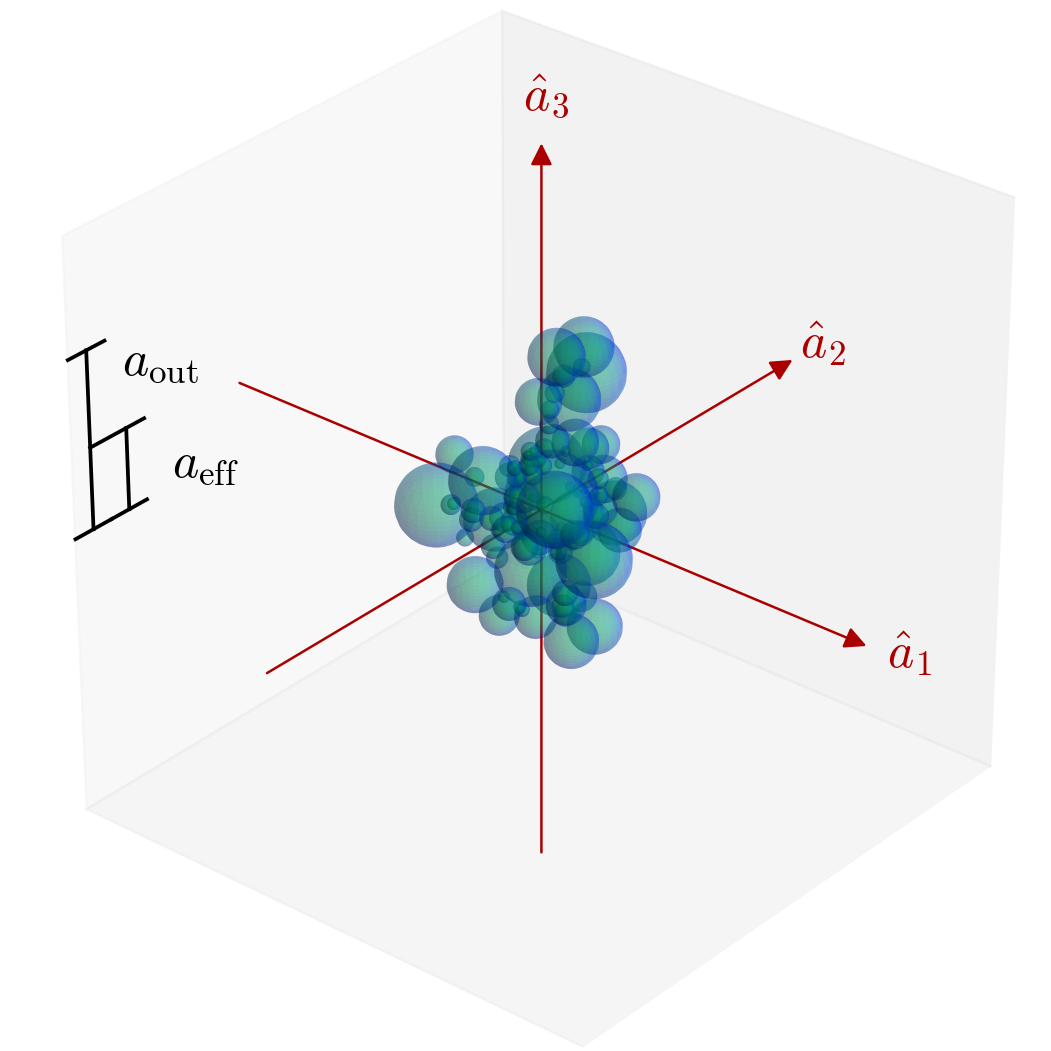}}

      & 
      \raisebox{-\totalheight}{\vspace{-10mm}\includegraphics[width=0.28\textwidth]{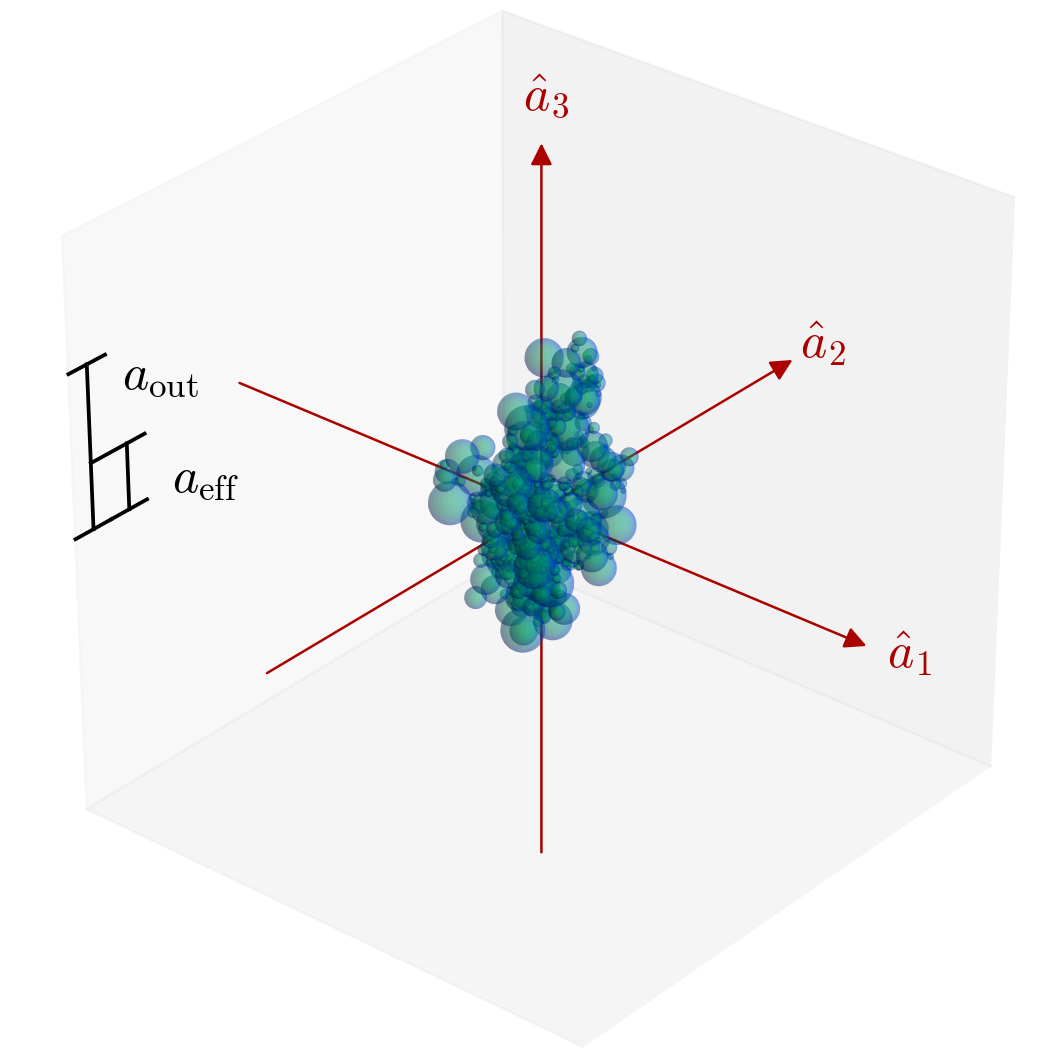}}
      & 
      \raisebox{-\totalheight}{\vspace{-10mm}\includegraphics[width=0.28\textwidth]{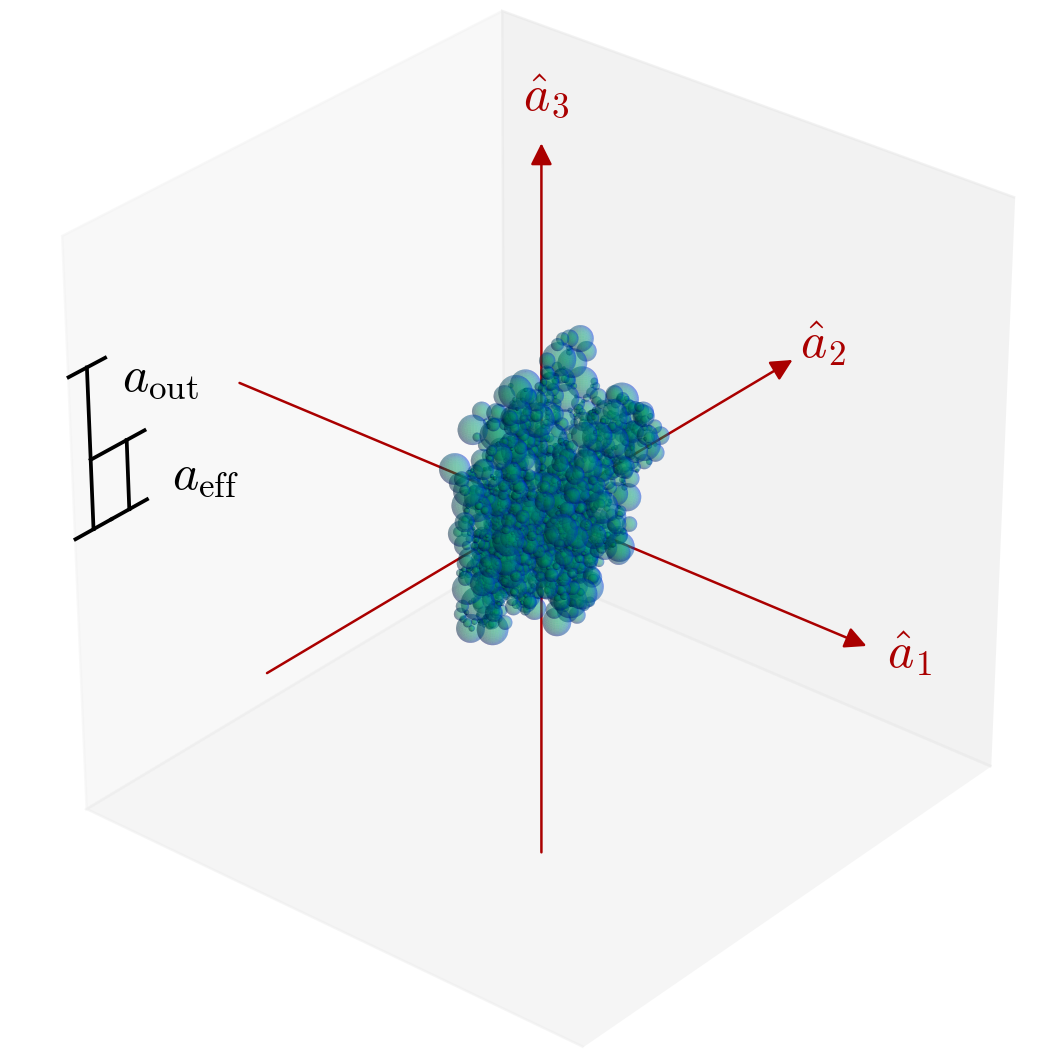}} \vspace{-1mm}
      \\
			& 
			$a_{\mathrm{out}}=457.4\ \mathrm{nm}$  & 
			$a_{\mathrm{out}}=901.2\ \mathrm{nm}$  & 
			$a_{\mathrm{out}}= 1214.7\ \mathrm{nm}$   \\
			
		    & 
			$N_{\mathrm{mon}}=117$, $N_{\mathrm{con}}=235$ & 
            $N_{\mathrm{mon}}=636$, $N_{\mathrm{con}}=1275$  & 
		    $N_{\mathrm{mon}}=1885$, $N_{\mathrm{con}}=3774$\\
		    
		    %& 
			%$N_{\mathrm{con}}=235$ & 
            %$N_{\mathrm{con}}=1275$  & 
		    %$N_{\mathrm{con}}=3774$\\
      \hline
\raisebox{-\totalheight}{$\mathrm{BAM2}$}
			&
      \raisebox{-\totalheight}{\vspace{-10mm}\includegraphics[width=0.28\textwidth]{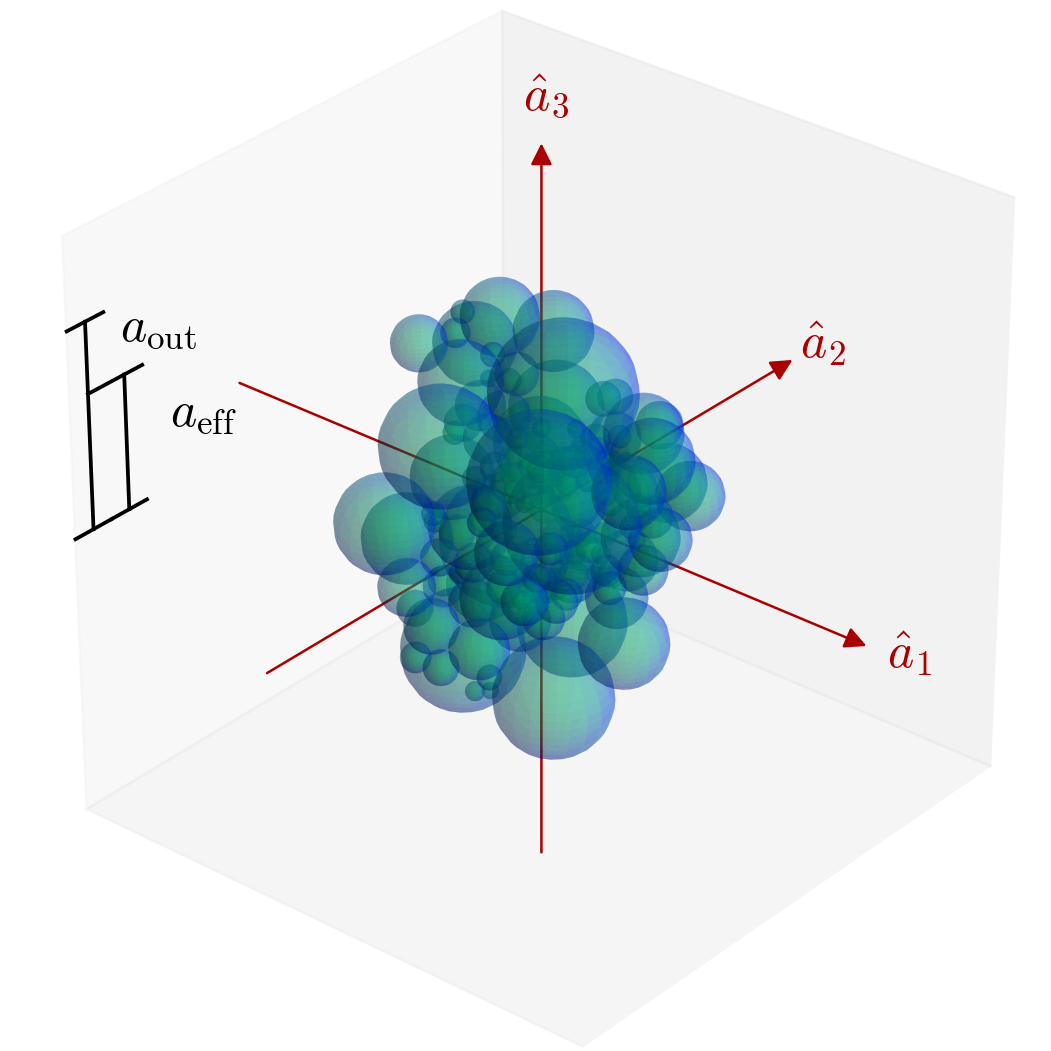}}

      & 
      \raisebox{-\totalheight}{\vspace{-10mm}\includegraphics[width=0.28\textwidth]{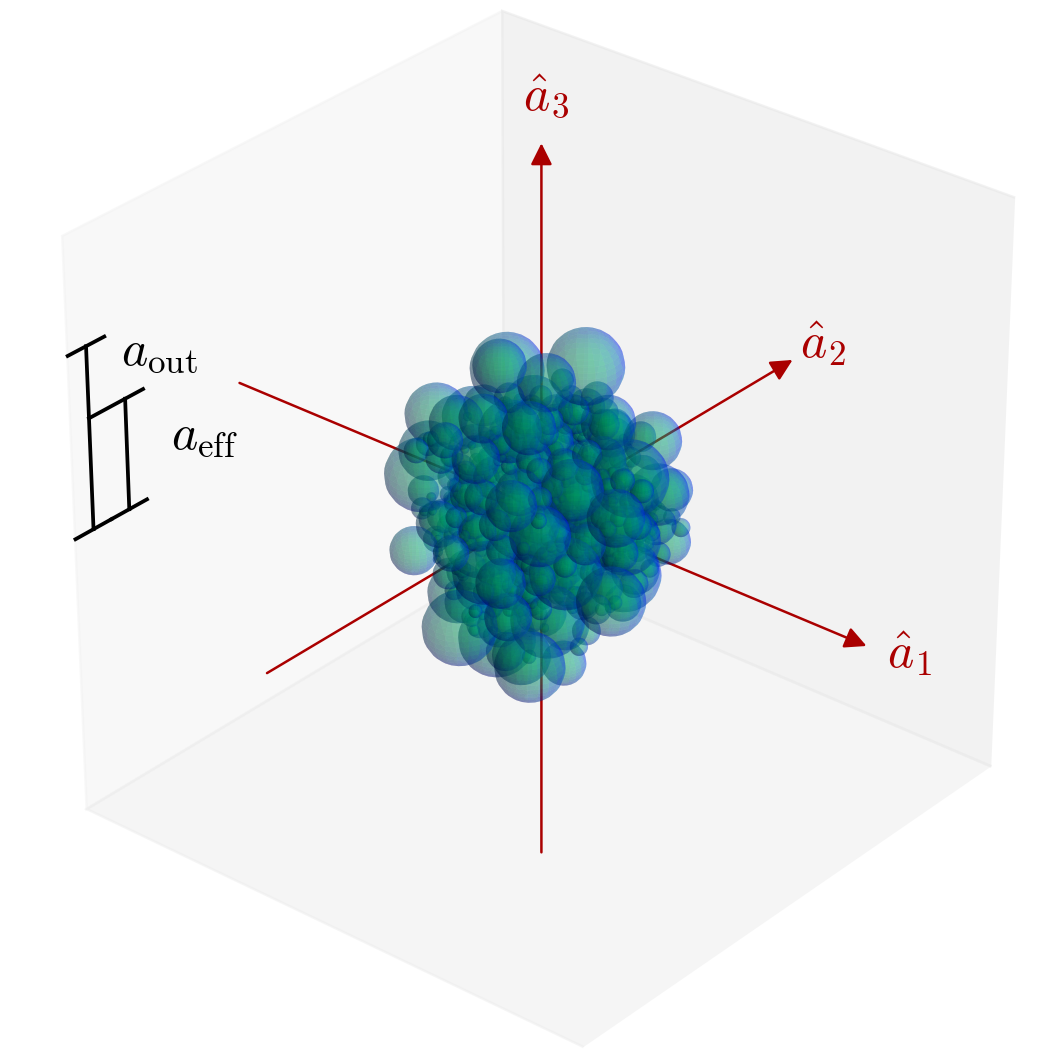}}
      & 
      \raisebox{-\totalheight}{\vspace{-10mm}\includegraphics[width=0.28\textwidth]{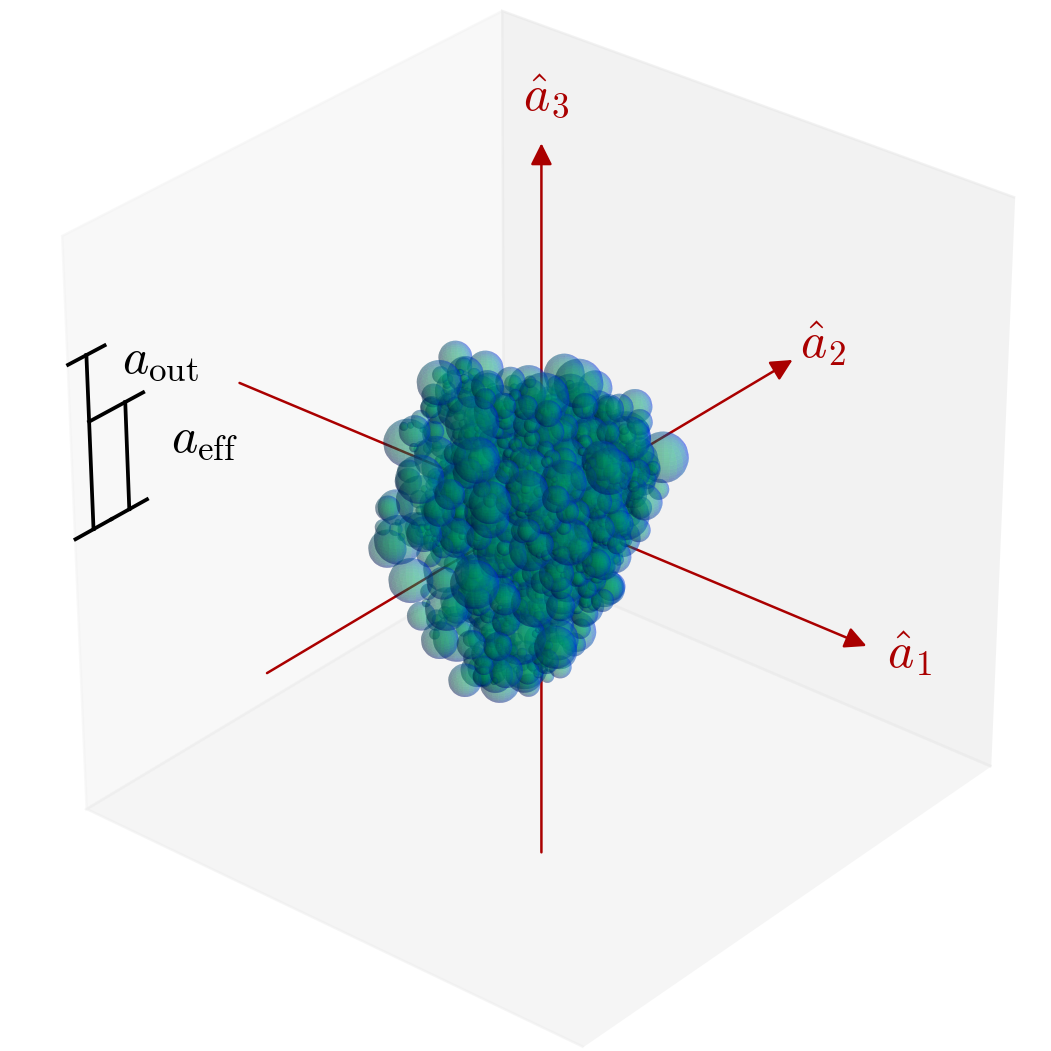}} \vspace{-1mm}
      \\
			& 
			$a_{\mathrm{out}}=320.1\ \mathrm{nm}$  & 
			$a_{\mathrm{out}}=603.3\ \mathrm{nm}$  & 
			$a_{\mathrm{out}}= 845.2\ \mathrm{nm}$   \\
			
		    & 
			$N_{\mathrm{mon}}=157$, $N_{\mathrm{con}}=475$ & 
            $N_{\mathrm{mon}}=649$, $N_{\mathrm{con}}=1950$  & 
		    $N_{\mathrm{mon}}=1814$, $N_{\mathrm{con}}=5546$\\
	    
		    %& 
			%$N_{\mathrm{con}}=475$ & 
            %$N_{\mathrm{con}}=1950$  & 
		    %$N_{\mathrm{con}}=5546$\\
			\hline
      \end{tabular}
      \vspace{0.0cm}
      \caption{Exemplary selection from the total ensemble of porous BA (top row), BAM1 (moddle row), and BAM2 (bottom row) aggregates for the effective grain radii $a_{\mathrm{eff}}=200\ \mathrm{nm}$ (left column), $a_{\mathrm{eff}}=350\ \mathrm{nm}$ (middle column), and $a_{\mathrm{eff}}=500\ \mathrm{nm}$ (right column) with the corresponding  numbers of monomers $N_{\mathrm{mon}}$ and neighbourhood connections $N_{\mathrm{con}}$. Monomers are sampled to guarantee an exact effective radius $a_{\mathrm{eff}}$. The radius $a_{\mathrm{out}}$ is associated to the smallest sphere enclosing the entire aggregate. The  axes  $\hat{a}_{\mathrm{1}}$, $\hat{a}_{\mathrm{2}}$, and $\hat{a}_{\mathrm{3}}$ are defined by the grain's moments of inertia ${ I_{\mathrm{a1}} > I_{\mathrm{a2}} > I_{\mathrm{a3}} }$, where $\hat{a}_{\mathrm{1}}$ is the designated axis of grain rotation. }
      \label{fig:Grains}
      \end{center}
\end{figure*}
\begin{figure*}[ht!]
	\begin{center}
	\includegraphics[width=0.47\textwidth]{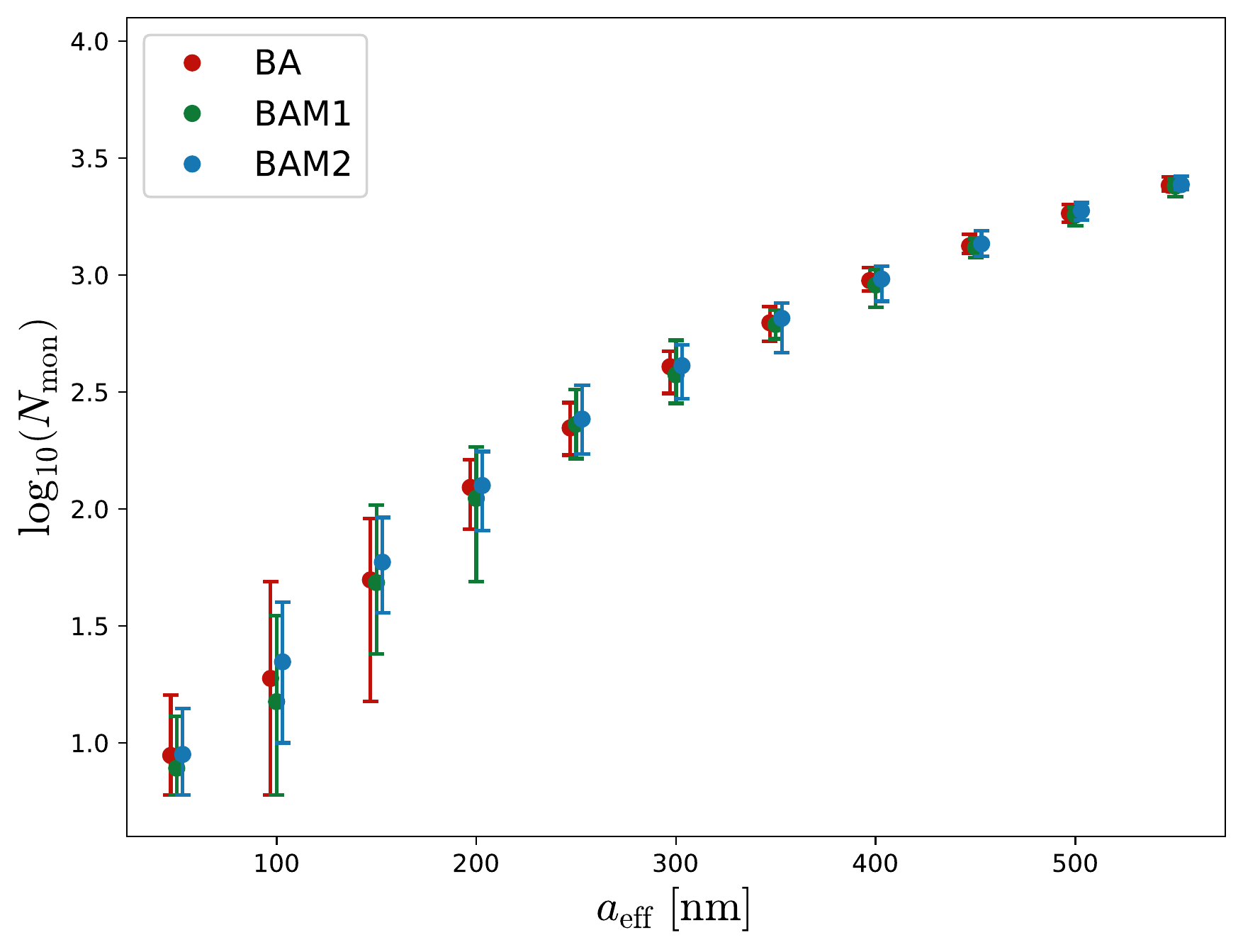}
	\includegraphics[width=0.47\textwidth]{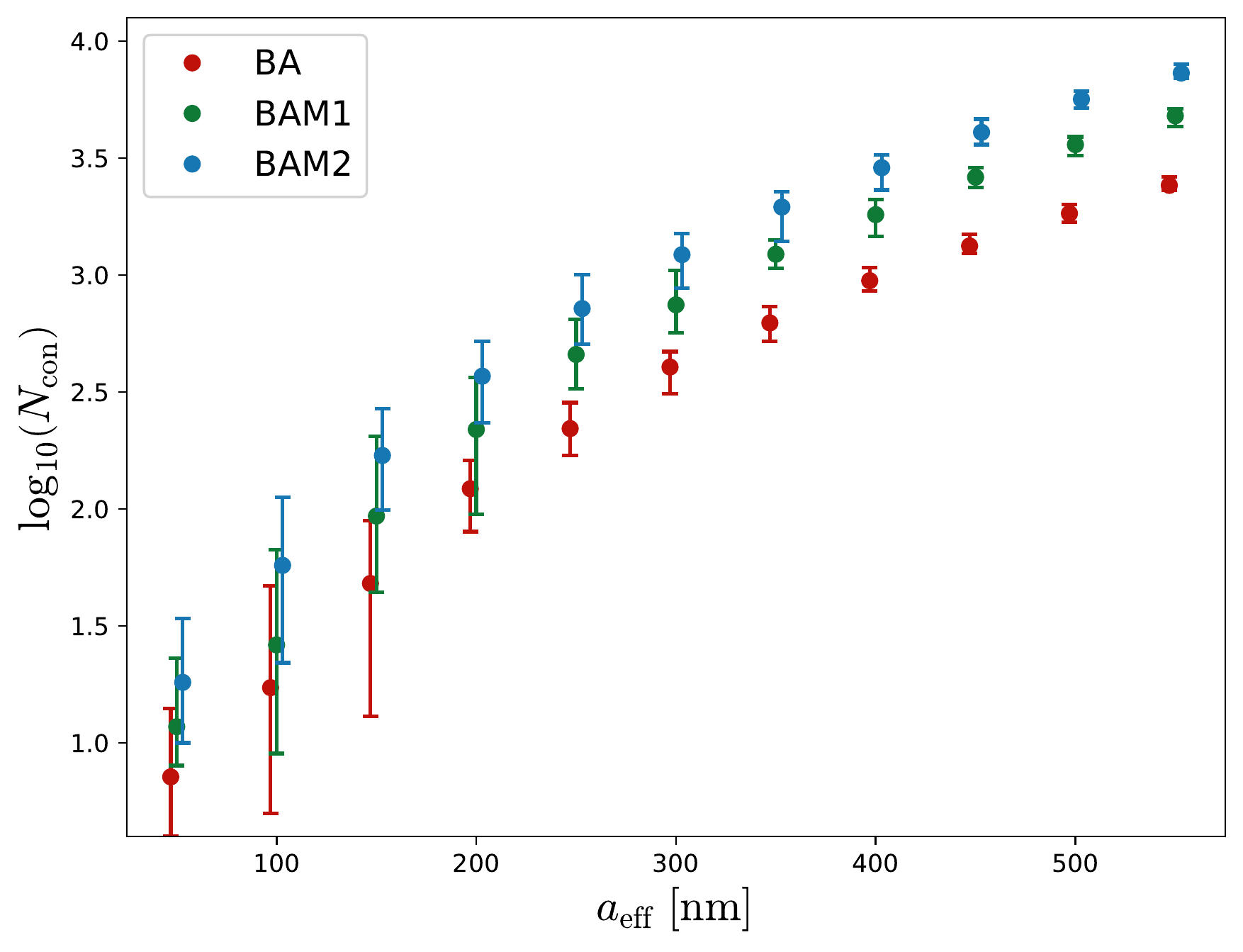}
	\end{center}
	
	\begin{center}
	\includegraphics[width=0.47\textwidth]{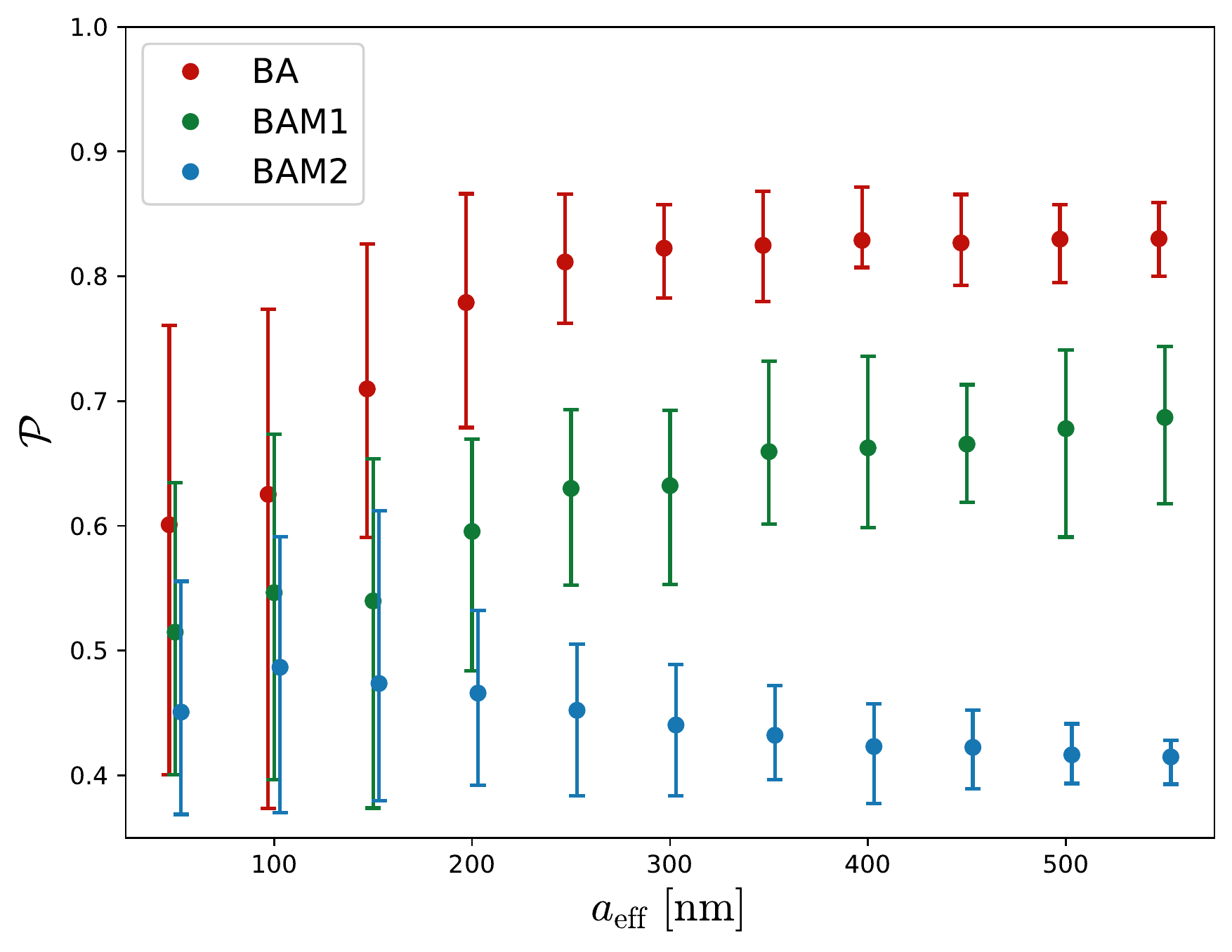}
	\includegraphics[width=0.47\textwidth]{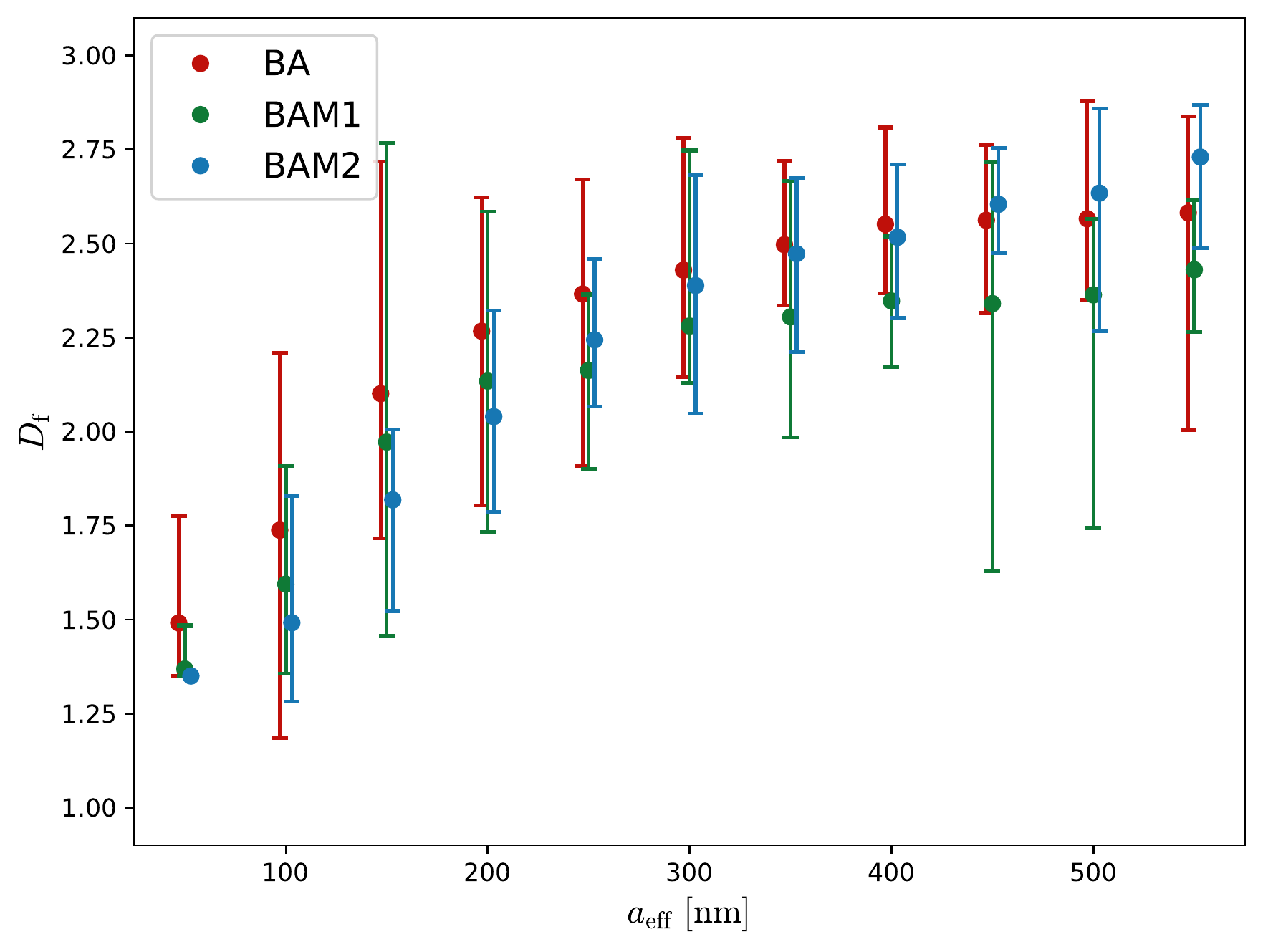}
	\end{center}
\caption{The distribution of the characteristic quantities of total number of monomers $N_{\mathrm{mon}}$ (top left), number of connected neighbours $N_{\mathrm{con}}$ (top right), porosity $\mathcal{P}$ (bottom left), and fractal dimension $D_{\mathrm{f}}$ (bottom right), for all BA (red), BAM1 (green), and BAM2 (blue) aggregates, respectively, dependent on effective radius $a_{\mathrm{eff}}$. Dots are the ensemble average over all shapes and materials while vertical bars represent the minima and maxima. Note that the data points have a small offset for better visibility.}
\label{fig:properties}
\end{figure*}
\section{Dust grain composition}
\label{sect:GrainComposition}
The exact composition of dust remains an open question, with a large variety of observed materials and sizes within the galaxy being similarly possible. Interstellar dust is usually modelled by grains with a spectrum of different sizes with silicate and carbonaceous components \citep[][]{Mathis1977, WeingartnerDraine2001X,Zhukovska2008,Voshchinnikov2010,Guillet2018}. Small spherical monomers condensate by depletion of  the most abundant elements  C, Mg, Si, Fe, and O from their immediate surrounding \citep[see e.g.][]{Kim2021}. Elements such Ti, Al, S Ca, Ni, respectively, are less abundant and may only contribute a few percent of the total dust mass \citep{Hensley2021}. The  abundance of molecules within the grains may be determined by observing the spectral dust absorption features. Silicate minerals from the olivine and pyroxene group are the most likely candidates to be present in order to account for the observed characteristic features \citep[][]{Mathis1977,Wada1999,LiDraine2001,Hensley2021}. 
However, it remains yet inconclusive if these minerals form dust in a crystalline or amorphous structure and to what extend iron is present in its pure form \citep[][]{DraineLi2007,Rogantini2019,DoDuy2020}. Carbonaceous grains may consist of a regular graphite lattice or amorphous structures and be partly hydrogenated \citep[][]{Wada1999,Goto2003,Mennella2006}. Possibly, carbonaceous and silicate grains are not even separate dust populations but baked into a single composite material \citep[][]{Draine2021}.\\
For mimicking the optical properties of dust grains models based on the refractive indices of various materials are developed and the data is publicly available\footnote{For the interested reader we refer to the website e.g. of Bruce~Draine {https://www.astro.princeton.edu/~draine/dust/dust.diel.html} and the THEMIS model
{https://www.ias.u-psud.fr/themis/}} \citep[see e.g.][]{DraineLi2007,Jones2012,Draine2021}. However, complementary well constrained models of the mechanical properties particularly for the composition of interstellar dust grain materials are still missing. Commonly, grain analogs consisting exclusively of icy or pure quartz ($\mathrm{SiO_2}$) materials are utilized to simulate grain growth processes \citep[e.g. ][]{Dominik1997,Wada2007,Seizinger2012}. This is simply due to the fact that quartz is an easily available material on the market with well constrained properties by laboratory experiments \citep{Kendall1987,Heim1999,Israelachvili2011,Krijt2013}.\\
In this paper, however, we explore the rotational disruption of three distinct grain materials labeled a-C, q-S, and co-S, respectively. Carbonaceous grains are represented by the mechanical parameters of amorphous carbon (a-C) and for compression silicate grains are considered built of pure quartz (q-S). In addition, we aim to approximate the mechanical properties of composite silicate (co-S) grains more precisely. Here, we assume the minerals forsterite ($\mathrm{Mg_2 SiO_4}$) and fayalite   ($\mathrm{Fe_2 SiO_4}$) of the olivine series as well as enstatite ($\mathrm{MgSiO_3}$) and  ferrosilite ($\mathrm{FeSiO_3}$) of the pyroxene series to be among the most likely ingredients of silicate grains \citep[][]{Petrovic2001,Zolensky2006,Zhukovska2008,Gail2009,Takigawa2012,Min2007,Kimura2015,Fogerty2016,Hoang2019,Escatllar2019,Kimura2020,Hensley2021,Draine2021}. In order to get an approximation of the material mixture, we match the abundance of individual elements within the considered minerals with the abundance of elements typical for the ISM \citep[][]{Min2007,Voshchinnikov2010,Compiegne2011,Hensley2021}.\\
In Table \ref{table:abundance} we present the relative abundances of elements within the ISM in comparison with the composition of our co-S model.  For our best fit co-S model we get that each individual silicate monomer consist of a mixture of $31\ \%$ forsterite, $29\ \%$ fayalite, $20\ \%$ enstatite, $20\ \%$ ferrosilite, respectively, and average the mechanical material properties accordingly. The abundance of elements within the co-S agrees well with the overall observations within the Milky Way but our co-S model shows a slight overabundance of Si. Naturally, the composition of co-S may locally be vastly different e.g. in the vicinity of oxygen or carbon rich AGB stars where dust grains are newly formed \cite[][]{Zhukovska2013}. Thus, we remain agnostic concerning the actual composition of interstellar dust but consider our co-S model to be an improvement compared to simulations using pure quartz monomers. 
% may be use packing index The role of fragment shapes in the simulations of asteroids as gravitational aggregates F. Ferrar in future studies
\section{Dust grain growth and aggregation}
\label{sect:GrainGrowth} %, and potentially self-gravity,
Dust grain aggregates may grow by ballistic hit-and-stick processes of monomers onto a grain's surface. This process is usually called ballistic particle-cluster aggregation (BCPA) \citep[][]{Kozasa1992,Bertini2007}. Newly formed grains may then grow to even larger aggregates by ballistic cluster-cluster aggregation (BCCA) \cite{Ossenkopf1993}. Such grain-grain collisions result significantly compressed aggregates and, subsequently, gas pressure may compress an aggregate even more \citep[see e.g][and references therein]{Dominik1997,Kataoka2013,Michoulier2022}.\\
In our study, we model such compression effects by the ballistic aggregation with migration (BAM) model introduced in \cite{Shen2008}. Here, grains simply grow by means of ballistic aggregation (BA) \footnote{BA and BCPA are used synonymous in literature.} of monomers hitting the aggregate from random directions. In the BAM model grain model monomers may migrate along the surface once  (BAM1) or twice (BAM2). Consequently, for BAM1 each monomer has
at least two connections with the aggregate and for BAM2 each
monomer has at least three \cite{Shen2008,Seizinger2013}.\\
Commonly, dust aggregation models utilize only a constant monomer size \citep[see e.g.][]{Kozasa1992,Shen2008,Wada2007,Bertini2007,Seizinger2012}. However, it seems unlikely that a condensation process of elements in nature would lead to exactly one monomer size. In fact, numerous laboratory experiments clearly indicate that grown aggregates consist of monomers with a variable radius \citep{Karasev2004,Chakrabarty2007,Slobodrian2010,Kandilian2015,Salameh2017,Paul2017,Baric2018,Kelesidis2018,Bauer2019,Wu2020,Zhang2020,Kim2021}. In this study we focus on a polydisperse system of monomers, where the exact distribution of the monomer radii within an aggregate may be approximated by a log-normal distribution \citep{Koylu1994,Lehre2003,Slobodrian2010,Bescond2014,Kandilian2015,ChaoLiu2015,Bauer2019,Wu2020,Zhang2020}.\\
We realize the BAM model with a Monte-Carlo approach in order to create an ensemble of pre-calculated grain analogs resembling the observed parameters of dust in the circumstellar and interstellar medium. The radius of the i-th monomer $a_{\mathrm{mon,i}}$ is sampled from a range of $a_{\mathrm{mon,i}}\in[10\ \mathrm{nm}, 100\ \mathrm{nm}]$. The log-normal distribution has a typical average of ${ \left< a_{\mathrm{mon}}  \right> = 20\ \mathrm{nm} }$ and a standard deviation of $1.65\ \mathrm{nm}$. Successively, each newly sampled monomer is shot on a random trajectory into the simulation domain until it hits the aggregate. For simplicity we assume that each monomer sticks onto the surface when colliding.  
For BAM1 and BAM2, respectively, the i-th monomer migrates along the  the surface of the initially hit monomer in a random direction to establish additional connections. In order to create an aggregate in equilibrium the monomer position $\vec{X}_{\mathrm{i}}$ is corrected in such a way that each overlap between connected monomers agrees with the material dependent equilibrium compression length $\delta_{\mathrm{0}}$ (see Sect. \ref{sect:MonomerContact} for details). The aggregation process is repeated until the dust aggregate reaches a certain volume of
\begin{equation}
    V_{\mathrm{agg}} =  \frac{4\pi}{3} \sum_{\mathrm{i}=1}^{N_{\mathrm{mon}}}   a_{\mathrm{mon,i}}^3\, .
\end{equation}
Ballistic aggregates are usually quantified by the total number of monomers $N_{\mathrm{mon}}$ \citep[see e.g.][]{Wada2007,Shen2008,Seizinger2012}.  However, dust observations are tightly connected to the effective size of the grains \citep[][]{Mathis1977,WeingartnerDraine2001X}. Hence, in our study we rather control for an exact effective radius of
\begin{equation}
    a_{\mathrm{eff}} = \left(  \frac{3 V_{\mathrm{agg}}}{4\pi}  \right)^{\frac{1}{3}}
\end{equation}
by an biased sampling of the last three monomer radii instead of getting the grain size indirectly from ${ a_{\mathrm{eff}}\approx N_{\mathrm{mon}}^{1/3} \left< a_{\mathrm{mon}}  \right> }$. Finally, we calculate the inertia tensor for each aggregate in order to determine the characteristic moments of inertia ${ I_{\mathrm{a1}} > I_{\mathrm{a2}} > I_{\mathrm{a3}} }$, along the unit vectors $\hat{a}_{\mathrm{1}}$, $\hat{a}_{\mathrm{2}}$, and $\hat{a}_{\mathrm{3}}$ (we refer t to RMK22 for the exact procedure).\\
In order to connect the rotational disruption of the grain ensemble to distinct quantities associated with the internal structure of fluffy dust grains we introduce the porosity $\mathcal{P}$, volume filling factor $\phi$, and fractal dimension $D_{\mathrm{f}}$ as well as the semi major axes $a<b<c$ unique for each individual aggregate.\\
The porosity quantifies the empty space within an aggregate where $\mathcal{P}=0$ is for a solid object and $\mathcal{P}=1$ for vacuum. Its value for an individual object depends on the definition of the surface that envelopes the aggregate. In our study we follow the procedure of determining $\mathcal{P}$ by utilizing the moments of inertia of an aggregate as outlined in \cite{Shen2008}. Here, the quantity
\begin{equation}
    \alpha_{\mathrm{i}} = \frac{5}{4\pi}\frac{I_{\mathrm{ai}}}{ \rho_{\mathrm{mat}} a_{\mathrm{eff}}^5  }
\end{equation}
is the ratio of the moment of inertia $I_{\mathrm{ai}}$ to that of an sphere with equivalent volume where the index ${\mathrm{i}  \in \{1,2,3\}   }$ denotes the three spatial directions. The corresponding semi major axes are
\begin{equation}
    a=a_{\mathrm{eff}}\sqrt{ \alpha_{\mathrm{2}} +\alpha_{\mathrm{3}} - \alpha_{\mathrm{1}}   }\, ,
\end{equation}
\begin{equation}
    b=a_{\mathrm{eff}}\sqrt{ \alpha_{\mathrm{3}} +\alpha_{\mathrm{1}} - \alpha_{\mathrm{2}}  }\, ,
\end{equation}
and
\begin{equation}
    c=a_{\mathrm{eff}}\sqrt{ \alpha_{\mathrm{1}} +\alpha_{\mathrm{2}} - \alpha_{\mathrm{3}}   }\, ,
\end{equation}
respectively. Finally, the porosity of an aggregate may then be written as 
\begin{equation}
    \mathcal{P}=1-\frac{a_{\mathrm{eff}}^3}{abc}\, ,
\end{equation}
whereas the complementary quantity $\phi=1-\mathcal{P}$ is the volume filling factor \cite{Shen2008}.\\
The fractal dimension is a measure of the shape of an aggregate where $D_{\mathrm{f}}=1$ represents a one dimensional line and $D_{\mathrm{f}}=3$ a compact sphere. We determine the fractal dimension $D_{\mathrm{f}}$ of each dust aggregate by the correlation function 
\begin{equation}
	C\chi^{D_{\mathrm{f}}-3} = \frac{n(\chi)}{4\pi \chi^2 l N_{\mathrm{mon}}}
\end{equation}
as outlined in \cite{Skorupski2014}. Here, $C$ is a scaling factor, $\chi$ is the distance from the center of mass, $l$ is a length with $l\ll a_{\mathrm{eff}}$, and $n(\chi)$ is the number density of connected monomers within the shell ${[\chi-l/2;\chi+l/2]}$.\\
We create dust grains with effective radii in the range ${ a_{\mathrm{eff}} = 50\ \mathrm{nm} - 550\ \mathrm{nm} }$ in steps of $50\ \mathrm{nm}$. For each $a_{\mathrm{eff}}$ we repeat the MC dust growth process with 30 random seeds for the BA, BAM1, and BAM2 configurations and the a-C, q-S, co-S materials. In total we pre-calculate an ensemble of 2970 individual grains as input for our three-dimensional N-body simulations.\\
We note that the grain growth by BA may heavily be impacted by grain charge \cite[][]{Matthews2012}, high impact velocities \citep[][]{Dominik1997,Ormel2009}, a preferential impact direction by means of grain alignment with the magnetic field \citep[][]{LazarianHoang2007,Hoang2022} or a gas-dust  drift (RMK22).  Hence, the resulting shapes in our grain ensemble may not be representative when compared for grain growth processes e.g. in 
the vicinity of AGB stars \citep[][]{Zhukovska2013} or in protostellar envelopes \citep[][]{Galametz2019}. However, our grain model cover a sufficiently broad variety of grain shapes to allow conclusions about their rotational stability even though each individual shape is not equally likely to be realize in nature.\\
An exemplary selection of the grains is shown in Fig.~\ref{fig:Grains}. In Fig.~\ref{fig:properties} we present the characteristic quantities of the entire grain ensemble as introduced above. The number of monomers $N_{\mathrm{mon}}$ as well as the number of connections $N_{\mathrm{con}}$ within each aggregate increase with effective radius $a_{\mathrm{eff}}$. By design BA grains have generally less connections compared to BAM grains. Compared to the BAM grains with a fixed monomer size  \citep[][]{Shen2008} the porosity $\mathcal{P}$ is not strictly increasing but stagnates for higher  $a_{\mathrm{eff}}$ because smaller monomers may easily migrate towards the center making our aggregates overall more compact. The fractal dimension $D_{\mathrm{f}}$ shown in Fig.~\ref{fig:properties} increases slightly towards larger grains but the resulting grain BA, BAM1, and BAM2 grains of different sizes shapes in this work are not well correlated with the fractal dimension $D_{\mathrm{f}}$. This is in contrast to the dust models of RMK22 where the grains are explicitly constructed to get an exact pre-determined  $D_{\mathrm{f}}$ and are not to be compared with the BAM grains as depicted in Fig.~\ref{fig:Grains}.

\section{Inter-monomer contact effects}
\label{sect:MonomerContact}
\begin{figure*}[ht!]
	\begin{center}
	\includegraphics[width=0.99\textwidth]{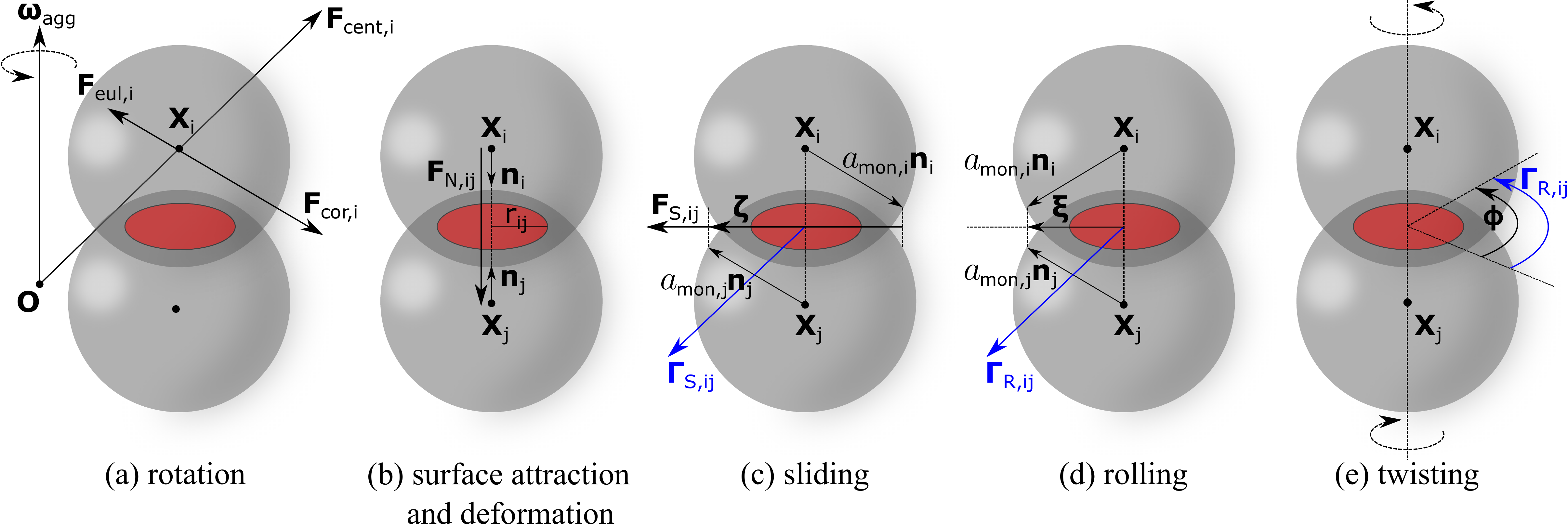}
	\end{center}
\caption{Schematic representation of the forces and torques acting in between the i-th and the j-th monomer at the positions $\vec{X}_{\mathrm{i}}$ and $\vec{X}_{\mathrm{j}}$, respectively. The common contact surface with radius $r_{\mathrm{ij}}$ is shaded in red.~~(a) Each individual monomer experiences external centrifugal $\vec{F}_{\mathrm{cent}}$, Coriolis $\vec{F}_{\mathrm{cor}}$, and Euler forces $\vec{F}_{\mathrm{eul}}$  as a result of the aggregate's accelerated rotation with angular velocity $\vec{\omega}_{\mathrm{agg}}$. ~~(b) The normal force $\vec{F}_{\mathrm{N,ij}}$ acts normal to the contact surface of two monomers because of surface attraction and mechanical deformation with the contact pointers remaining anti-parellel ${ \vec{n}_{\mathrm{i}} = -\vec{n}_{\mathrm{j}} }$. ~~(c) The sliding force $\vec{F}_{\mathrm{S,ij}}$ is parallel to the sliding displacement $\zeta$. Both $\vec{F}_{\mathrm{S,ij}}$ and the corresponding sliding torque $\vec{\Gamma}_{\mathrm{S,ij}}$ work tangential to the contact surface. ~~(d) The same for the rolling of monomers withing the aggregate: The rolling displacement $\vec{\zeta}$, the force $\vec{F}_{\mathrm{R,ij}}$, and the $\vec{\Gamma}_{\mathrm{R,ij}}$ are tangential to the contact surface. ~~(e) The twisting of monomers in contact is not associated with a net force. The resulting torque $\vec{\Gamma}_{\mathrm{T,ij}}$ points is parallel to the normal vector $\vec{\phi}$ of the twisting.  We note that the depicted quantities are not to scale since $r_{\mathrm{ij}}\ll a_{\mathrm{mon}}$.}
\label{fig:scetch}
\end{figure*}
\begin{table*}[]
\centering
 \begin{tabular}{|c|c|c|ccccc|} 
 \hline
  & carbon:  & silicate: &  silicate: & forsterite & fayalite & enstatite & ferrosilite\\
  & amorphous (a-C)  & quartz (q-S)  &  composite (co-S)  & $31\ \%$ & $29\ \%$ & $20\ \%$  & $20\ \%$\\
 \hline
 $\gamma\ [\mathrm{mJ\ m}^{-2}]$ & $50_{\ (11)}$ & $20_{\ (1)}$ & 70 & $70_{\ (4)}$ & $70_{\ (4)}$ & $70_{\ (4)}$ & $70_{\ (4)}$\\
 $E\ [\mathrm{GPa}]$ & $168_{\ (9,10)}$ & $54_{\ (1,3)}$ & 169 & $204_{\ (6,8)}$ & $140_{\ (6)}$ & $180_{\ (6)}$ & $142_{\ (7)}$\\
 $\nu$ & $0.16_{\ (9)}$ & $0.17_{\ (1,3)}$ & 0.27 & $0.24_{\ (2,6,8)}$ & $0.32_{\ (2,6)}$ & $0.21_{\ (2,6)}$ & $0.31_{\ (7)}$\\
 $\rho_{\mathrm{mat}}\ [\mathrm{kg\ m}^{-3}]$ & $2368_{\ (9,11)}$ & $2650_{\ (1,3)}$ & 3707 & $3213_{\ (2,5, 14)}$ & $4393_{\ (2,5,14)}$ & $3209_{\ (2,5, 14)}$ & $4014_{\ (5,7,14)}$\\
% $\mathcal{S}_{\mathrm{mon}}\ [\mathrm{MJ\ m}^{-3}]$ & 130 & 13.1 & 19.1 o. 0.95? & 19.1 & X & X & X\\
% $\xi_{crit}\ [\mathrm{nm}]$ & 0.2 & 0.2 & 0.2 & 0.2 & 0.2 & 0.2 & 0.2\\
% $T_{\mathrm{vis}}\ [\mathrm{ps}]$ & 12.5 & 12.5 & 12.5 & 12.5 & 12.5 & 12.5& 12.5\\
 \hline
\end{tabular}
\caption{Material parameters of the surface energy $\gamma$, Young’s modulus $E$, Poisson number $\nu$, and material density $\rho_{\mathrm{mat}}$ for the monomer materials of amorphous carbon (a-C), pure quartz (q-S), and composite silicate (co-S) considered in our N-body simulations. The co-S material is assumed to consist of a mixture of the different minerals of forsterite, fayalite, enstatite, and ferrosilite, respectively. For all the materials we assume a critical rolling displacement of $\xi_{\mathrm{crit}}=0.2\ \mathrm{nm}_{\ (1,3)}$ and a viscous dumping time of  $T_{\mathrm{vis}} = 5\ \mathrm{ps}_{\ (1,12,13)}$.\vspace{3mm}  \\ References: {(1) \cite{Seizinger2012}}, {(2) \cite{Christensen1996}}, {(3) \cite{Dominik1997}}, {(4) \cite{Bogdan2020}}, {(5) \cite{Williams1980}}, {(6) \cite{Petrovic2001}}, {(7) \cite{Mitchell2004}}, {(8) \cite{Gouriet2019}}, {(9) \cite{Jensen2015}}, {(10) \cite{Remediakis2007}}, {(11) \cite{Zebda2008}}, {(12) \cite{Krijt2013}}, {(13) \cite{Seizinger2013}}, {(14) \cite{Cardarelli2008}} }
\label{table:MaterialProperties}
\end{table*}
In this section we outline the forces and torques acting between the monomers of an aggregate in detail. In Fig. \ref{fig:scetch} we provide a schematic illustration of all considered monomer interactions. Two individual monomers in physical contact establish a common contact surface area and experience an attraction because of the van der Waals force. For some materials stronger attractions such as Coulomb forces between charged monomers, dipole-dipole interaction within ices, or metallic binding between iron pallets may become of relevance.\\
The attraction is quantified by the material dependent energy per surface area $\gamma$. Assuming monomers act like elastic spheres with an radius of $a_{\mathrm{mon,i}}$ and $a_{\mathrm{mon,j}}$, respectively,  their elastic deformation causes an repulsive force. An analytical description of these forces was first presented in \cite{Johnson1971} with the so called JKR model \citep[see also][]{Hertz1896} where the equilibrium radius of the contact surface is 
\begin{equation}
    r_{\mathrm{0}} = \left( \frac{9\pi \gamma R^2}{E^*} \right)^{1/3}
\end{equation}
when no external force are acting between monomers i.e. the attractive and repulsive forces are in balance. Here, the quantity ${ R=a_{\mathrm{mon,i}}a_{\mathrm{mon,j}} / (a_{\mathrm{mon,i}}+a_{\mathrm{mon,j}}) }$ is the reduced monomer radius whereas the elastic parameter ${ E^{*}=  E / (2-2\nu^2 ) }$ is determined by the material specific constants of the Poisson number $\nu$ and Young's modulus $E$, respectively \citep[we refer to][for further details]{Johnson1987}. Utilizing that a monomer contact breaks for the characteristic pulling force of 
\begin{equation}
  F_{\mathrm{C}}=3\pi\gamma R\, ,
\end{equation}
as outlined in \citep{Johnson1971}, the normal force along the unit vector 
\begin{equation}
    \vec{n}_{\mathrm{c}} =  \frac{\vec{X}_{\mathrm{i}}-\vec{X}_{\mathrm{j}}}{\left|  \vec{X}_{\mathrm{i}}-\vec{X}_{\mathrm{j}} \right|}\, .
\end{equation}
 may be written as
\begin{equation}
\vec{F}_{\mathrm{N,ij}} = 4 F_{\mathrm{C}} \left[ \left( \frac{r}{r_{\mathrm{0}}} \right)^3 -  \left( \frac{r}{r_{\mathrm{0}}} \right)^{3/2} \right]\vec{n}_{\mathrm{c}} 
\label{eq:Fcontact}
\end{equation}
Consequently, for $F_{\mathrm{N,ij}} > 0$ the i-th monomer is exerting a pushing force onto its j-th neighbour. Otherwise, for $F_{\mathrm{N,ij}} < 0$, the j-th monomer is pulling on the i-th monomer. \\
Later, the JKR model was extended by \cite{Dominik1997}
 considering the mechanics of rolling \citep{Dominik1995,Dominik1997}, sliding \citep{Dominik1996,Dominik1997} , and twisting motions \citep{Dominik1997} in between connected monomers. In order to track the relative motion of monomers over we use the formulation of the contact pointers $\vec{n}_{\mathrm{i}}$ and $\vec{n}_{\mathrm{j}}$ as outlined in \cite{Dominik2002}. These vectors point initially towards the centers of neighboring monomers when a new contact is established (see Fig. \ref{fig:scetch}).\\
The force acting on the i-th monomer because of the sliding motion of the j-th monomer may the be written as 
\begin{equation}
\vec{F}_{\mathrm{S,ij}} = -8 r_{\mathrm{0}} G^{*} \vec{\zeta} \frac{ \left( a_{\mathrm{mon,j}} \vec{n}_{\mathrm{j}} + a_{\mathrm{mon,i}} \vec{n}_{\mathrm{i}}  \right)\vec{n}_{\mathrm{c}}}{\left|  \vec{X}_{\mathrm{i}}-\vec{X}_{\mathrm{j}}  \right|}
\end{equation}
where
\begin{equation}
\vec{\zeta} = a_{\mathrm{mon,i}} \vec{n}_{\mathrm{i}} + a_{\mathrm{mon,j}} \vec{n}_{\mathrm{j}}  - \left( a_{\mathrm{mon,i}} \vec{n}_{\mathrm{i}} \vec{n}_{\mathrm{c}} - a_{\mathrm{mon,j}} \vec{n}_{\mathrm{c}}  \vec{n}_{\mathrm{i}}  \right)\vec{n}_{\mathrm{c}}
\label{eq:zeta}
\end{equation}
is the sliding displacement and 
\begin{equation}
\vec{\Gamma}_{\mathrm{S,ij}} = - 8 r_{\mathrm{0}} G^{*} a_{\mathrm{mon,i}}  \vec{n}_{\mathrm{i}} \times \vec{\zeta}
\end{equation}
is the associated sliding torque \citep[][]{Johnson1987,Dominik1996,Dominik1997}. Here, the material constant 
$ { G^{*}=G / ( 2-2\nu_{\mathrm{i}}^2 ) } $ depends on 
the shear modulus ${G = E/(2 +  2\nu) }$ \citep[see e.g.][]{Cardarelli2008}.\\
In contrast to sliding a rolling motion does not result in a force \citep[see e.g.][]{Dominik1997,Wada2007}  but only in a torque of 
\begin{equation}
\vec{\Gamma}_{\mathrm{R,ij}} = -4 F_{\mathrm{C}} \vec{n}_{\mathrm{i}} \times \vec{\xi}\, ,
\label{eq:xi}
\end{equation}
where the rolling displacement depends on the contact pointers via ${ \vec{\xi} = R(\vec{n}_{\mathrm{i}} + \vec{n}_{\mathrm{j}} ) }$.\\
The same for the twisting between monomers with a torque of
\begin{equation}
\vec{\Gamma}_{\mathrm{T,ij}} = \frac{16}{3} G^{*}r_0^3 \vec{\phi}\, .
\end{equation}
whereas the motion of the twisting follows the direction of the vector
\begin{equation}
\vec{\phi} = \vec{n}_{\mathrm{c}}(t) \int_{0}^t {   \left( \vec{\omega}_{\mathrm{i}}(t') - \vec{\omega}_{\mathrm{j}}(t')  \right)\vec{n}_{\mathrm{c}}(t')\mathrm{d}t'}\, .
\label{eq:phi}
\end{equation}
Here, the angular velocities $\vec{\omega}_{\mathrm{i}}(t)$ and $\vec{\omega}_{\mathrm{j}}(t)$, respectively, describe the relative rotation of two monomers in contact \citep[][]{Dominik1997}. \\
In this study we extend the monomer contact physics pioneered by \cite{Johnson1971}, \cite{Dominik1995,Dominik1996,Dominik1997}, and \cite{Wada2007} by introducing additional forces emerging from the accelerated rotation of the grain aggregate. In the notation of our grain model the centrifugal force acting on each monomer may be write as
\begin{equation}
\vec{F}_{\mathrm{cent,i}} = -m_{\mathrm{mon,i}}\  \vec{\omega}_{\mathrm{agg}} \times \left( \vec{\omega}_{\mathrm{agg}} \times \vec{X}_{\mathrm{i}} \right)
\label{eq:Fcent}
\end{equation}
where ${ m_{\mathrm{mon,i}} = 4\pi/3\ \rho_{\mathrm{mat}} a_{\mathrm{mon,i}}^3 }$ is the mass of the i-th monomer.\\
We emphasize that our numerical setup operates in a co-rotating coordinate system with its origin  coinciding with the center of mass of the most massive fragment. Hence, the Coriolis effect acts as an additional fictive force on each individual monomers via 
\begin{equation}
\vec{F}_{\mathrm{cor,i}} = -2m_{\mathrm{mon,i}}\   \vec{\omega}_{\mathrm{agg}} \times \frac{\mathrm{d} \vec{X}_{\mathrm{i}}}{\mathrm{d} t}\, .
\end{equation}
Furthermore, our aggregates are initially at rest and are gradually spun-up. Consequently, the acceleration of each monomer within the aggregate leads to an Euler force of
\begin{equation}
\vec{F}_{\mathrm{eul,i}} = -m_{\mathrm{mon,i}}  \frac{\mathrm{d} \vec{\omega}_{\mathrm{agg}}}{\mathrm{d} t} \times \vec{X}_{\mathrm{i}}
\end{equation}
acting on each monomer where ${\mathrm{d} \vec{\omega}_{\mathrm{agg}} / \mathrm{d} t  }$ is the angular acceleration.\\
When monomers establish a new contact they start to oscillate around the equilibrium position driven by the surface attraction and monomer deformation. In nature it is expected that the oscillation becomes dampened because the deformation dissipates energy. A weak damping force may be introduced based on the relative velocity of the monomers \citep[][]{Kataoka2013X,Seizinger2012} to artificially reduce oscillations. However, as shown in \cite{Seizinger2012}, such a force may come with numerical instabilities. However, the damping model presented in \cite{Krijt2013} shows that the elastic dampening can significantly increase the dissipation of energy. In this study we apply the viscoelastic damping force
\begin{equation}
\vec{F}_{\mathrm{D,ij}} = \frac{2 E^{*} }{\nu_{\mathrm{i}}^2  } \frac{\mathrm{d} \delta}{\mathrm{d} t} r_{\mathrm{ij}}T_{\mathrm{vis}} \vec{n}_{\mathrm{c}}\, ,
\end{equation}
as introduced by \cite{Seizinger2013}. Here, the dampening depends not on the relative velocity, but instead on the time evolution of the compression lengths
\begin{equation}
\delta = a_{\mathrm{i}}+a_{\mathrm{j}}-\left| \vec{X}_{\textit{i}} - \vec{X}_{\textit{j}} \right|\, ,
\end{equation}
i.e. the overlap between two spherical monomers with a distance of $| \vec{X}_{\textit{i}} - \vec{X}_{\textit{j}} |$ apart from each other. For an aggregate in equilibrium the compression length is ${ \delta_{\mathrm{0}} = r_{\mathrm{0}}/(3R) }$. The characteristic viscoelastic timescale $T_{\mathrm{vis}}$ in the order of  ${ 1\ \mathrm{ps} - 10\ \mathrm{ps} }$  \citep[][]{Krijt2013} but poorly constrained for specific grain materials. For smaller values
of  $T_{\mathrm{vis}}$ the material effectively behaves elastically and no damping is expected. For simplicity, we adapt a fixed value of $T_{\mathrm{vis}}=5\ \mathrm{ps}$ for all considered grains independent of material.\\
The full set of material parameters applied in our simulations is listed in Table \ref{table:MaterialProperties}. We emphasize that all parameters are best estimates for ISM conditions i.e. assuming a cold and dry environment. In general, these  parameters (especially the surface energy $\gamma$), have a wide range because they are highly sensitive to temperature \citep[][]{Bogdan2020}, monomer size \citep[][]{Bauer2019}, and humidity \citep[][]{Heim1999,Fuji1999,Kimura2015,Steinpilz2019,Bogdan2020}. Subsequent studies on this matter must complement the mechanical parameters utilized in this paper. 

\section{Radiative torque disruption (RATD)}
\label{sect:RATD}
In this section we briefly outline the radiative torques (RAT) that lead to rapid grain rotation and potentially to the disruption by centrifugal forces. It is well established that grains with an irregular shape acquire a certain amount of angular velocity when exposed to directed radiation \citep{Dolginov1976,DraineWeingartner1996,DraineWeingartner1997,WeingartnerDraine2003,LazarianHoang2007}. Here, the gain of angular momentum over time follows 
\begin{equation}
\omega_{\mathrm{agg}}(t)=\omega_{\mathrm{RAT}}\left[ 1-\exp\left( -\frac{t}{\tau_{\mathrm{drag}}}   \right) \right]\, ,
\label{eq:OmegaAgg}
\end{equation}
where $\tau_{\mathrm{drag}}$ is the characteristic timescale of the rotational drag by means of gas collisions \citep{Draine1996} and photon emission \citep{DraineLazarian1998} and 
\begin{equation}
\omega_{\mathrm{RAT}}=\frac{\Gamma_{\mathrm{RAT}}\tau_{\mathrm{drag}} }{I_{\mathrm{a1}}}
\label{eq:OmegaRAT}
\end{equation}
is the terminal angular velocity. The torque $\Gamma_{\mathrm{RAT}}$ is characteristic for individual grains and  depends on the spectrum of the radiation field as well as the shape of a particular grain and its material composition \citep[see e.g.][]{Hoang2014}. A solution of $\Gamma_{\mathrm{RAT}}$ may be calculated by numerical approximations \citep{Draine1996,DraineFlatau2013,Herranen2019} or analytical toy models \citep{LazarianHoang2007}. However, we emphasize that we do not evaluate $\omega_{\mathrm{RAT}}$  explicitly within the scope of this paper but assume a maximal grain rotation to guarantee the disruption of all aggregates. For the corresponding angular acceleration follows then
\begin{equation}
\frac{\mathrm{d} \omega_{\mathrm{agg}}(t)}{\mathrm{d} t} = \frac{ \omega_{\mathrm{RAT}}}{\tau_{\mathrm{drag}} }\exp\left( -\frac{t}{\tau_{\mathrm{drag}}}   \right)\, .
\label{eq:OmegaAcceleration}
\end{equation}
In principle, grains would also spin-up when exposed to an gaseous flow \citep[][, RMK22]{Lazarian2007Mech,DasWeingartner2016,Hoang2018AMech} leading to an mechanical torque (MET) $\Gamma_{\mathrm{MET}}$. However, the time evolution and acceleration of the angular velocity  would still follow the same curves governed by Eq.~\ref{eq:OmegaAgg} and Eq.~\ref{eq:OmegaAcceleration}, respectively, but evaluated with $\Gamma_{\mathrm{MET}}$ instead of $\Gamma_{\mathrm{RAT}}$.\\
A dust aggregate cannot simply be modelled as a rigid body. Internal relaxation processes such as Barnett relaxation \citep{Purcell1979,Lazarian1997}, nuclear relaxation \citep[][]{LazarianDraine1999A} or inelastic relaxation \citep[][]{Purcell1979,LazarianEfroimsky1999} would dissipate rotational energy. Subsequently, the dissipation would re-orient the direction of the angular velocity $\omega_{\mathrm{agg}}$. For typical interstellar conditions the total relaxation time is much smaller than the drag time $\tau_{\mathrm{drag}}$ \citep{WeingartnerDraine2003} and the grain axis $\hat{a}_{\mathrm{1}}$ becomes the most likely axis of grain rotation (compare Fig. \ref{fig:Grains}). For simplicity we assume in this study that the angular velocity points always in the direction of $\hat{a}_{\mathrm{1}}$ i.e.
${ \vec{\omega}_{\mathrm{agg}}(t) = \hat{a}_{\mathrm{1}} \omega(t) }$ for all of our three-dimensional N-body simulations.\\
Potentially, the aggregate becomes ripped apart by centrifugal forces long before reaching the terminal angular velocity $\omega_{\mathrm{RAT}}$ \citep[][]{Silsbee2016,Hoang2019}. A mathematical framework for the radiative torque disruption (RATD) of grains was presented in \cite{Hoang2019}. Here, the critical angular velocity for grain suggested for RATD is
\begin{equation}
    \omega_{\mathcal{S}} = \frac{2}{a_{\mathrm{eff}}}  \left(  \frac{\mathcal{S}_{\mathrm{max}}  }{ \rho_{\mathrm{mat}}  }  \right)^{1/2}\,.
\label{eq:OmegaRATD}
\end{equation}
Any dust aggregate exceeding the rotational limit $\omega_{\mathrm{agg}}>\omega_{\mathcal{S}}$ becomes inevitably destroyed.
This theoretical upper limit is derived from the material density $\rho_{\mathrm{mat}}$, the effective radius $a_{\mathrm{eff}}$, and the maximal tensile strength $\mathcal{S}_{\mathrm{max}}$ of the aggregate \citep[see][for further details]{Hoang2020Galaxy}. However, the tensile strength of porous dust aggregates is not well constrained and may be lower by several orders of magnitude compared to solid bodies because individual substructures are not fully connected. In \cite{Greenberg1995} a means to estimate the tensile strength of aggregates was provided by evaluating
\begin{equation}
    \mathcal{S}_{\mathrm{max}} = \frac{3}{2} \left< N_{\mathrm{con}} \right>  \frac{ \phi E}{ a_{\mathrm{mon}}   h}
\label{eq:TensileStrength}
\end{equation}
\citep[see also][]{LiGreenberg1997}. Here, $E$ is the binding energy, $h$ is related to the overlap of monomers, $\left< N_{\mathrm{con}} \right>$ is the average number of connections between all monomers, and $ a_{\mathrm{mon}} $ is the  monomer size of a monodisperse aggregate.\\
Complementary N-body simulations by \cite{Seizinger2013Tensile} suggest that the tensile strength is not directly related to the initial volume filling factor $\phi$ as assumed e.g. by  \cite{Greenberg1995} or \cite{Blum2006}.  Instead, the tensile strength is ${ \mathcal{S}_{\mathrm{max}} \propto \phi^{1.6}}$  or ${ \mathcal{S}_{\mathrm{max}} \propto \phi^{1.9}}$, respectively, for different quartz BA, BAM, and hexagonal aggregates. Comparable results are presented by \cite{TatsuumaKataoka2019} where the relation ${ \mathcal{S}_{\mathrm{max}} \propto \phi^{1.8}}$ was suggested in particular for icy and quartz BCCA aggregates and ${ \mathcal{S}_{\mathrm{max}} \propto \phi^{2/(3-D_{\mathrm{f}} )} }$ for arbitrary grain shapes with fractal dimension $D_{\mathrm{f}}$. We emphasize that in these simulations the aggregates got merely stretched or compressed, respectively, upon breaking, but do not experience any effects associated with rotation at all.

\section{Numerical setup}
\label{sect:NumericalSetup}
We implement the physical effects as outlined above into a C++ code that combines the physics of \cite{Seizinger2012,Seizinger2013} with the code presented in RMK22 for the growth of aggregates. The time evolution of the net force acting on each monomer,
\begin{equation}
m_{\mathrm{mon,i}} \frac{\mathrm{d} \vec{\varv}_{\mathrm{i}}}{\mathrm{d} t} = \vec{F}_{\mathrm{cent,i}} + \vec{F}_{\mathrm{cor,i}} + \vec{F}_{\mathrm{eul,i}} + \sum_{\mathrm{j=1, i \neq j}}^{N_{\mathrm{mon}}}  {           \left( \vec{F}_{\mathrm{N,ij}} + \vec{F}_{\mathrm{S,ij}} \right) }\, ,
\end{equation}
and that of the corresponding torque
\begin{equation}
I_{\mathrm{mon,i}} \frac{\mathrm{d} \vec{\omega}_{\mathrm{i}}}{\mathrm{d} t} = \sum_{\mathrm{j=1, i \neq j}}^{N_{\mathrm{mon}}} { \left( 
\vec{\Gamma}_{\mathrm{S,ij}} + \vec{\Gamma}_{\mathrm{R,ij}} + \vec{\Gamma}_{\mathrm{T,ij}} \right) }\, ,
\end{equation}
is calculated with second order accuracy by an symplectic Leap-Frog integration scheme. Here, the quantity $\vec{\varv}_{\mathrm{i}}$ is the velocity of an individual monomer within the simulation domain, and $\vec{\omega}_{\mathrm{i}}$ is the angular velocity caused by sliding, rolling, and twisting motions of its neighbors, whereas ${ I_{\mathrm{mon,i}}= 2/5\ m_{\mathrm{mon,i}}    a_{\mathrm{mon,i}}^2 }$ is the moment of inertia of a particular monomer.\\
In order to account for the dissipation of energy we assume that sliding, rolling, and twisting operate in the elastic limit until the corresponding displacement reaches some characteristic critical limit. The theoretical limits are
${ \zeta_{\mathrm{c}} = r_0(1-\nu)/(16\pi) }$ for sliding and
${ \phi_{\mathrm{c}} = 1 /(16\pi) }$ for twisting (compare Eq.~\ref{eq:zeta} and Eq.~\ref{eq:phi}). However, the critical limit of rolling $\xi_{\mathrm{c}}$ (see Eq.~\ref{eq:xi}) is still highly debated. For silicate monomers of different sizes this limit is within the range of ${ \xi_{\mathrm{c}} \in [0.2\ \mathrm{nm}, 3.2\ \mathrm{nm}] }$  \citep[][]{Dominik1997,Heim1999,Paszun2008}. In this study we apply the conservative value of ${ \xi_{\mathrm{c}} = 0.2\ \mathrm{nm} }$ for the entire ensemble of aggregates independent of size and material. When the displacements because of the relative motion of monomers in contact exceed their critical values, energy is dissipated and our code modifies the contact pointers  $\vec{n}_{\mathrm{i}}$ and $\vec{n}_{\mathrm{j}}$ to ensure that ${ | \vec{\zeta} | =  \zeta_{\mathrm{c}} }$, ${ | \vec{\phi} | =  \phi_{\mathrm{c}} }$, and ${ | \vec{\xi} | =  \xi_{\mathrm{c}} }$, respectively \citep[see][for details]{Dominik2002,Wada2007}.\\
Individual monomers operate on a characteristic time scale  that may be write as \citep{Wada2007}
\begin{equation}
    t_{\mathrm{dis,i}}  = \frac{1}{6^{2/3}}\sqrt{  \frac{ m_{\mathrm{mon,i}} r_{\mathrm{0}}^2 }{\pi\ \gamma R^2} }\, .
\end{equation}
Furthermore, it is not allowed for an monomer to move a distance larger than its own radius limiting the time step further by
${  t_{\mathrm{vel,i}} = a_{\mathrm{mon,i}}/\varv_{\mathrm{i}} }$. The final time step for our Leap-Frog integration scheme is then the minimum of these characteristic times ${ \Delta t=0.02\ \min_{\mathrm{i}}\left(t_{\mathrm{dis,i}}, t_{\mathrm{vel,i}}\right) }$ of all the monomers as well as monomer connections within the simulations domain. The factor of $\Delta t$ is a little smaller than the one suggested by \cite{Wada2007} that guarantees conservation of energy below an error of $10^{-3}$. However, monomers may roll along the surface of the rapidly rotating aggregate and choosing a smaller time step allows for a more accurate spatial resolution of such displacement of monomer within the aggregate.\\
The rotational damping acts usually on short timescales  $\tau_{\mathrm{drag}}$ in the order of days up to several hundred years compared to typical astronomical processes in the ISM \citep[see e.g.][]{WeingartnerDraine2003,Tazaki2017,Hoang2022}. A computational burden arises, because integration time $\Delta t$ is only a few ns. Consequently, some simulations may take up to $10^{18}$ time steps to terminate. Hence, a much smaller drag time $\tau_{\mathrm{drag}}$ in the order of hours instead of years is selected to reduce the total run-time of the code. However, this does not impact the result of the rotational disruption simulations as long as the change in centrifugal forces per time step remains much smaller than the displacement processes for individual monomers to reach a new equilibrium position within the rotating aggregate. Even for a $\tau_{\mathrm{drag}}$ of a few hours it is still guaranteed that centrifugal forces are the dominant cause of grains disruption because the spin-up remains slow enough that the shear within the aggregate by Euler forces remains negligible. In detail, a $\tau_{\mathrm{drag}}$ is selected such that the aggregate reaches an angular velocity where the destruction is guaranteed within a few hours of simulation time but the Centrifugal force remains always the dominant force exerted on each monomer i.e. ${ F_{\mathrm{eul,i}} \ll  F_{\mathrm{cent,i}} }$.
The simulations setup starts at ${ \omega_{\mathrm{agg}}=0\ \mathrm{rad\ s^{-1}} }$ and follows the curve of Eq.\ref{eq:OmegaAgg} up to a given terminal angular velocity  ${ \omega_{\mathrm{RAT}} }$ by default. Alternatively the code terminates for the condition $N_{\mathrm{con}}=1$ i.e. only two connected monomers remain within the simulation domain.\\
For detecting monomer collisions with variable sizes and the tracking of contact breaking events we use a loose octree data structure \citep[see e.g.][for further details]{Ulrich2000,Raschdorf2009} in order to optimize the runtime of the code even more. A new contact is established when two moving monomers touch each other, which means $\delta \leq 0$. When individual monomers or even entire sub-structures of the initial aggregate become unconnected we only track the evolution and spin-up process of the most massive remaining fragment while smaller fragments are allowed to leave the simulation domain. After each time step the remaining connected fragments are identified with a recursive flood fill algorithm on an undirected graph that represents the monomer connection relations. For every thousandth time step, the affiliation of monomer to a certain fragment, the number of monomers $N_{\mathrm{mon}}$ remaining within the simulation domain and the monomer positions $\vec{X}_{\mathrm{i}}$ as well as the forces and connections in between monomers are recorded.\\
In order to quantify the total stress in between all connected monomers within the largest fragment we introduce the quantity of the total stress
\begin{equation}
\sigma_{\Sigma} = \sum_{\mathrm{i=1}}^{N_{\mathrm{mon}}} \sum_{\mathrm{j>i}}^{N_{\mathrm{mon}}}  { \frac{\left|F_{\mathrm{N,ij}}\right|}{\pi r_{\mathrm{ij}}^2} }\, ,
\end{equation}
where $F_{\mathrm{N,ij}}=0$ for non-connected monomers.
\section{Results and discussion}
\label{sect:ResultsDiscussion}
We perform numerical N-body simulations for each individual pre-calculated grain aggregate with our numerical N-body setup upon rotational disruption. All grains are disrupted at an angular velocity of $\omega_{\mathrm{disr}}$ before the overall spin-up process reaches the terminal angular velocity of ${ \omega_{\mathrm{RAT}}=3\cdot 10^{10}\ \mathrm{rad\ s^{-1}} }$. We emphasize once again that the exact values of $\omega_{\mathrm{RAT}}$ is of minor relevance for our N-body simulations as long as the disruption of each aggregate is guaranteed and the centrifugal force and the Euler force follow strictly the relation ${ F_{\mathrm{eul,i}} \ll  F_{\mathrm{cent,i}} }$ for the entire spin-up process as governed by Eq.~\ref{eq:OmegaAgg} and Eq.~\ref{eq:OmegaAcceleration}, respectively.

\subsection{Time evolution of the rotational disruption process}
\begin{figure}[ht!]
	\includegraphics[width=0.49\textwidth]{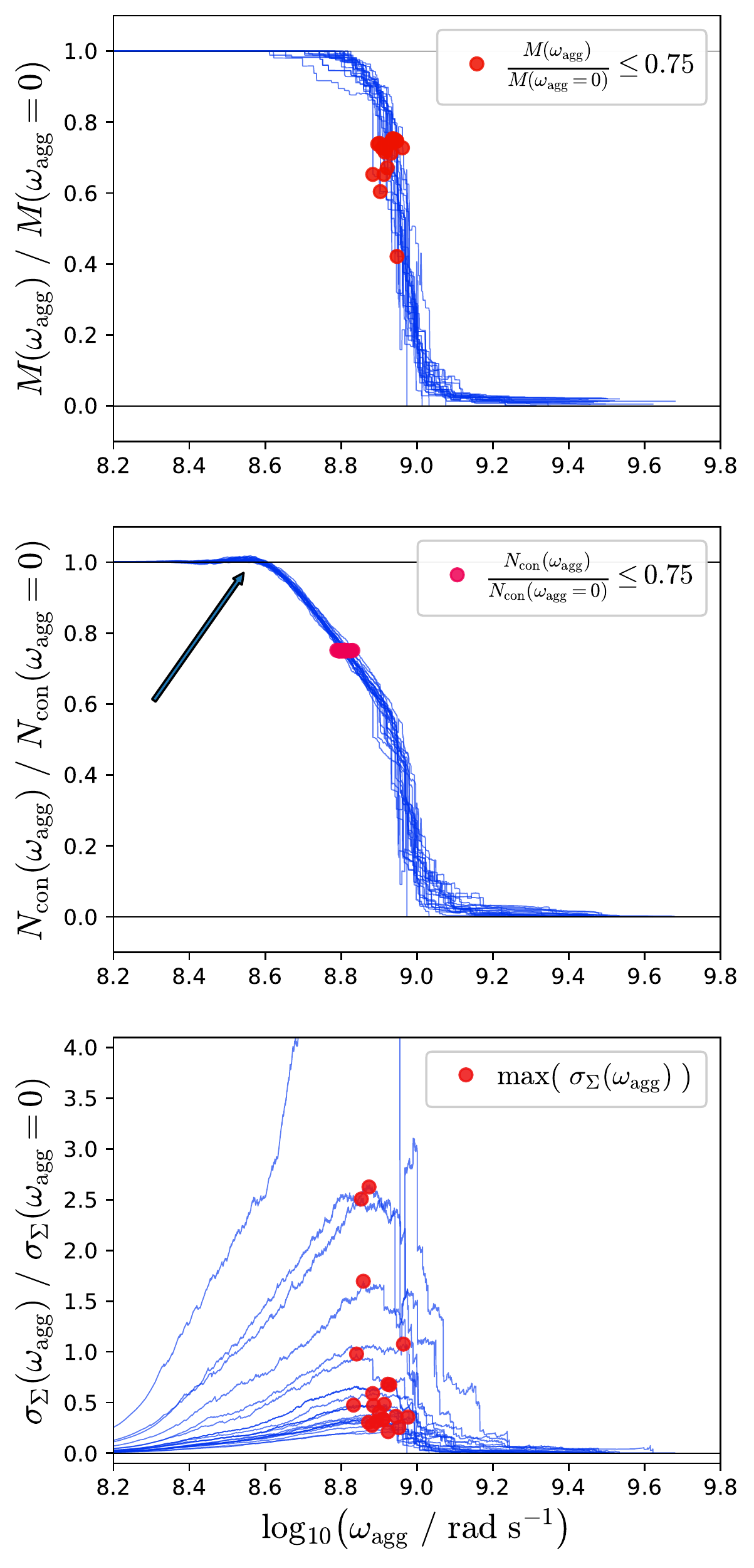}
\caption{Evolution of the mass  $M(\omega_{\mathrm{agg}})$ (top panel), number of connections $N_{\mathrm{con}}(\omega_{\mathrm{agg}})$ (middle panel), and total stress $\sigma_\Sigma(\omega_{\mathrm{agg}})$ (bottom panel) within the largest fragment dependent on the increasing angular velocity $\omega_{\mathrm{agg}}$. The blue lines represent the exemplary ensemble of q-S BAM2 grains with ${ a_{\mathrm{eff}}=350\ \mathrm{nm} }$. An arrow points to the peak value of the number of connections $N_{\mathrm{con}}$. Red dots indicate the characteristics angular velocity $\omega_{\mathrm{agg}}$ up to which an individual aggregate has lost $25\ \%$ of its initial mass, initial number of connections, or reached its peak stress. }
\label{fig:evolution}
\end{figure}
\begin{figure*}[ht!]
	\begin{center}
	\includegraphics[width=0.89\textwidth]{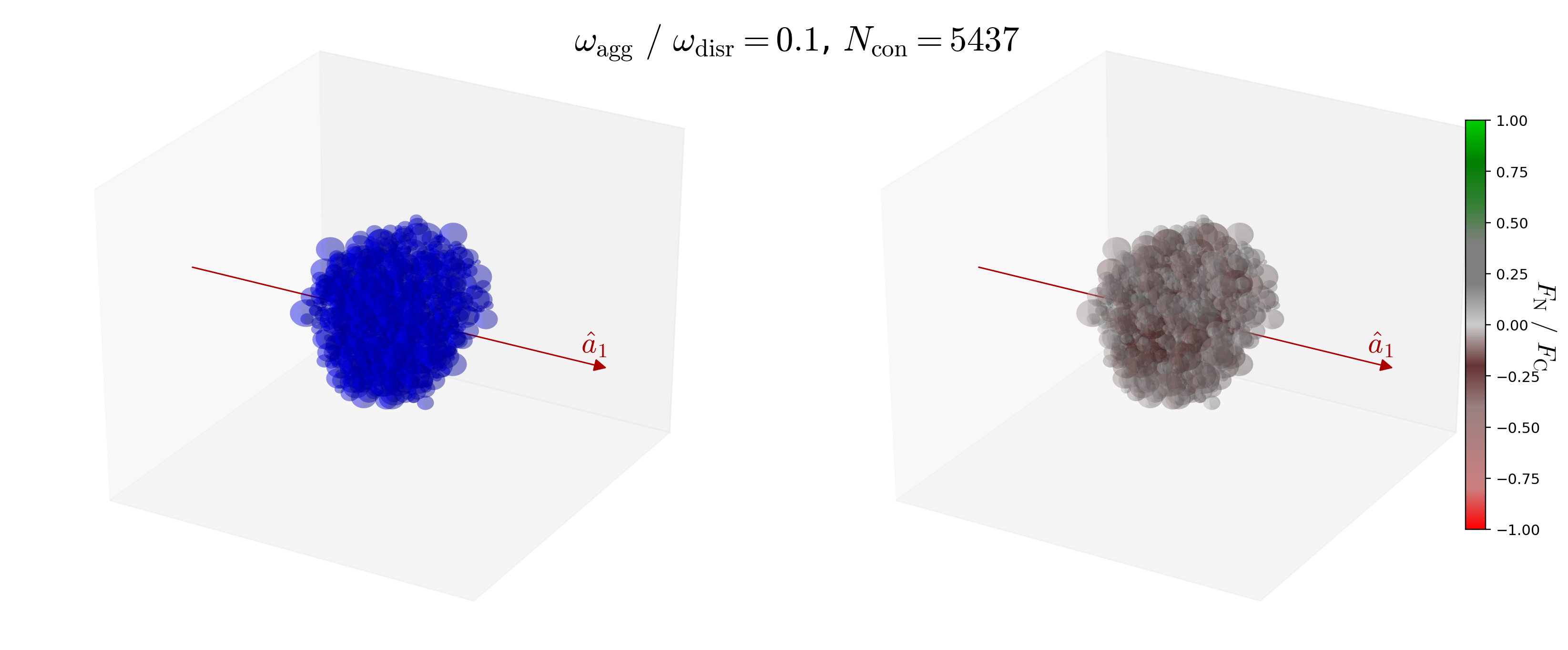}\
	\end{center}
\vspace{-10mm}	
	\begin{center}
	\includegraphics[width=0.89\textwidth]{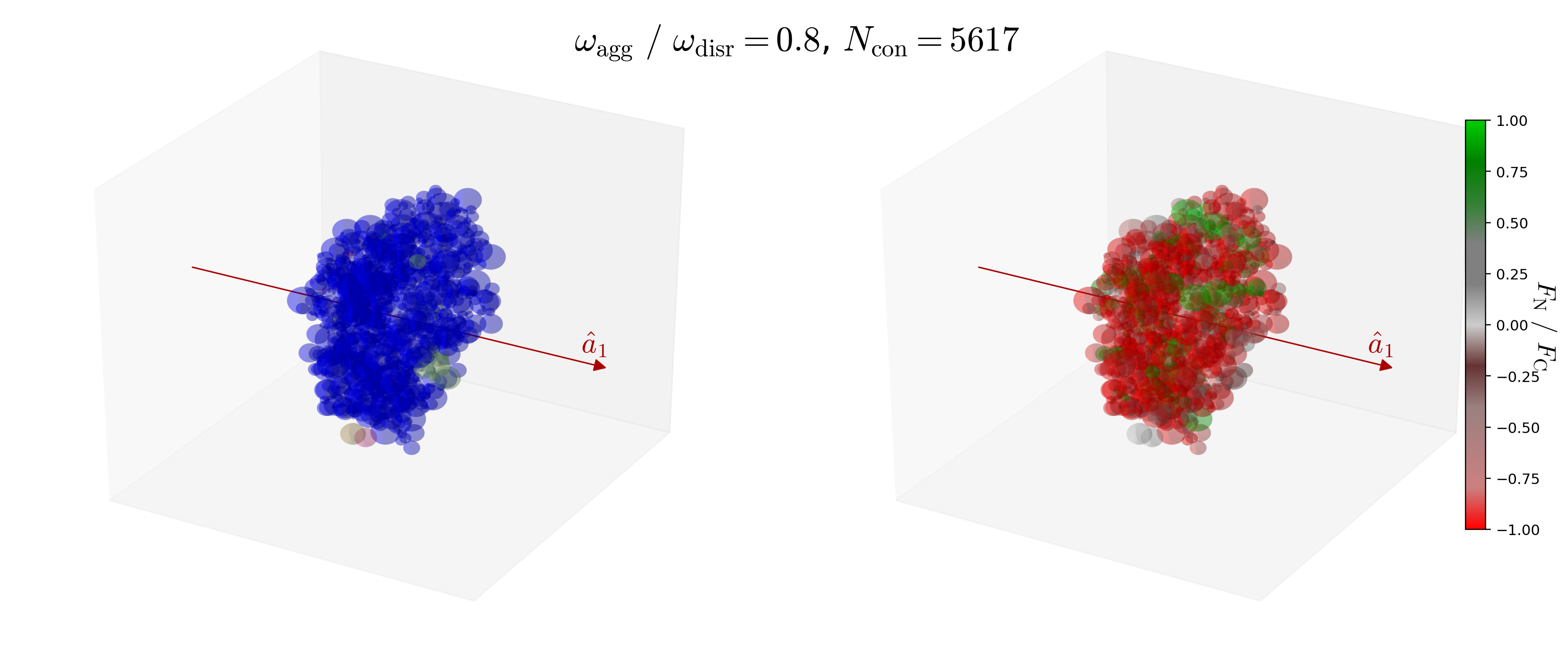}
    \end{center}
\vspace{-10mm}    
    \begin{center}
    \includegraphics[width=0.89\textwidth]{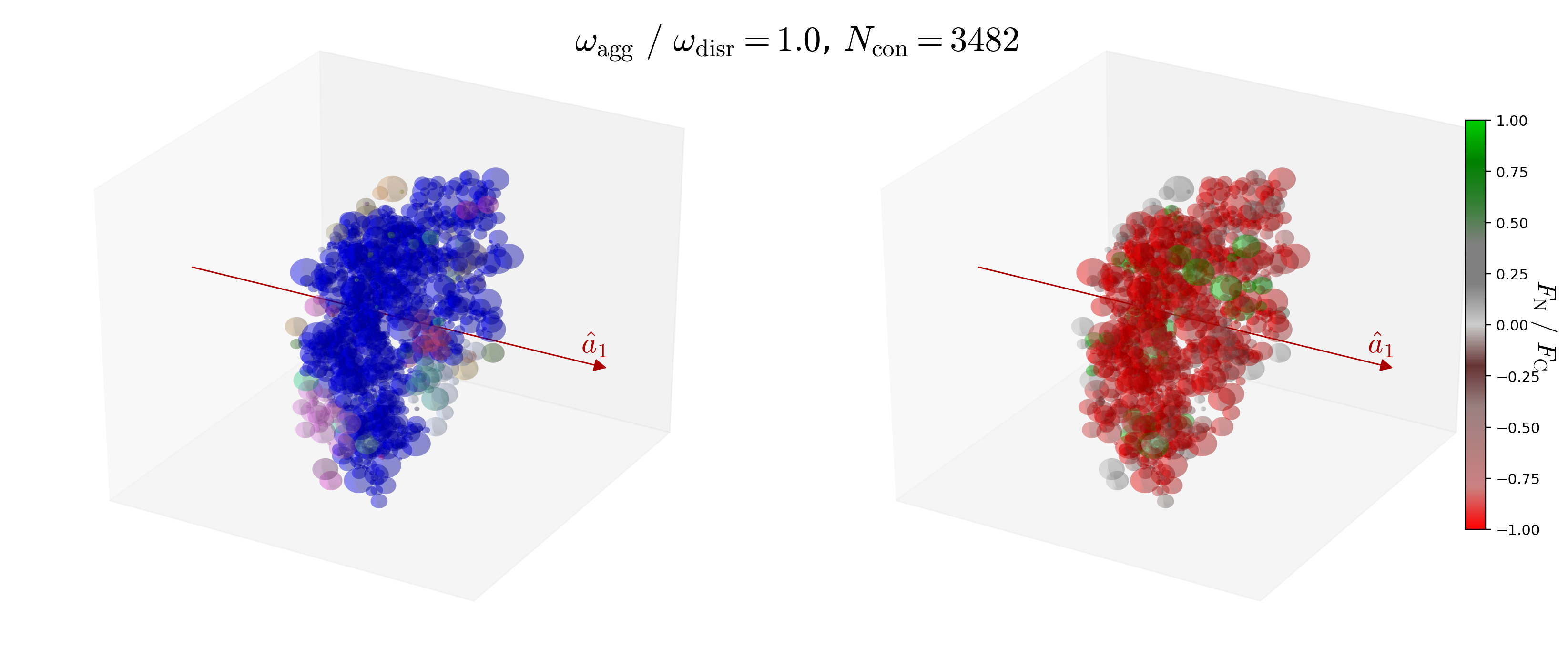}
	\end{center}
\vspace{-10mm}
\caption{Three exemplary snapshots of a BAM2 grain aggregate with $a_{\mathrm{eff}}=500\ \mathrm{nm}$ of an particular rotational disruption simulation with the corresponding number of connection $N_{\mathrm{con}}$ for the angular velocities of ${\omega/\omega_{\mathrm{disr}}=0.1}$ (top row), ${\omega/\omega_{\mathrm{disr}}=0.8}$ (bottom row), ${\omega/\omega_{\mathrm{disr}}=1.0}$ (bottom row). We emphasize that the exemplary grain is identical to the one in the lower right corner of Fig. \ref{fig:Grains}. \textit{Left column:} Color coded is the affiliation of the distinct connected fragments. Smaller fragments are arbitrarily colored whereas the most massive fragment is always depicted in blue. \textit{Right column:} The monomers are colored according to the largest magnitude of the normal force $F_{\mathrm{N}}$ exerted from its connected neighbours where $F_{\mathrm{N}}>0$ (green) represents pushing forces while $F_{\mathrm{N}}<0$ (red) are pulling forces, respectively. }
\label{fig:SimSteps}
\end{figure*}
\begin{figure*}[h!]
	\begin{flushleft}
	\includegraphics[width=0.99\textwidth]{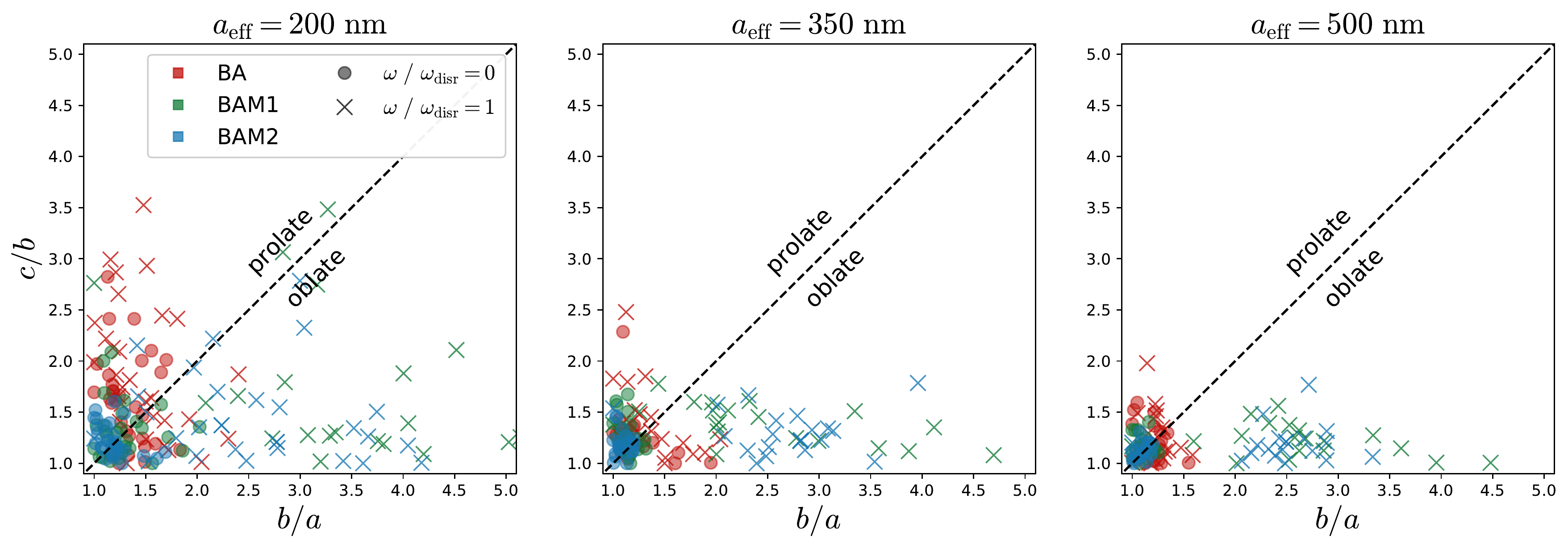}
	\end{flushleft}
\caption{Evolution of grain shapes for a-C BA (red), BAM1 (green), and BAM2 (blue) grains with an effective radius of ${ a_{\mathrm{eff}} = 200\ \mathrm{nm} }$ (left panel), ${ a_{\mathrm{eff}} = 350\ \mathrm{nm} }$ (middle panel), and ${ a_{\mathrm{eff}} = 500\ \mathrm{nm} }$ (right panel), respectively. Dots represent the initial grain shape i.e. ${ \omega_{\mathrm{agg}}/\omega_{\mathrm{disr}} = 0 }$ whereas crosses are the grain shape at disruption for ${ \omega_{\mathrm{agg}}/\omega_{\mathrm{disr}}=1 }$.  We note the clear tendency of the porous grains to become oblate in shape with increasing $\omega_{\mathrm{agg}}$. }
\label{fig:shape}
\end{figure*}
\begin{figure}[h!]
	\begin{center}
	\includegraphics[width=0.47\textwidth]{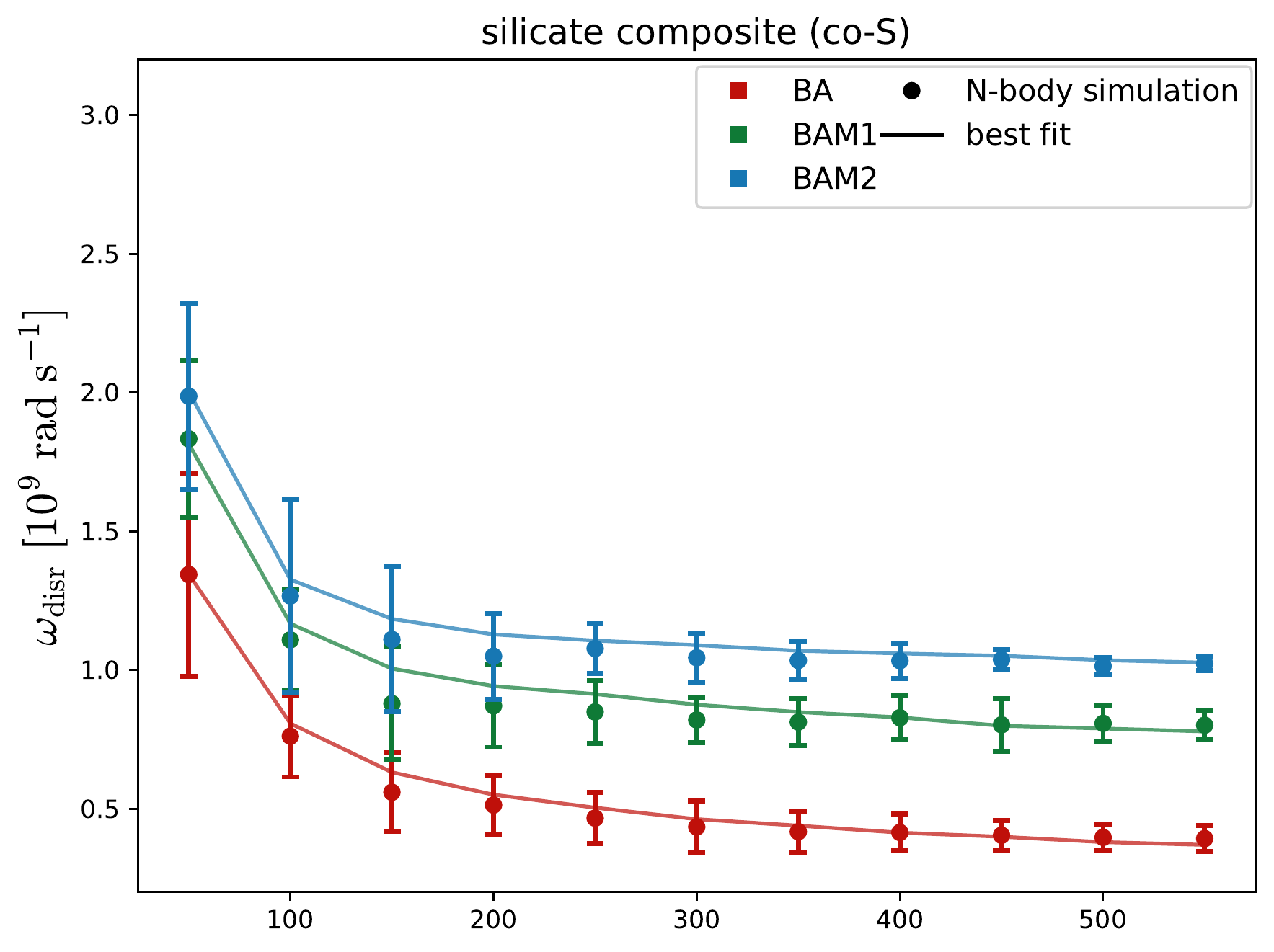}\\
	\includegraphics[width=0.47\textwidth]{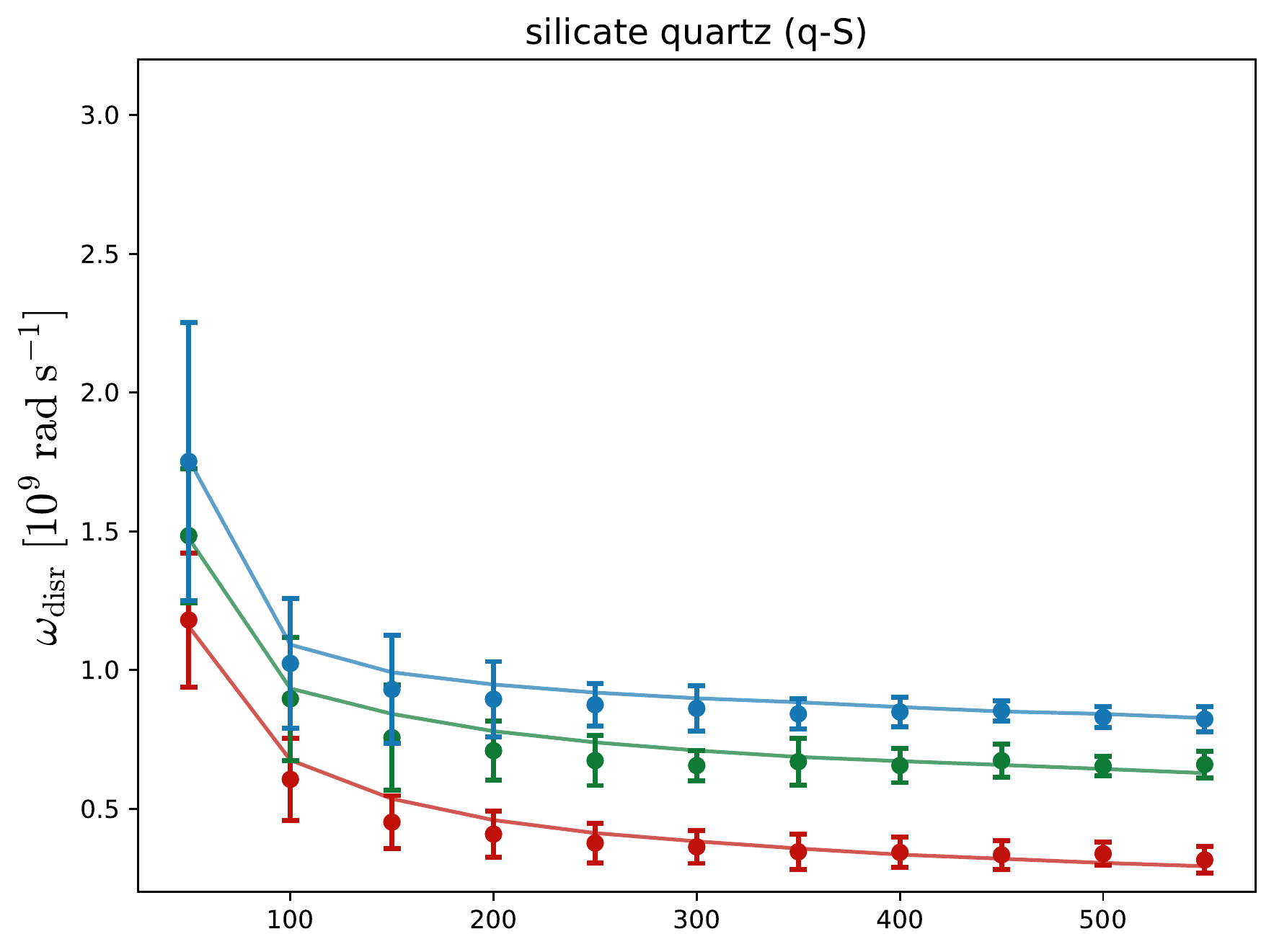}\\
    \includegraphics[width=0.47\textwidth]{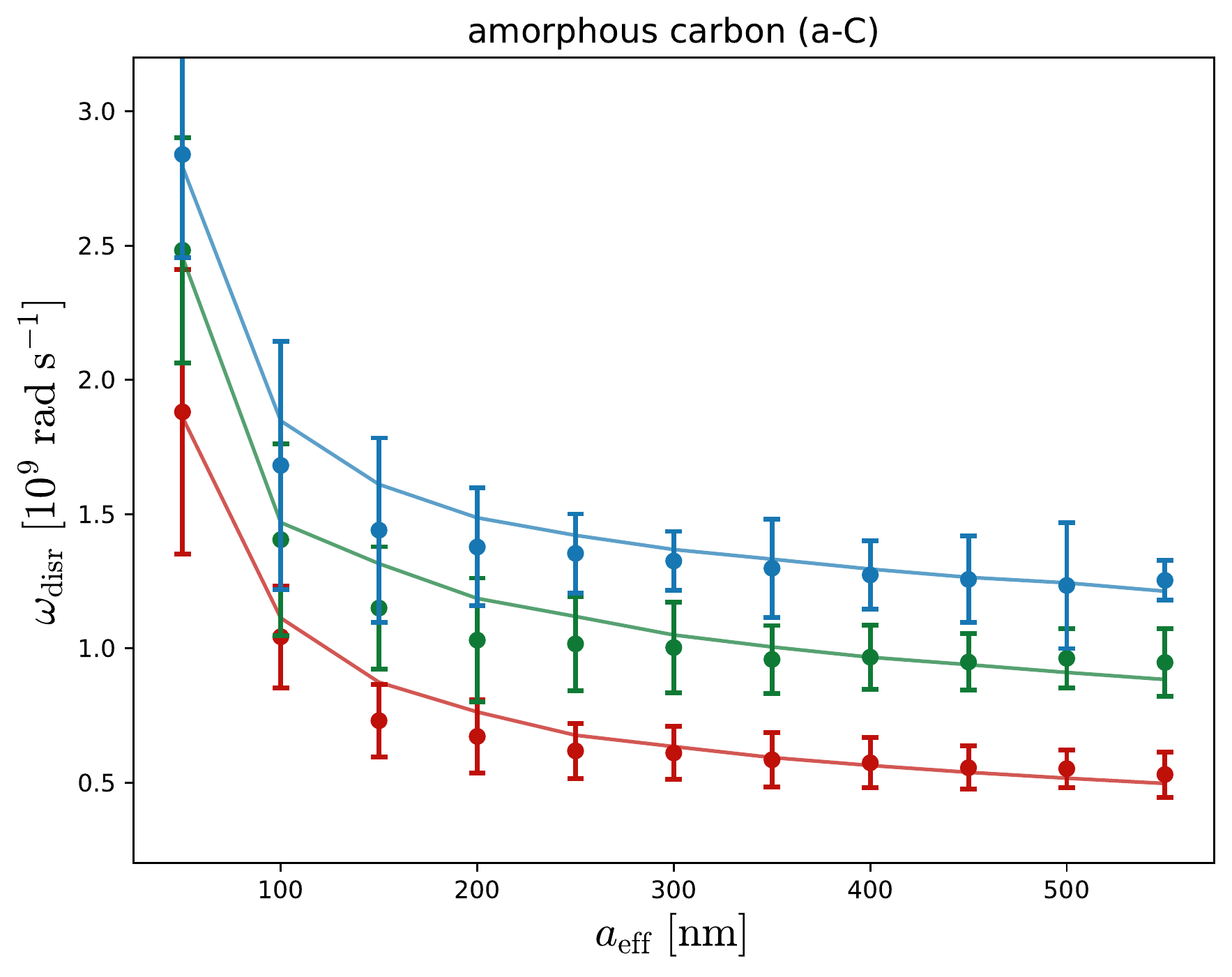}
	\end{center}
\caption{ The angular velocity of disruption $\omega_{\mathrm{disr}}$ dependent on grain size for the ensembles of co-S (top panel),  q-S (biddle panel), and a-C (bottom panel) grain materials dependent on the effective radius $a_{\mathrm{eff}}$. Color coded are the BA (red), BAM1 (green), and BAM2 (blue) grains. Vertical bars are the range between minimal and maximal values of $\omega_{\mathrm{disr}}$ resulting from our N-body disruption simulations while solid lines represent the best fit model.}
\label{fig:FinalFit}
\end{figure}
\begin{figure}[ht!]
	\begin{center}
	\includegraphics[width=0.49\textwidth]{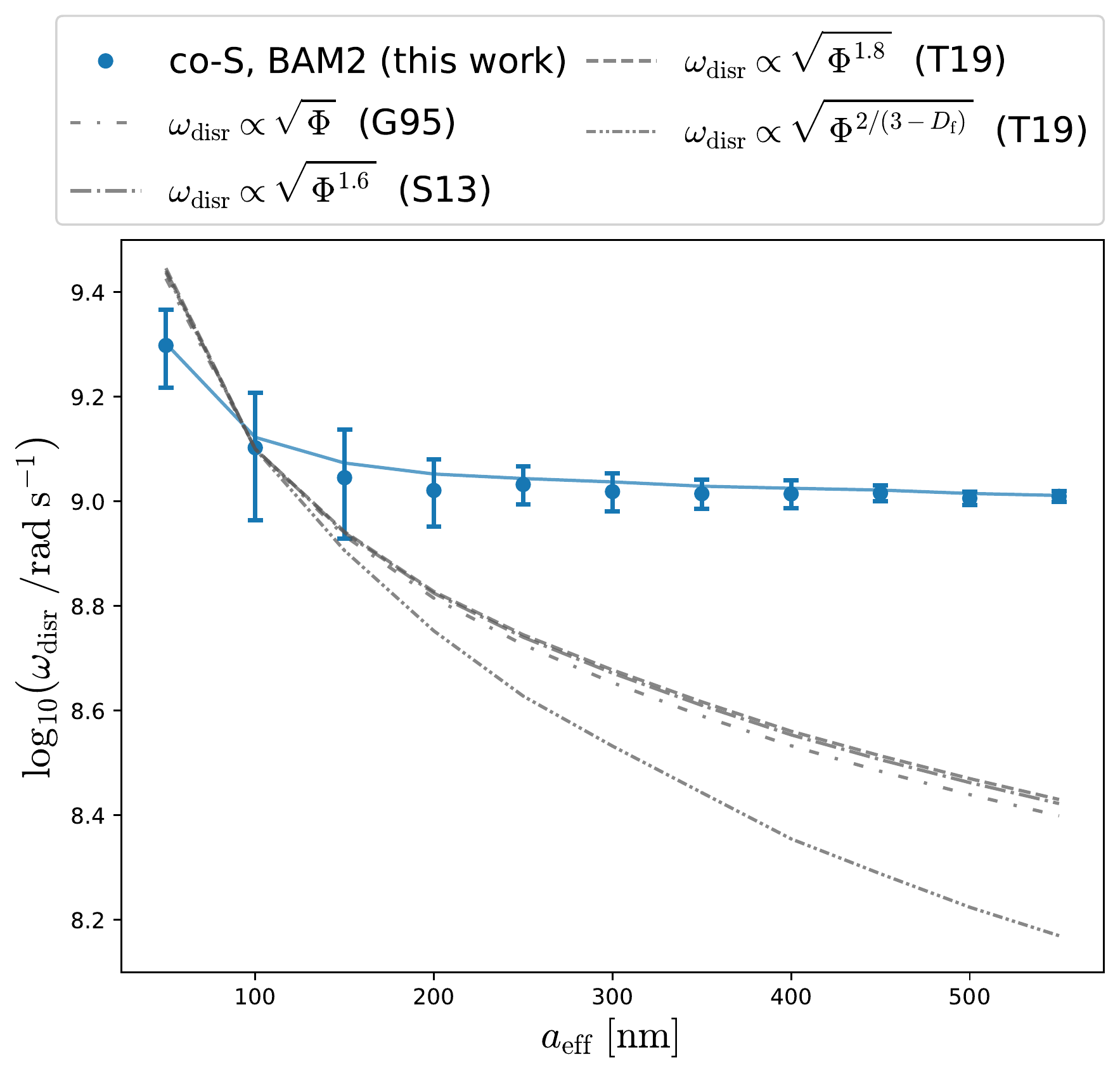}
	\end{center}
\caption{The same as Fig. \ref{fig:FinalFit} but only for co-S BAM2 grains (blue) in comparison with the predictions of $\omega_{\mathrm{agg}}$ based on the tensile strengths models presented in \cite{Greenberg1995} (G95), \cite{Seizinger2013Tensile} (S13), and \cite{TatsuumaKataoka2019} (T19), respectively (gray). The latter are re-scaled to match our simulation results for ${ a_{\mathrm{eff}}=100\ \mathrm{nm} }$.  }
\label{fig:FinalComp}
\end{figure} 
A typical simulation result for a q-S BAM2 grain with an effective radius of $a_{\mathrm{eff}}=350\ \mathrm{nm}$ is shown in Fig.~\ref{fig:evolution}. The evolution of the mass $M(\omega_{\mathrm{agg}})$ during the spin up process remains constant up to ${ \omega_{\mathrm{agg}} \approx 3\cdot 10^8\ \mathrm{rad\ s^{-1}} }$. Once the grain spins even faster, contacts break and individual monomers as well as smaller fragments start to break from the aggregates surface. Simultaneously, monomers wander outwards driven by centrifugal forces. For a small period the total number connections exceeds even the initial value and drops then steadily. This features is most pronounced for BAM2 grains. Eventually, the simulations reach the condition of $N_{\mathrm{con}}=1$ and terminate. The breaking of contacts  and the subsequent mass loss is continuous in nature while the angular velocity $\omega_{\mathrm{agg}}$ increases by roughly one order of magnitude. To designate a characteristic angular velocity $\omega_{\mathrm{disr}}$ of rotational disruption based on mass loss or broken connections would be arbitrary. In Fig.~\ref{fig:evolution} we show also the time evolution of the total stress $\sigma_\Sigma(\omega_{\mathrm{agg}})$ of each aggregate. In contrast to the mass loss and the braking of contacts the stress $\sigma_\Sigma(\omega_{\mathrm{agg}})$ rises steadily but drops then abruptly. Hence, we define this  unique feature in the time evolution of each aggregate  to be associated with the angular velocity of $\omega_{\mathrm{disr}}$ of rotational disruption. The value of  $\omega_{\mathrm{disr}}$ is close to but not identical with a mass loss of $25\ \%$.

\subsection{The fragmentation of rotating grains}
In Fig.~\ref{fig:SimSteps} we show snapshots of a typical spin-up process and rotational disruption event for one representative grain. The grain is identical to the BAM2 with $a_{\mathrm{eff}}=500\ \mathrm{nm}$ depicted to the bottom right of Fig.~\ref{fig:Grains}. Once the grain has spun up to an angular velocity of ${ \omega_{\mathrm{agg}} / \omega_{\mathrm{disr}} =0.1}$ the aggregate starts to experience a stretching force and first connections break. However, all monomers are still connected to the same aggregate. Compared to its original configuration as depicted in Fig.~\ref{fig:Grains} the aggregate changed its shape already, because monomers are re-arranged within the aggregate by means of rolling. The aggregate fragments for ${ \omega_{\mathrm{agg}} / \omega_{\mathrm{disr}} =0.8}$ and new connections are established as monomers move into a new equilibrium position driven by pulling forces and pushing forces between connected neighbours. We report that in this phase small disconnected fragments may rarely reconnect to the most massive fragment. For an angular velocity of ${ \omega_{\mathrm{agg}} / \omega_{\mathrm{disr}} =1.0}$ larger fragments are separated from the most massive cluster. The remaining cluster goes through a phase of relaxation, where its total stress $\sigma_\Sigma$ declines while simultaneously the mass loss is still an ongoing process. We note that if the spin-up process of the grain were to stop at an angular velocity ${ \omega_{\mathrm{agg}} / \omega_{\mathrm{disr}}  \leq 1.0}$ the aggregate would still fragment, but the mass loss would stop eventually as the remaining most massive fragment reaches a new equilibrium configuration.\\
Dust destruction processes such as shattering efficiently redistribute larger grains sizes towards smaller ones \citep{Dwek1980,Tielens1994}. In modelling this process it is usually assumed that the new size distribution of the fragments follows a power-law \citep{Hellyer1970,Hirashita2009A,Kirchschlager2019}. A similar power-law is assumed in models of the grain redistribution by means of RATD \citep[e.g.][]{Giang2020}. However, the fragmentation process of rapidly rotating porous dust aggregates remains poorly constrained. In fact, our N-body simulation results so far indicate that the considered BAM aggregates almost completely break down into their individual building blocks. Hence, the resulting size distribution of the fragments is expected to be almost identical to the initial size distribution of the monomers. However, we note that our numerical setup does not track smaller fragments once they become separated from the most massive fragment. Furthermore, each individual fragment will experience its own own individual spin-up and drag (see Sect.~\ref{sect:RATD}). Smaller fragments may also eventually become too small for an RAT effective for a further spin-up upon the disruption point \citep[][]{LazarianHoang2007}. Such effects need to be taken into account in forthcoming studies to answer question about the resulting size distribution of an ensemble of rotationally disrupted grains conclusively.

\subsection{Rotational deformation of grain shapes}
In Fig.~\ref{fig:shape} we present the time evolution of the shapes of a-C grains with different sizes up to the breaking point. Initially, 
for ${ \omega_{\mathrm{agg}} / \omega_{\mathrm{disr}} =0.0}$ the ensemble of grains with an effective radius of $a_{\mathrm{eff}}=200\ \mathrm{nm}$ are almost  oblate and prolate shapes in equal parts. Larger grains are with $a_{\mathrm{eff}}=350\ \mathrm{nm}$ and $a_{\mathrm{eff}}=500\ \mathrm{nm}$ are slightly more prolate where most of the axis ratios are $c/b<1.7$. At the breaking point i.e. ${ \omega_{\mathrm{agg}} / \omega_{\mathrm{disr}} =1.0}$ the a-C grains become deformed where a oblate shapes are the most likely outcome with an axis ratio up to $b/a \leq 5.0$ for BAM grains. The exception is the ensemble of small BA grains with  $a_{\mathrm{eff}}=200\ \mathrm{nm}$ where a prolate shape seem to be the more favourable configuration. We note that  BA grains reach a smaller  axis ratios  because they break more easily with a much shorter deformation phase compared to BAM1 and BAM2 grains. We report similar trends for the deformation of grains with q-S and co-S materials, respectively, where oblate grains are the most likely shape of rotational disruption.\\
\begin{table*}[ht!]
\centering
\begin{tabular}{|l|ccc|ccc|ccc|}
\hline
%{} &  \multicolumn{3}{c}{CA} & \multicolumn{3}{c}{SQ} & \multicolumn{3}{c}{SC} \\

{} & {} & a-C &{}  & {} & q-S & {} & {} & co-S & {} \\
\cline{3-3}
\cline{6-6}
\cline{9-9}
%\midrule
{}   & BA   & BAM1 & BAM2 & BA   & BAM1 & BAM2 & BA   & BAM1 & BAM2\\
\cline{2-10}
%\cline{5-7}
%#\cline{8-10}

A         &  1.97 & 2.28 & 2.25 &  2.13 & 2.18 & 2.22 &  1.50 & 1.65 & 1.57 \\
$\alpha$  &  0.10 & 0.12 & 0.13 &  0.10 & 0.14 & 0.14 &  0.10  & 0.14 & 0.15\\
%$\beta$   &  A1 & A2 & A3&  A1 & A2 & A3&  A1 & A2 & A3\\
\hline
\end{tabular}
\caption{Best fit parameters of $\omega_{\mathrm{disr}}$ based on our N-body disruption simulation results.}
\end{table*}
\begin{figure*}[ht!]
	\begin{flushleft}
	\includegraphics[width=0.99\textwidth]{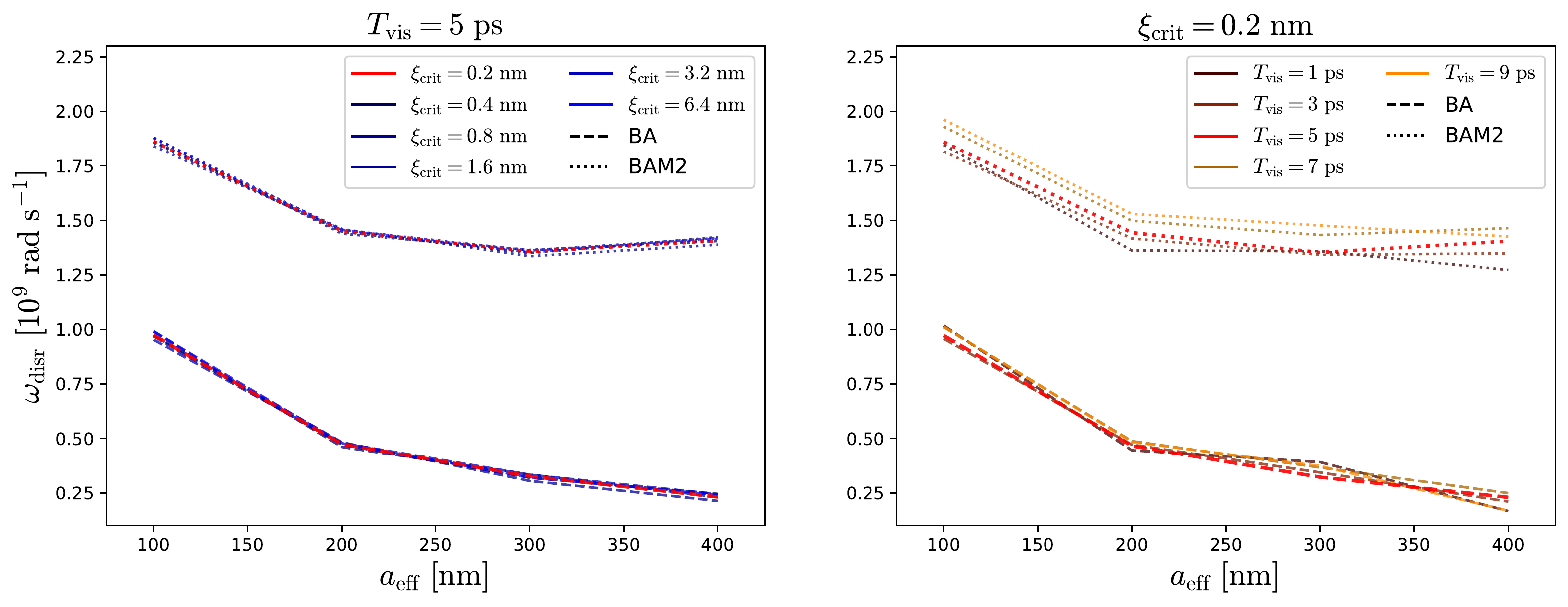}
	\end{flushleft}
 \vspace{0mm}
\caption{The same as Fig. \ref{fig:FinalFit} but only for a-C BA (deshed lines) and BAM2 (dotted lines) grains. The left panel shows the impact of critical rolling displacement $\xi_{\mathrm{crit}} \in [0.2\ \mathrm{nm}, 6.4\ \mathrm{nm}]$ on $\omega_{\mathrm{agg}}$ while the viscous dumping time $T_{\mathrm{vis}}$ remains constant. The right panel is for a constant $\xi_{\mathrm{crit}}$ but with $T_{\mathrm{vis}} \in [1\ \mathrm{ps}, 9\ \mathrm{ps}]$. Red lines represent the default parameters of our N-body disruption simulations.  }
\label{fig:ParamTest}
\end{figure*}
Quantifying the grain shape by the fractal dimension $D_{\mathrm{f}}$ during the spin-up process reveals no clear trend. As depicted in Fig.~\ref{fig:properties} the fractal dimension is not well correlated with the grain shapes of the initial ensemble. Up to the braking point the variation of $D_{\mathrm{f}}$ becomes even larger. The same for the porosity $\mathcal{P}$. At the beginning of the spin up process $\mathcal{P}$ starts to increase slightly for BAM1 and BAM2 grains while the BA ones break without an increase in   $\mathcal{P}$. Close to the breaking point the porosity distribution has a large variation and becomes virtually identical for BAM1 and BAM2 grains. Eventually, the analysis of the fractal dimension $D_{\mathrm{f}}$ as well as the porosity $\mathcal{P}$ does no longer apply because the calculation of these quantities fails as the grain aggregates break down into its individual monomers  (see Sect.~\ref{sect:GrainGrowth}).\\
The shape and porosity of interstellar dust is still a matter of debate. For example BA grain growth processes favor roundish grain aggregates with a fractal dimension of about $D_{\mathrm{f}}=2.0$ where the principle axis are ${a  \approx b\approx  c}$. \cite{Guillet2018} developed a grain model based on spheroids of amorphous silicate and amorphous carbon to reproduce both starlight polarization and polarized sub-millimeter emission. The best fit model suggest prolate grains with an axes ratio of ${ b/a = 1/3 }$ and a porosity of $\mathcal{P}=0.2$. More recently, \cite{DraineHensley2021} suggest 
prolate grains with an axis ratio of $b/a=0.6$ or oblate grains with  
$b/a=1.5$ with a porosity of about $\mathcal{P}=0.4$. Comparing the initial porosity of our  ensemble grown by BAM (see Fig.~\ref{fig:properties}) reveals that only the BAM2 ensemble would match the limitation in porosity  with values $\mathcal{P}=0.4-0.5$.\\
Elongated grains may be the result of hit and stick processes of BCPA with preferential direction, e.g. for magneto-hydrodynamic turbulence \citep[][]{Yan2003} or for grain aggregates aligned with the magnetic field direction \citep[][]{Hoang2022}. However, the latter effect requires rapid grain rotation. Subsequently, the spin-up process would increase the relative velocity between the surface of the rotating aggregate and impinging monomers may potentially destroy the entire aggregate in a catastrophic disruption event \citep{Benz19995,Morris2017,Schwartz2018}. What we find with our N-body simulations is that accelerated rotation deforms initially more roundish aggregates preferentially into oblate shapes.\\
For RATs the axis ration of the grains is linked to the radiation field via the maximal angular velocity $\omega_{\mathrm{RAT}}$. The more extreme cases, as depicted in Fig.~\ref{fig:shape} with ${b/a = 4-5}$ would require for the grains to be in close vicinity of a supernova or\\ active galactic nucleus \citep{Hoang2019,Giang2020,Giang2021}. We emphasize that the torques $\Gamma_{\mathrm{RAT}}$ and  $\Gamma_{\mathrm{MET}}$  (see Sect.~\ref{sect:RATD}) are tightly connected to the grain shape \citep[][RMK22]{LazarianHoang2007,DasWeingartner2016}. Consequently, the terminal angular velocity $\omega_{\mathrm{RAT}}$ would be marginally increased during the deformation phase accelerating the grain rotation even more (see Eq. \ref{eq:OmegaAcceleration}). However, this effect would be continuous over a long time span and not considerably impact the internal grain dynamics as far as the final value of $\omega_{\mathrm{disr}}$ is concerned. More severe is the change in $\omega_{\mathrm{RAT}}$ during the fragmentation phase. Here, the RAT would decrease as the aggregate fragments and subsequently the rotation would slow down and the grain stabilizes by reaching a new equilibrium configuration. However, considering a dynamical spin-up process in our N-body simulation would not only require to calculate the torque $\Gamma_{\mathrm{RAT}}$ but also the rotational drag timescale $\tau_{\mathrm{drag}}$ for each time step by means of time consuming numerical approximate methods \citep[][RMK22]{DraineFlatau2013}. Potentially, both quantities $\Gamma_{\mathrm{RAT}}$ and $\tau_{\mathrm{drag}}$ may be parameterized based on the shape and material parameters of each individual BAM grain. For now a dynamical spin-up goes beyond the scope of this paper.

\subsection{Average rational disruption of the grain ensemble}
In Fig.~\ref{fig:FinalFit} we present the characteristic angular velocity $\omega_{\mathrm{disr}}$ for the entire set of simulation results. The magnitude of $\omega_{\mathrm{disr}}$ is roughly in the range of about  ${ 5\cdot 10^8 - 5\cdot 10^9\ \mathrm{rad\ s^{-1}} }$ for different materials and grain sizes. This result agrees well with the predictive model presented in \cite{Hoang2019}. However, the exact value of $\omega_{\mathrm{disr}}$ depends on the exact material properties and the internal structure of the initial grain. While rapidly rotating solid bodies break into few fragments by driven by centrifugal forces, a porous material may fragment at lower angular velocity and into a larger number of pieces. The maximal tensile strength $\mathcal{S}_{\mathrm{max}}$ is usually utilized to quantify the response of a material during stretching. However, what we find that utilizing the maximal tensile strength $\mathcal{S}_{\mathrm{max}}$ as introduced in Eq.~\ref{eq:TensileStrength} to calculate the critical angular velocity $\omega_{\mathcal{S}}$ with  Eq.~\ref{eq:OmegaRATD} does not reproduce the average rotational disruption $\omega_{\mathrm{disr}}$ resulting from our N-body simulations. We argue that $\mathcal{S}_{\mathrm{max}}$ is insufficient to describe the dynamical behavior of rotating aggregates because in a stretching aggregate each monomers experience only the local forces in between its connected neighbours. In a rotating aggregate, however, each individual monomer experiences also the centrifugal force. The maximal tensile strength   $\mathcal{S}_{\mathrm{max}}$ is determined by stretching of granular materials performed along a distinct axis, whereas the centrifugal force is radial with respect to the rotation axis. Furthermore, considering only stretching produces acting on an aggregate does not not lead to a large displacement of monomers compared to the aggregate's scale. It is important to note that, our N-body simulations reveal that monomers are rolling outwards within the grain aggregates. This movement of $N_{\mathrm{mon}}$ monomers allows potentially  to establish up to ${N_{\mathrm{mon}}(N_{\mathrm{mon}}-1) \approx N_{\mathrm{mon}}^2}$ new connections. Naturally, some monomers are already connected and only a small fraction of all unconnected monomers is close enough within the aggregate to newly connect. Hence, the relation cannot be quadratic but with an much smaller exponent $\alpha \ll 2$. In order to match our data we suggest to extend the definition of Eq. \ref{eq:TensileStrength} by the number of monomers $N_{\mathrm{mon}}$. By utilizing Eq.~\ref{eq:OmegaRATD} the characteristic angular velocity of rotation disruption for a polydisperse grain aggregate reads then 
\begin{equation}
    \omega_{\mathrm{disr}} =   \frac{A}{a_{\mathrm{eff}}}  \sqrt{     \frac{ \gamma}{ \rho_{\mathrm{mat}}\left< a_{\mathrm{mon}}  \right>  }  } \Biggl(\ \left< N_{\mathrm{con}}  \right> N_{\mathrm{mon}} \Phi\   \Biggr)^{\alpha}
\label{eq:omegadisr}
\end{equation}
where $\left< a_{\mathrm{mon}}  \right>$ is the average monomer radius and $A$ and $\alpha$, respectively, are fit parameters\footnote{Note that ${ \left< N_{\mathrm{con}}  \right>  = N_{\mathrm{con}} / N_{\mathrm{mon}} }$ and thus Eq.~\ref{eq:omegadisr} may also be written in an equivalent form with a factor ${(\left< N_{\mathrm{con}}  \right> N_{\mathrm{mon}} \Phi\  )^{\alpha} = ( N_{\mathrm{con}}  \Phi\  )^{\alpha} }$  } to match the simulation results. Assigning a separate exponent to each of the individual quantities $\left< N_{\mathrm{con}}  \right>$, $N_{\mathrm{mon}}$, and $\Phi$, respectively, does not improve the accuracy of the fit. Best fit results of $\omega_{\mathrm{disr}}$ are plotted in Fig.~\ref{fig:FinalFit}. The fit matches very well with the ensemble average of the different considered grain sizes.\\
In Fig.~\ref{fig:FinalComp} we show a comparison of our best fit of $\omega_{\mathrm{disr}}$ for {co-S} BAM 2 grains with that calculated with different parameterizations of the tensile strength $\mathcal{S}_{\mathrm{max}}$ given in literature (see also Sect. \ref{sect:RATD}). These parameterizations of $\mathcal{S}_{\mathrm{max}}$ may not necessarily be evaluated with the parameters provided by our grain models. Hence, we scale the resulting to match our results of $\omega_{\mathrm{disr}}$ at for an effective radius of ${ a_{\mathrm{eff}}=100\ \mathrm{nm} }$. The comparison reveals that previous attempts of modelling $\omega_{\mathrm{disr}}$ by a volume filling factor $\phi$ dependent tensile strength $\mathcal{S}_{\mathrm{max}}$ cannot reproduce the asymptotic behavior of our simulation results towards larger grain radii $a_{\mathrm{eff}}$.\\
Finally, we emphasize that the presented results of rotational disruption are calculated for aggregates of carbonaceous and silicate monomers loosely connected by van der Waals force. However, materials other than carbon or silicates may form much stronger bonds \citep{Dominik1997}. For instance, pure iron in the form of small pallets may be present in the interstellar dust that would create metallic bonds between monomers \citep{Dominik1997,Draine2021}. The same for ices covering the surface of the grain aggregate where dipole-dipole interactions may increase the resistance against rotational disruption. Furthermore, high impact collisions of monomers or compressive stress acting on the aggregate may lead to sintering, an effect where the monomers fuse at the contact surface. Subsequently, sintering leads to a neck between monomers \citep{Maeno1983,Blackford2007,Sirono2017} strengthening the connection between monomers as well but at the same time it makes the aggregate a whole more brittle, because as a neck would not allow for rolling motions between monomers. Moreover, an effect that may weaken monomer connections is dust heating. A higher dust temperature decreases the surface energy $\gamma$ up to one order of magnitude \citep[][]{Bogdan2020}. Any luminous environment with a radiation field strong enough to drive grain rotation up to $\approx 10^9\ \mathrm{rad\ s^{-1}}$ would inevitably heat the dust grains up to several hundredth of Kelvin. Thus $\gamma$ and subsequently $\omega_{\mathrm{disr}}$ would decrease. Altogether, such effects need to be taken into consideration in forthcoming studies in order to complement our numerical setup of the rotational disruption of of porous dust aggregates. 
%The size distribution in the ISM follows a power law where most of the dust mass is in smaller grains \citep{Mathis1977,DraineLi2007}.
% \section{Discussion}
%\cite{Guillet2018}
%we speculate that the growth process is alread limited by grain rotation as the relative velcoity in colldiing roating aggregates is fastly increased as in ...
%However, ... goes beyond the scope of this study ... tensile strength is not a good proxy parameter

\subsection{Impact of critical rolling displacement and viscous damping}
The critical rolling displacement $\xi_{\mathrm{crit}}$ and the viscous damping time scale $T_{\mathrm{vis}}$ are the least constrained parameters in our N-body simulations. Laboratory data indicates a huge variation in these parameters of about one order of magnitude for silicates \citep[][]{Dominik1995,Heim1999,Krijt2013}.\\
In order to quantify the impact $\xi_{\mathrm{crit}}$ and $T_{\mathrm{vis}}$, respectively, on rational disruption we repeat our N-body simulation for a subset of the pre-calculated grain aggregates
varying  $\xi_{\mathrm{crit}}$ in the range $0.2\ \mathrm{nm}$ to $6.4\ \mathrm{nm}$ and $T_{\mathrm{vis}}$ from $1\ \mathrm{ps}$ to $9\ \mathrm{ps}$. In Fig.~\ref{fig:ParamTest} we present the exemplary results of the parameter test of a-C BA and BAM2 grains. Variations in the rolling displacement $\xi_{\mathrm{crit}}$ show little effect. The resulting angular velocity of rotational disruption $\omega_{\mathrm{disr}}$ shows small fluctuations but no clear trend with an increase in $\xi_{\mathrm{crit}}$. Hence, we attribute the fluctuation to numerical effects (see also appendix~\ref{sect:BreakingTest}). In contrast to $\xi_{\mathrm{crit}}$ an increase in $T_{\mathrm{vis}}$ leads to a clear increase of $\omega_{\mathrm{disr}}$. This effect is most evident for BAM2 grains with $\omega_{\mathrm{disr}}$ being about $11\ \%$ higher for the largest applied viscous time of $T_{\mathrm{vis}} = 9\ \mathrm{ps}$. Similar trends can be reported for the q-S and co-S materials. Further improvements in the predictive accuracy of our rational disruption simulations cannot be achieved upon forthcoming laboratory data especially for the viscoelastic damping of oscillations within carbonaceous aggregates.

\section{Summary}
\label{sect:Summary}
We aimed to  study the rotational disruption of interstellar dust aggregate analogs. An ensemble of porous dust aggregates by means of ballistic aggregation and migration (BAM) is pre-calculated where for BA, BAM1, and BAM2 grains each monomer has at least one, two, or three connections, respectively, with its neighbors. Here, we modified the original BAM algorithm presented in \cite{Shen2008} to work with variable monomer sizes. We estimated the composition of the grain aggregates based on the abundance of elements in the ISM to approximate their mechanical properties. Numerical three-dimensional N-body simulations are performed to determine characteristic angular velocity of rational disruption $\omega_{\mathrm{disr}}$ for each aggregate individually. The numerical setup is based on the work of \cite{Dominik1997}, \cite{Wada2007},  \cite{Seizinger2012}, and RMK22, respectively. We modified their setup by intruding additional forces associated with an accelerated rotation of an aggregate. A subsequent  analysis of the disruption event allows to describe the average rotational disruption of porous grain ensembles.\\
The findings of this study are summarized as follows:
\begin{itemize}
    \item Compared to the original BAM algorithm considering only single a monomer size we report that a grain growth by BAM with polydisperse monomers results in aggregates that are more compact and less porous as smaller monomer tend to migrate deeper into the aggregate. For the same reason the porosity $\mathcal{P}$ stagnates with an increasing effective radius $a_{\mathrm{eff}}$ with values up to  $\mathcal{P}=0.4-0.85$ for BA whereas BAM1 and BAM2 grains are less porous with $\mathcal{P}=0.35-0.7$ and $\mathcal{P}=0.35-0.5$, respectively.
    \item Our simulation results reveal that rotating porous  dust aggregates are not disrupted by a single abrupt event characteristic for a brittle material but rather by a continuous process where  aggregates experience a continuous mass loss. Subsequently, the initial aggregate breaks ultimately down into fragments not larger than a few monomers. 
    \item The initial distribution of the pre-calculated polydisperse BAM grains have oblate and prolate shapes, respectively, almost in equal parts. However, under accelerated rotation the grain aggregates enter a phase of deformation and, subsequently, the grain shapes are finally preferentially redistributed towards oblate shapes. 
    \item We introduce the quantity of the total stress $\sigma_\Sigma$ as a measure for the internal aggregate dynamics and time evolution. We report that $\sigma_\Sigma$ increases up to a   peak value characteristic for individual aggregates while the rotation accelerates and subsequently starts to loose mass. This peak roughly coincides with a mass loss of $25\ \%$ with respect to the initial grain mass. For higher angular velocities $\sigma_\Sigma$ drops sharply while the mass loss continues. We utilize this peak in $\sigma_\Sigma$ to define the breaking point on a physical basis and finally to determine the characteristic angular velocity $\omega_{\mathrm{disr}}$ of rotational disruption.
    \item In the deformation phase of BAM aggregates individual monomers are moving along the grain surface driven by the centrifugal force and new connections may be established. Consequently, the additional connections stabilize the BAM aggregates against the disruptive centrifugal forces acting on the monomers. 
    \item Our N-body simulations reveal that the angular velocity $\omega_{\mathrm{disr}}$ reaches an asymptotic limit towards larger grain sizes of $a_{\mathrm{eff}} \gtrapprox 300\ \mathrm{nm}$. This finding is in contrast to previous attempts  to describe the rational disruption of porous aggregates analytically  based on the maximal tensile strength where $\omega_{\mathrm{disr}}$ would continuously decrease for an increasing $a_{\mathrm{eff}}$.
\end{itemize}

\appendix
\section{Connection breaking test}
\label{sect:BreakingTest}
\begin{figure}[h]
	\begin{center}
	\includegraphics[width=0.49\textwidth]{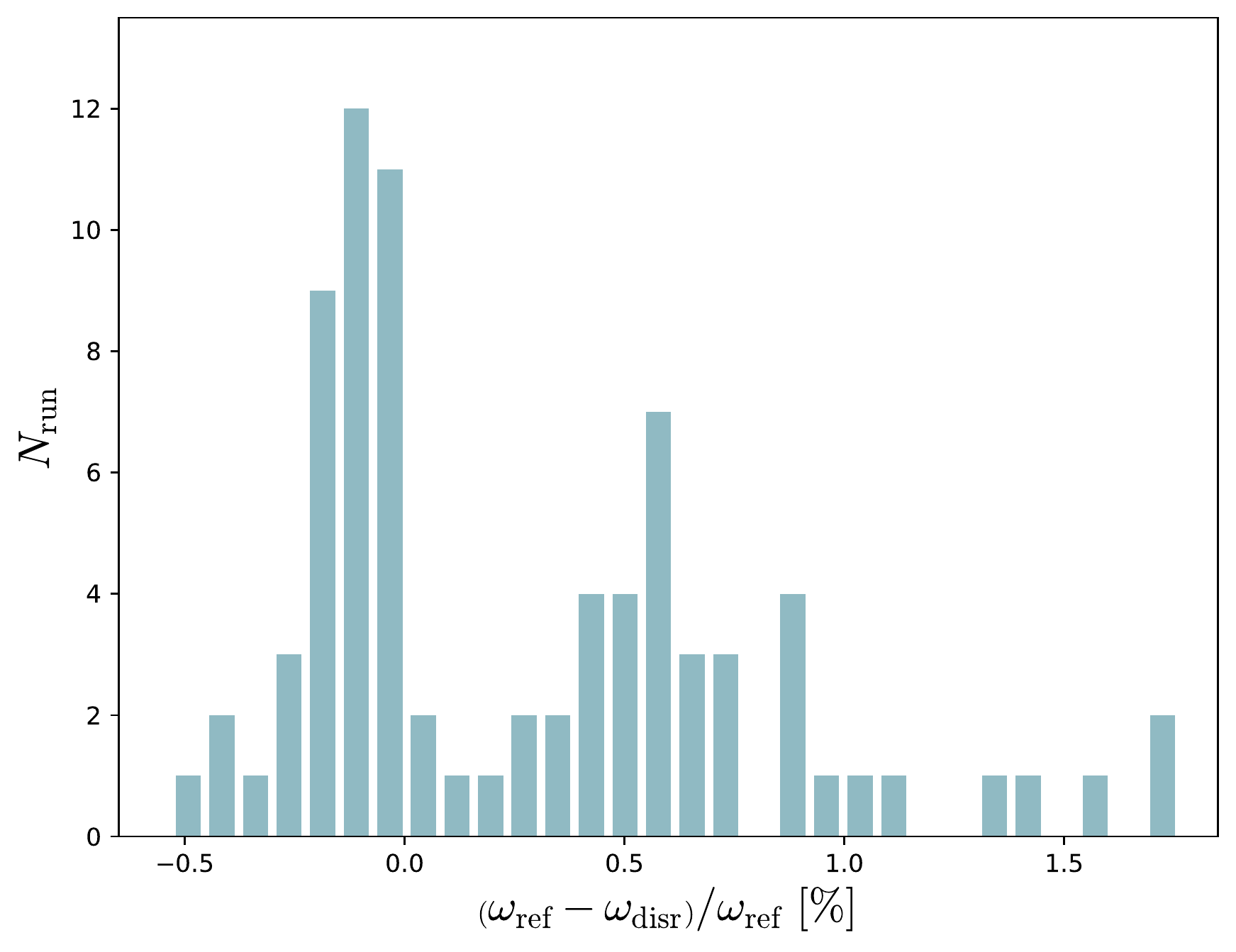}
	\end{center}
\caption{The distribution of the angular velocity $\omega_{\mathrm{disr}}$ resulting from a number of runs $N_{\mathrm{run}}$ of N-body simulations in compression with the corresponding exact analytical solution $\omega_{\mathrm{disr}}$.}
\label{fig:error}
\end{figure}

In this sections we provide a test scenario for the accuracy of our code. Due to the complexity of N-body simulation it is not feasible to derive an analytical solution for an entire aggregate. However, the problem can exactly be solved for two connected monomer with of equal size i.e. ${ a_{\mathrm{mon,i}}=a_{\mathrm{mon,j}} }$ where the reduced radius simply becomes $R=a_{\mathrm{mon,i}}/2$. In this case the centrifugal force  (see Eq.~\ref{eq:Fcent}) may be written as
\begin{equation}
    F_{\mathrm{cent,i}} = -m_{\mathrm{i}}\omega_{\mathrm{agg}}^2 a_{\mathrm{mon,i}}
\end{equation}
by using the criterion of a broken contact  $F_{\mathrm{N,ij}}=-F_{\mathrm{C}}$ the corresponding critical contact radius becomes
$r_{\mathrm{C}} = ( 1/6 )^{2/3}r_{0}$. 
Putting $r_{\mathrm{C}}$  into Eq.~\ref{eq:Fcontact}  allows the solve for the exact angular velocity 
\begin{equation}
    \omega_{\mathrm{ref}} =  \sqrt{ \frac{5\pi}{6}\frac{ \gamma} {m_{\mathrm{mon,i}} }  }
\label{eq:OmegaRef}
\end{equation}
where the two equally sized monomers become separated by centrifugal forces. We use this quantity for reference to evaluate the  accuracy of our code. A number of 80 benchmark runs are perform with randomly selected monomer raddii ${ a_{\mathrm{mon,i}}\in [10\ \mathrm{nm},100\ \mathrm{nm}] }$. In Fig.~\ref{fig:error} we compare the resulting angular velocity $\omega_{\mathrm{disr}}$ of our numerical setup introduced in Sect.~\ref{sect:NumericalSetup} with Eq.~\ref{eq:OmegaRef}. The resulting error ${ (\omega_{\mathrm{ref}} -\omega_{\mathrm{disr}})  / \omega_{\mathrm{ref}} }$ is below $1.8\ \%$  with a trend of underpredicting  $\omega_{\mathrm{ref}}$. An error of $1.8\ \%$ is comparable to the range we observe for the parameters test of the rolling displacement $\xi_{\mathrm{crit}}$ as presented in Fig.~\ref{fig:ParamTest}. Hence, we assume a numerical accuracy of about $1.8\ \%$ for our numerical N-body simulations. Consequently, the uncertainties in our reported results of the angular velocity of rotational disruption  $\omega_{\mathrm{ref}}$ is most impacted by the lack of exact laboratory material parameters rather than numerical errors.

%\section{List of applied physical quantities}
%In Table \ref{tab:Quantities} we briefly list all physical quantities utilized in this paper.

\begin{acknowledgements}
Special thanks goes to  Wilhelm Kley for numerous fruitfull discussion about N-body simulations of aggregates. The authors thank Thiem Hoang, Cornelis P. Dullemond, and Bruce T. Draine for usefull insights into the topic of dust composition and dynamics. S.R., P.N., and R.S.K. acknowledge financial support from the Heidelberg cluster of excellence (EXC 2181 - 390900948) “{\em STRUCTURES}: A unifying approach to emergent phenomena in the physical world, mathematics, and complex data”, specifically via the exploratory project EP 4.4. S.R. and R.S.K. also thank for support from Deutsche Forschungsgemeinschaft (DFG) via the Collaborative Research Center (SFB 881, Project-ID 138713538) 'The Milky Way System' (subprojects A01, A06, B01, B02, and B08). And we thanks for funding form the European Research Council in the ERC synergy grant “{\em ECOGAL} – Understanding our Galactic ecosystem: From the disk of the Milky Way to the formation sites of stars and planets” (project ID 855130). The project made use of computing resources provided by {\em The L\"{a}nd} through bwHPC and by DFG through grant INST 35/1134-1 FUGG. Data are in part stored at SDS@hd supported by the Ministry of Science, Research and the Arts and by DFG through grant INST 35/1314-1 FUGG.
\end{acknowledgements}

\bibliographystyle{aa}
\bibliography{./bibtex}

\begin{thebibliography}{137}
\expandafter\ifx\csname natexlab\endcsname\relax\def\natexlab#1{#1}\fi

\bibitem[{Baric {et~al.}(2018)Baric, Grossmann, Koch, \& Mädler}]{Baric2018}
Baric, V., Grossmann, H.~K., Koch, W., \& Mädler, L. 2018, Particle \&
  Particle Systems Characterization, 35, 1800177

\bibitem[{{Barlow} {et~al.}(2010){Barlow}, {Krause}, {Swinyard}, {Sibthorpe},
  {Besel}, {Wesson}, {Ivison}, {Dunne}, {Gear}, {Gomez}, {Hargrave}, {Henning},
  {Leeks}, {Lim}, {Olofsson}, \& {Polehampton}}]{Barlow2010}
{Barlow}, M.~J., {Krause}, O., {Swinyard}, B.~M., {et~al.} 2010, \aap, 518,
  L138

\bibitem[{Bauer {et~al.}(2019)Bauer, Daun, Huber, \& Will}]{Bauer2019}
Bauer, F., Daun, K., Huber, F., \& Will, S. 2019, Appl. Phys. B, 125, 109

\bibitem[{Benz \& Asphaug(1999)}]{Benz19995}
Benz, W. \& Asphaug, E. 1999, Icarus, 142, 5

\bibitem[{{Bertini} {et~al.}(2007){Bertini}, {Thomas}, \&
  {Barbieri}}]{Bertini2007}
{Bertini}, I., {Thomas}, N., \& {Barbieri}, C. 2007, \aap, 461, 351

\bibitem[{Bescond {et~al.}(2014)Bescond, Yon, Ouf, Ferry, Delhaye, Gaffié,
  Coppalle, \& Rozé}]{Bescond2014}
Bescond, A., Yon, J., Ouf, F.~X., {et~al.} 2014, Aerosol Science and
  Technology, 48, 831

\bibitem[{{Bevan} \& {Barlow}(2016)}]{Bevan2016}
{Bevan}, A. \& {Barlow}, M.~J. 2016, \mnras, 456, 1269

\bibitem[{{Birnstiel} {et~al.}(2010){Birnstiel}, {Dullemond}, \&
  {Brauer}}]{Birnstiel2010}
{Birnstiel}, T., {Dullemond}, C.~P., \& {Brauer}, F. 2010, \aap, 513, A79

\bibitem[{Blackford(2007)}]{Blackford2007}
Blackford, J.~R. 2007, Journal of Physics D: Applied Physics, 40, R355

\bibitem[{{Blum} {et~al.}(2006){Blum}, {Schr{\"a}pler}, {Davidsson}, \&
  {Trigo-Rodr{\'\i}guez}}]{Blum2006}
{Blum}, J., {Schr{\"a}pler}, R., {Davidsson}, B. J.~R., \&
  {Trigo-Rodr{\'\i}guez}, J.~M. 2006, \apj, 652, 1768

\bibitem[{{Bogdan} {et~al.}(2020){Bogdan}, {Pillich}, {Landers}, {Wende}, \&
  {Wurm}}]{Bogdan2020}
{Bogdan}, T., {Pillich}, C., {Landers}, J., {Wende}, H., \& {Wurm}, G. 2020,
  \aap, 638, A151

\bibitem[{Cardarelli(2008)}]{Cardarelli2008}
Cardarelli, F. 2008, Materials handbook: a concise desktop reference, 2nd edn.
  (London: Springer)

\bibitem[{Chakrabarty {et~al.}(2007)Chakrabarty, Moosm\"{u}ller, Arnott, Garro,
  Slowik, Cross, Han, Davidovits, Onasch, \& Worsnop}]{Chakrabarty2007}
Chakrabarty, R.~K., Moosm\"{u}ller, H., Arnott, W.~P., {et~al.} 2007, Appl.
  Opt., 46, 6990

\bibitem[{{Chokshi} {et~al.}(1993){Chokshi}, {Tielens}, \&
  {Hollenbach}}]{Chokshi1993}
{Chokshi}, A., {Tielens}, A.~G.~G.~M., \& {Hollenbach}, D. 1993, \apj, 407, 806

\bibitem[{Christensen(1996)}]{Christensen1996}
Christensen, N.~I. 1996, Journal of Geophysical Research: Solid Earth, 101,
  3139

\bibitem[{{Compi{\`e}gne} {et~al.}(2011){Compi{\`e}gne}, {Verstraete}, {Jones},
  {Bernard}, {Boulanger}, {Flagey}, {Le Bourlot}, {Paradis}, \&
  {Ysard}}]{Compiegne2011}
{Compi{\`e}gne}, M., {Verstraete}, L., {Jones}, A., {et~al.} 2011, \aap, 525,
  A103

\bibitem[{{Das} \& {Weingartner}(2016)}]{DasWeingartner2016}
{Das}, I. \& {Weingartner}, J.~C. 2016, \mnras, 457, 1958

\bibitem[{{Do-Duy} {et~al.}(2020){Do-Duy}, {Wright}, {Fujiyoshi}, {Glasse},
  {Siebenmorgen}, {Smith}, {Stecklum}, \& {Sterzik}}]{DoDuy2020}
{Do-Duy}, T., {Wright}, C.~M., {Fujiyoshi}, T., {et~al.} 2020, \mnras, 493,
  4463

\bibitem[{{Dolginov} \& {Silantev}(1976)}]{Dolginov1976}
{Dolginov}, A.~Z. \& {Silantev}, N.~A. 1976, \apss, 43, 337

\bibitem[{{Dominik} \& {N{\"u}bold}(2002)}]{Dominik2002}
{Dominik}, C. \& {N{\"u}bold}, H. 2002, \icarus, 157, 173

\bibitem[{{Dominik} \& {Tielens}(1995)}]{Dominik1995}
{Dominik}, C. \& {Tielens}, A.~G.~G.~M. 1995, Philosphical Magazine A, 72, 783

\bibitem[{{Dominik} \& {Tielens}(1996)}]{Dominik1996}
{Dominik}, C. \& {Tielens}, A.~G.~G.~M. 1996, Philosophical Magazine, Part A,
  73, 1279

\bibitem[{{Dominik} \& {Tielens}(1997)}]{Dominik1997}
{Dominik}, C. \& {Tielens}, A.~G.~G.~M. 1997, \apj, 480, 647

\bibitem[{{Dorschner} \& {Henning}(1995)}]{Dorschner1995}
{Dorschner}, J. \& {Henning}, T. 1995, \aapr, 6, 271

\bibitem[{{Draine}(1996)}]{Draine1996}
{Draine}, B.~T. 1996, in Astronomical Society of the Pacific Conference Series,
  Vol.~97, Polarimetry of the Interstellar Medium, ed. W.~G. {Roberge} \&
  D.~C.~B. {Whittet}, 16

\bibitem[{{Draine} \& {Flatau}(2013)}]{DraineFlatau2013}
{Draine}, B.~T. \& {Flatau}, P.~J. 2013, arXiv e-prints, arXiv:1305.6497

\bibitem[{{Draine} \& {Hensley}(2021{\natexlab{a}})}]{Draine2021}
{Draine}, B.~T. \& {Hensley}, B.~S. 2021{\natexlab{a}}, \apj, 909, 94

\bibitem[{{Draine} \& {Hensley}(2021{\natexlab{b}})}]{DraineHensley2021}
{Draine}, B.~T. \& {Hensley}, B.~S. 2021{\natexlab{b}}, \apj, 919, 65

\bibitem[{{Draine} \& {Lazarian}(1998)}]{DraineLazarian1998}
{Draine}, B.~T. \& {Lazarian}, A. 1998, \apj, 508, 157

\bibitem[{{Draine} \& {Lee}(1984)}]{Draine1984}
{Draine}, B.~T. \& {Lee}, H.~M. 1984, \apj, 285, 89

\bibitem[{{Draine} \& {Li}(2007)}]{DraineLi2007}
{Draine}, B.~T. \& {Li}, A. 2007, \apj, 657, 810

\bibitem[{{Draine} \& {Weingartner}(1996)}]{DraineWeingartner1996}
{Draine}, B.~T. \& {Weingartner}, J.~C. 1996, ApJ, 470, 551

\bibitem[{{Draine} \& {Weingartner}(1997)}]{DraineWeingartner1997}
{Draine}, B.~T. \& {Weingartner}, J.~C. 1997, \apj, 480, 633

\bibitem[{{Dwek} \& {Scalo}(1980)}]{Dwek1980}
{Dwek}, E. \& {Scalo}, J.~M. 1980, \apj, 239, 193

\bibitem[{Escatllar {et~al.}(2019)Escatllar, Lazaukas, Woodley, \&
  Bromley}]{Escatllar2019}
Escatllar, A.~M., Lazaukas, T., Woodley, S.~M., \& Bromley, S.~T. 2019, ACS
  Earth and Space Chemistry, 3, 2390, pMID: 32055761

\bibitem[{{Fogerty} {et~al.}(2016){Fogerty}, {Forrest}, {Watson}, {Sargent}, \&
  {Koch}}]{Fogerty2016}
{Fogerty}, S., {Forrest}, W., {Watson}, D.~M., {Sargent}, B.~A., \& {Koch}, I.
  2016, \apj, 830, 71

\bibitem[{Fuji {et~al.}(1999)Fuji, Machida, Takei, Watanabe, \&
  Chikazawa}]{Fuji1999}
Fuji, M., Machida, K., Takei, T., Watanabe, T., \& Chikazawa, M. 1999,
  Langmuir, 15, 4584

\bibitem[{{Gail} {et~al.}(2009){Gail}, {Zhukovska}, {Hoppe}, \&
  {Trieloff}}]{Gail2009}
{Gail}, H.~P., {Zhukovska}, S.~V., {Hoppe}, P., \& {Trieloff}, M. 2009, \apj,
  698, 1136

\bibitem[{{Galametz} {et~al.}(2019){Galametz}, {Maury}, {Valdivia}, {Testi},
  {Belloche}, \& {Andr{\'e}}}]{Galametz2019}
{Galametz}, M., {Maury}, A.~J., {Valdivia}, V., {et~al.} 2019, \aap, 632, A5

\bibitem[{{Giang} \& {Hoang}(2021)}]{Giang2021}
{Giang}, N.~C. \& {Hoang}, T. 2021, \apj, 922, 47

\bibitem[{{Giang} {et~al.}(2020){Giang}, {Hoang}, \& {Tram}}]{Giang2020}
{Giang}, N.~C., {Hoang}, T., \& {Tram}, L.~N. 2020, \apj, 888, 93

\bibitem[{{Goto} {et~al.}(2003){Goto}, {Gaessler}, {Hayano}, {Iye}, {Kamata},
  {Kanzawa}, {Kobayashi}, {Minowa}, {Saint-Jacques}, {Takami}, {Takato}, \&
  {Terada}}]{Goto2003}
{Goto}, M., {Gaessler}, W., {Hayano}, Y., {et~al.} 2003, \apj, 589, 419

\bibitem[{{Gouriet} {et~al.}(2019){Gouriet}, {Carrez}, \&
  {Cordier}}]{Gouriet2019}
{Gouriet}, K., {Carrez}, P., \& {Cordier}, P. 2019, Minerals, 9, 787

\bibitem[{{Greenberg} {et~al.}(1995){Greenberg}, {Mizutani}, \&
  {Yamamoto}}]{Greenberg1995}
{Greenberg}, J.~M., {Mizutani}, H., \& {Yamamoto}, T. 1995, \aap, 295, L35

\bibitem[{{Guillet} {et~al.}(2018){Guillet}, {Fanciullo}, {Verstraete},
  {Boulanger}, {Jones}, {Miville-Desch{\^e}nes}, {Ysard}, {Levrier}, \&
  {Alves}}]{Guillet2018}
{Guillet}, V., {Fanciullo}, L., {Verstraete}, L., {et~al.} 2018, \aap, 610, A16

\bibitem[{{Heim} {et~al.}(1999){Heim}, {Blum}, {Preuss}, \& {Butt}}]{Heim1999}
{Heim}, L.-O., {Blum}, J., {Preuss}, M., \& {Butt}, H.-J. 1999, \prl, 83, 3328

\bibitem[{{Hellyer}(1970)}]{Hellyer1970}
{Hellyer}, B. 1970, \mnras, 148, 383

\bibitem[{{Hensley} \& {Draine}(2021)}]{Hensley2021}
{Hensley}, B.~S. \& {Draine}, B.~T. 2021, \apj, 906, 73

\bibitem[{{Herranen} {et~al.}(2019){Herranen}, {Lazarian}, \&
  {Hoang}}]{Herranen2019}
{Herranen}, J., {Lazarian}, A., \& {Hoang}, T. 2019, \apj, 878, 96

\bibitem[{Hertz(1896)}]{Hertz1896}
Hertz, H. 1896, Miscellaneous papers (Macmillan)

\bibitem[{{Hirashita} \& {Yan}(2009)}]{Hirashita2009A}
{Hirashita}, H. \& {Yan}, H. 2009, \mnras, 394, 1061

\bibitem[{{Hoang}(2020)}]{Hoang2020Galaxy}
{Hoang}, T. 2020, Galaxies, 8, 52

\bibitem[{{Hoang}(2022)}]{Hoang2022}
{Hoang}, T. 2022, \apj, 928, 102

\bibitem[{{Hoang} {et~al.}(2018){Hoang}, {Cho}, \& {Lazarian}}]{Hoang2018AMech}
{Hoang}, T., {Cho}, J., \& {Lazarian}, A. 2018, \apj, 852, 129

\bibitem[{{Hoang} {et~al.}(2014){Hoang}, {Lazarian}, \& {Martin}}]{Hoang2014}
{Hoang}, T., {Lazarian}, A., \& {Martin}, P.~G. 2014, \apj, 790, 6

\bibitem[{{Hoang} {et~al.}(2019){Hoang}, {Tram}, {Lee}, \& {Ahn}}]{Hoang2019}
{Hoang}, T., {Tram}, L.~N., {Lee}, H., \& {Ahn}, S.-H. 2019, Nature Astronomy,
  3, 766

\bibitem[{Israelachvili(2011)}]{Israelachvili2011}
Israelachvili, J.~N. 2011, Intermolecular and surface forces (Academic press)

\bibitem[{{Jensen} {et~al.}(2015){Jensen}, {Wise}, \& {Odegard}}]{Jensen2015}
{Jensen}, B.~D., {Wise}, K.~E., \& {Odegard}, G.~M. 2015, Journal of Physical
  Chemistry A, 119, 9710

\bibitem[{{Johnson}(1987)}]{Johnson1987}
{Johnson}, K.~L. 1987, {Contact Mechanics}

\bibitem[{{Johnson} {et~al.}(1971){Johnson}, {Kendall}, \&
  {Roberts}}]{Johnson1971}
{Johnson}, K.~L., {Kendall}, K., \& {Roberts}, A.~D. 1971, Proceedings of the
  Royal Society of London Series A, 324, 301

\bibitem[{{Jones}(2012)}]{Jones2012}
{Jones}, A.~P. 2012, \aap, 542, A98

\bibitem[{Kandilian {et~al.}(2015)Kandilian, Heng, \& Pilon}]{Kandilian2015}
Kandilian, R., Heng, R.-L., \& Pilon, L. 2015, Journal of Quantitative
  Spectroscopy and Radiative Transfer, 151, 310

\bibitem[{Karasev {et~al.}(2004)Karasev, Onischuk, Glotov, Baklanov, Maryasov,
  Zarko, Panfilov, Levykin, \& Sabelfeld}]{Karasev2004}
Karasev, V., Onischuk, A., Glotov, O., {et~al.} 2004, Combustion and Flame,
  138, 40

\bibitem[{{Karovicova} {et~al.}(2013){Karovicova}, {Wittkowski}, {Ohnaka},
  {Boboltz}, {Fossat}, \& {Scholz}}]{Karovicova2013}
{Karovicova}, I., {Wittkowski}, M., {Ohnaka}, K., {et~al.} 2013, \aap, 560, A75

\bibitem[{{Kataoka} {et~al.}(2013{\natexlab{a}}){Kataoka}, {Tanaka}, {Okuzumi},
  \& {Wada}}]{Kataoka2013}
{Kataoka}, A., {Tanaka}, H., {Okuzumi}, S., \& {Wada}, K. 2013{\natexlab{a}},
  \aap, 557, L4

\bibitem[{{Kataoka} {et~al.}(2013{\natexlab{b}}){Kataoka}, {Tanaka}, {Okuzumi},
  \& {Wada}}]{Kataoka2013X}
{Kataoka}, A., {Tanaka}, H., {Okuzumi}, S., \& {Wada}, K. 2013{\natexlab{b}},
  \aap, 554, A4

\bibitem[{Kelesidis {et~al.}(2018)Kelesidis, Furrer, Wegner, \&
  Pratsinis}]{Kelesidis2018}
Kelesidis, G.~A., Furrer, F.~M., Wegner, K., \& Pratsinis, S.~E. 2018,
  Langmuir, 34, 8532, pMID: 29940739

\bibitem[{{Kendall} {et~al.}(1987){Kendall}, {McN. Alford}, \&
  {Birchall}}]{Kendall1987}
{Kendall}, K., {McN. Alford}, N., \& {Birchall}, J.~D. 1987, Proceedings of the
  Royal Society of London Series A, 412, 269

\bibitem[{{Kim} {et~al.}(2021){Kim}, {Takigawa}, {Tsuchiyama}, {Matsuno},
  {Enju}, {Kawano}, \& {Komaki}}]{Kim2021}
{Kim}, T.~H., {Takigawa}, A., {Tsuchiyama}, A., {et~al.} 2021, \aap, 656, A42

\bibitem[{Kimura {et~al.}(2015)Kimura, Wada, Senshu, \& Kobayashi}]{Kimura2015}
Kimura, H., Wada, K., Senshu, H., \& Kobayashi, H. 2015, The Astrophysical
  Journal, 812, 67

\bibitem[{{Kimura} {et~al.}(2020){Kimura}, {Wada}, {Yoshida}, {Hong}, {Senshu},
  {Arai}, {Hirai}, {Kobayashi}, {Ishibashi}, \& {Yamada}}]{Kimura2020}
{Kimura}, H., {Wada}, K., {Yoshida}, F., {et~al.} 2020, \mnras, 496, 1667

\bibitem[{{Kirchschlager} {et~al.}(2019){Kirchschlager}, {Schmidt}, {Barlow},
  {Fogerty}, {Bevan}, \& {Priestley}}]{Kirchschlager2019}
{Kirchschlager}, F., {Schmidt}, F.~D., {Barlow}, M.~J., {et~al.} 2019, \mnras,
  489, 4465

\bibitem[{K\"oyl\"u \& Faeth(1994)}]{Koylu1994}
K\"oyl\"u, U.~O. \& Faeth, G.~M. 1994, Journal of Heat Transfer, 116, 971

\bibitem[{{Kozasa} {et~al.}(1992){Kozasa}, {Blum}, \& {Mukai}}]{Kozasa1992}
{Kozasa}, T., {Blum}, J., \& {Mukai}, T. 1992, \aap, 263, 423

\bibitem[{{Krijt} {et~al.}(2013){Krijt}, {G{\"u}ttler}, {Hei{\ss}elmann},
  {Dominik}, \& {Tielens}}]{Krijt2013}
{Krijt}, S., {G{\"u}ttler}, C., {Hei{\ss}elmann}, D., {Dominik}, C., \&
  {Tielens}, A.~G.~G.~M. 2013, Journal of Physics D Applied Physics, 46, 435303

\bibitem[{{Lazarian} \& {Draine}(1999)}]{LazarianDraine1999A}
{Lazarian}, A. \& {Draine}, B.~T. 1999, \apjl, 520, L67

\bibitem[{{Lazarian} \& {Efroimsky}(1999)}]{LazarianEfroimsky1999}
{Lazarian}, A. \& {Efroimsky}, M. 1999, \mnras, 303, 673

\bibitem[{{Lazarian} \& {Hoang}(2007{\natexlab{a}})}]{LazarianHoang2007}
{Lazarian}, A. \& {Hoang}, T. 2007{\natexlab{a}}, \mnras, 378, 910

\bibitem[{{Lazarian} \& {Hoang}(2007{\natexlab{b}})}]{Lazarian2007Mech}
{Lazarian}, A. \& {Hoang}, T. 2007{\natexlab{b}}, \apjl, 669, L77

\bibitem[{{Lazarian} \& {Roberge}(1997)}]{Lazarian1997}
{Lazarian}, A. \& {Roberge}, W.~G. 1997, \apj, 484, 230

\bibitem[{Lehre {et~al.}(2003)Lehre, Jungfleisch, Suntz, \&
  Bockhorn}]{Lehre2003}
Lehre, T., Jungfleisch, B., Suntz, R., \& Bockhorn, H. 2003, Appl. Opt., 42,
  2021

\bibitem[{{Li} \& {Draine}(2001)}]{LiDraine2001}
{Li}, A. \& {Draine}, B.~T. 2001, \apj, 554, 778

\bibitem[{{Li} \& {Greenberg}(1997)}]{LiGreenberg1997}
{Li}, A. \& {Greenberg}, J.~M. 1997, \aap, 323, 566

\bibitem[{Liu {et~al.}(2015)Liu, Yin, Hu, Jin, \& Sorensen}]{ChaoLiu2015}
Liu, C., Yin, Y., Hu, F., Jin, H., \& Sorensen, C.~M. 2015, Aerosol Science and
  Technology, 49, 928

\bibitem[{Maeno \& Ebinuma(1983)}]{Maeno1983}
Maeno, N. \& Ebinuma, T. 1983, The Journal of Physical Chemistry, 87, 4103

\bibitem[{{Mathis} {et~al.}(1977){Mathis}, {Rumpl}, \&
  {Nordsieck}}]{Mathis1977}
{Mathis}, J.~S., {Rumpl}, W., \& {Nordsieck}, K.~H. 1977, \apj, 217, 425

\bibitem[{{Matsuura}(2011)}]{Matsuura2011}
{Matsuura}, M. 2011, in Astronomical Society of the Pacific Conference Series,
  Vol. 445, Why Galaxies Care about AGB Stars II: Shining Examples and Common
  Inhabitants, ed. F.~{Kerschbaum}, T.~{Lebzelter}, \& R.~F. {Wing}, 531

\bibitem[{{Matthews} {et~al.}(2012){Matthews}, {Land}, \&
  {Hyde}}]{Matthews2012}
{Matthews}, L.~S., {Land}, V., \& {Hyde}, T.~W. 2012, \apj, 744, 8

\bibitem[{{Mennella}(2006)}]{Mennella2006}
{Mennella}, V. 2006, \apjl, 647, L49

\bibitem[{{Michoulier} \& {Gonzalez}(2022)}]{Michoulier2022}
{Michoulier}, S. \& {Gonzalez}, J.-F. 2022, \mnras, 517, 3064

\bibitem[{{Min} {et~al.}(2007){Min}, {Waters}, {de Koter}, {Hovenier},
  {Keller}, \& {Markwick-Kemper}}]{Min2007}
{Min}, M., {Waters}, L.~B.~F.~M., {de Koter}, A., {et~al.} 2007, \aap, 462, 667

\bibitem[{Mitchell(2004)}]{Mitchell2004}
Mitchell, R.~H. 2004, 8th International Kimberlite Conference: The J. Barry
  Hawthorne volume, Vol.~2 (Gulf Professional Publishing)

\bibitem[{Morris \& Burchell(2017)}]{Morris2017}
Morris, A. \& Burchell, M. 2017, Icarus, 296, 91

\bibitem[{{Nozawa} {et~al.}(2007){Nozawa}, {Kozasa}, {Habe}, {Dwek}, {Umeda},
  {Tominaga}, {Maeda}, \& {Nomoto}}]{Nozawa2007}
{Nozawa}, T., {Kozasa}, T., {Habe}, A., {et~al.} 2007, \apj, 666, 955

\bibitem[{{O'Donnell} \& {Mathis}(1997)}]{ODonnell1997}
{O'Donnell}, J.~E. \& {Mathis}, J.~S. 1997, \apj, 479, 806

\bibitem[{{Ormel} {et~al.}(2009){Ormel}, {Paszun}, {Dominik}, \&
  {Tielens}}]{Ormel2009}
{Ormel}, C.~W., {Paszun}, D., {Dominik}, C., \& {Tielens}, A.~G.~G.~M. 2009,
  \aap, 502, 845

\bibitem[{{Ossenkopf}(1993)}]{Ossenkopf1993}
{Ossenkopf}, V. 1993, \aap, 280, 617

\bibitem[{{Paszun} \& {Dominik}(2008)}]{Paszun2008}
{Paszun}, D. \& {Dominik}, C. 2008, \aap, 484, 859

\bibitem[{Paul {et~al.}(2017)Paul, Silverstein, \& Krekeler}]{Paul2017}
Paul, K.~C., Silverstein, J., \& Krekeler, M.~P. 2017, Environmental Earth
  Sciences, 76, 1

\bibitem[{{Petrovic}(2001)}]{Petrovic2001}
{Petrovic}, J.~J. 2001, Journal of Materials Science, 36, 1573

\bibitem[{{Purcell}(1979)}]{Purcell1979}
{Purcell}, E.~M. 1979, \apj, 231, 404

\bibitem[{Raschdorf \& Kolonko(2009)}]{Raschdorf2009}
Raschdorf, S. \& Kolonko, M. 2009, Clausthal University of Technology

\bibitem[{{Reissl} {et~al.}(2022){Reissl}, {Meehan}, \& {Klessen}}]{Reissl2022}
{Reissl}, S., {Meehan}, P., \& {Klessen}, R.~S. 2022, arXiv e-prints,
  arXiv:2201.03694

\bibitem[{Remediakis {et~al.}(2007)Remediakis, Fyta, Mathioudakis, Kopidakis,
  \& Kelires}]{Remediakis2007}
Remediakis, I.~N., Fyta, M.~G., Mathioudakis, C., Kopidakis, G., \& Kelires,
  P.~C. 2007, Diamond and Related Materials, 16, 1835, proceedings of the 6th
  Specialists Meeting in Amorphous Carbon

\bibitem[{{Rogantini} {et~al.}(2019){Rogantini}, {Costantini}, {Zeegers}, {de
  Vries}, {Mehdipour}, {de Groot}, {Mutschke}, {Psaradaki}, \&
  {Waters}}]{Rogantini2019}
{Rogantini}, D., {Costantini}, E., {Zeegers}, S.~T., {et~al.} 2019, \aap, 630,
  A143

\bibitem[{Salameh {et~al.}(2017)Salameh, van~der Veen, Kappl, \& van
  Ommen}]{Salameh2017}
Salameh, S., van~der Veen, M.~A., Kappl, M., \& van Ommen, J.~R. 2017,
  Langmuir, 33, 2477, pMID: 28186771

\bibitem[{{Schwartz} {et~al.}(2018){Schwartz}, {Michel}, {Jutzi}, {Marchi},
  {Zhang}, \& {Richardson}}]{Schwartz2018}
{Schwartz}, S.~R., {Michel}, P., {Jutzi}, M., {et~al.} 2018, Nature Astronomy,
  2, 379

\bibitem[{{Seizinger} {et~al.}(2013{\natexlab{a}}){Seizinger}, {Krijt}, \&
  {Kley}}]{Seizinger2013}
{Seizinger}, A., {Krijt}, S., \& {Kley}, W. 2013{\natexlab{a}}, \aap, 560, A45

\bibitem[{{Seizinger} {et~al.}(2012){Seizinger}, {Speith}, \&
  {Kley}}]{Seizinger2012}
{Seizinger}, A., {Speith}, R., \& {Kley}, W. 2012, \aap, 541, A59

\bibitem[{{Seizinger} {et~al.}(2013{\natexlab{b}}){Seizinger}, {Speith}, \&
  {Kley}}]{Seizinger2013Tensile}
{Seizinger}, A., {Speith}, R., \& {Kley}, W. 2013{\natexlab{b}}, \aap, 559, A19

\bibitem[{{Shen} {et~al.}(2008){Shen}, {Draine}, \& {Johnson}}]{Shen2008}
{Shen}, Y., {Draine}, B.~T., \& {Johnson}, E.~T. 2008, \apj, 689, 260

\bibitem[{{Silsbee} \& {Draine}(2016)}]{Silsbee2016}
{Silsbee}, K. \& {Draine}, B.~T. 2016, \apj, 818, 133

\bibitem[{{Sirono} \& {Ueno}(2017)}]{Sirono2017}
{Sirono}, S.-i. \& {Ueno}, H. 2017, \apj, 841, 36

\bibitem[{{Skorupski} {et~al.}(2014){Skorupski}, {Mroczka}, {Wriedt}, \&
  {Riefler}}]{Skorupski2014}
{Skorupski}, K., {Mroczka}, J., {Wriedt}, T., \& {Riefler}, N. 2014, Physica A
  Statistical Mechanics and its Applications, 404, 106

\bibitem[{Slobodrian {et~al.}(2011)Slobodrian, Rioux, \&
  Piche}]{Slobodrian2010}
Slobodrian, R., Rioux, C., \& Piche, M. 2011, Open Journal of Metal, 1, 7

\bibitem[{{Spitzer} \& {Arny}(1978)}]{Spitzer1978}
{Spitzer}, L. \& {Arny}, T.~T. 1978, American Journal of Physics, 46, 1201

\bibitem[{{Steinpilz} {et~al.}(2019){Steinpilz}, {Teiser}, \&
  {Wurm}}]{Steinpilz2019}
{Steinpilz}, T., {Teiser}, J., \& {Wurm}, G. 2019, \apj, 874, 60

\bibitem[{{Takigawa} \& {Tachibana}(2012)}]{Takigawa2012}
{Takigawa}, A. \& {Tachibana}, S. 2012, \apj, 750, 149

\bibitem[{{Tatsuuma} \& {Kataoka}(2021)}]{Tatsuuma2021}
{Tatsuuma}, M. \& {Kataoka}, A. 2021, \apj, 913, 132

\bibitem[{{Tatsuuma} {et~al.}(2019){Tatsuuma}, {Kataoka}, \&
  {Tanaka}}]{TatsuumaKataoka2019}
{Tatsuuma}, M., {Kataoka}, A., \& {Tanaka}, H. 2019, \apj, 874, 159

\bibitem[{{Tazaki} {et~al.}(2017){Tazaki}, {Lazarian}, \&
  {Nomura}}]{Tazaki2017}
{Tazaki}, R., {Lazarian}, A., \& {Nomura}, H. 2017, \apj, 839, 56

\bibitem[{{Tielens} {et~al.}(1994){Tielens}, {McKee}, {Seab}, \&
  {Hollenbach}}]{Tielens1994}
{Tielens}, A.~G.~G.~M., {McKee}, C.~F., {Seab}, C.~G., \& {Hollenbach}, D.~J.
  1994, \apj, 431, 321

\bibitem[{Ulrich(2000)}]{Ulrich2000}
Ulrich, T. 2000, in Game Programming Gems, ed. M.~DeLoura (Charles River
  Media), 444--453

\bibitem[{{Voshchinnikov} \& {Henning}(2010)}]{Voshchinnikov2010}
{Voshchinnikov}, N.~V. \& {Henning}, T. 2010, \aap, 517, A45

\bibitem[{{Wada} {et~al.}(2007){Wada}, {Tanaka}, {Suyama}, {Kimura}, \&
  {Yamamoto}}]{Wada2007}
{Wada}, K., {Tanaka}, H., {Suyama}, T., {Kimura}, H., \& {Yamamoto}, T. 2007,
  \apj, 661, 320

\bibitem[{{Wada} {et~al.}(1999){Wada}, {Kaito}, {Kimura}, {Ono}, \&
  {Tokunaga}}]{Wada1999}
{Wada}, S., {Kaito}, C., {Kimura}, S., {Ono}, H., \& {Tokunaga}, A.~T. 1999,
  \aap, 345, 259

\bibitem[{{Weingartner} \& {Draine}(2001)}]{WeingartnerDraine2001X}
{Weingartner}, J.~C. \& {Draine}, B.~T. 2001, \apj, 548, 296

\bibitem[{{Weingartner} \& {Draine}(2003)}]{WeingartnerDraine2003}
{Weingartner}, J.~C. \& {Draine}, B.~T. 2003, \apj, 589, 289

\bibitem[{Williams \& Jadwick(1980)}]{Williams1980}
Williams, R.~J. \& Jadwick, J.~J. 1980, Handbook of lunar materials (Houston,
  TX, United States: NASA Reference Publication)

\bibitem[{Wu {et~al.}(2020)Wu, Faccinetto, Grimonprez, Batut, Yon, Desgroux, \&
  Petitprez}]{Wu2020}
Wu, J., Faccinetto, A., Grimonprez, S., {et~al.} 2020, Atmospheric Chemistry
  and Physics, 20, 4209

\bibitem[{{Yan} \& {Lazarian}(2003)}]{Yan2003}
{Yan}, H. \& {Lazarian}, A. 2003, \apjl, 592, L33

\bibitem[{{Zebda} {et~al.}(2008){Zebda}, {Sabbah}, {Ababou-Girard}, {Solal}, \&
  {Godet}}]{Zebda2008}
{Zebda}, A., {Sabbah}, H., {Ababou-Girard}, S., {Solal}, F., \& {Godet}, C.
  2008, Applied Surface Science, 254, 4980

\bibitem[{Zhang {et~al.}(2020)Zhang, Qi, Shi, Gao, \& Ren}]{Zhang2020}
Zhang, J.-Y., Qi, H., Shi, J.-W., Gao, B.-H., \& Ren, Y.-T. 2020, Opt. Express,
  28, 37249

\bibitem[{{Zhukovska} {et~al.}(2008){Zhukovska}, {Gail}, \&
  {Trieloff}}]{Zhukovska2008}
{Zhukovska}, S., {Gail}, H.~P., \& {Trieloff}, M. 2008, \aap, 479, 453

\bibitem[{{Zhukovska} \& {Henning}(2013)}]{Zhukovska2013}
{Zhukovska}, S. \& {Henning}, T. 2013, \aap, 555, A99

\bibitem[{{Zhukovska} {et~al.}(2015){Zhukovska}, {Petrov}, \&
  {Henning}}]{Zhukovska2015}
{Zhukovska}, S., {Petrov}, M., \& {Henning}, T. 2015, \apj, 810, 128

\bibitem[{{Zolensky} {et~al.}(2006){Zolensky}, {Zega}, {Yano}, {Wirick},
  {Westphal}, {Weisberg}, {Weber}, {Warren}, {Velbel}, {Tsuchiyama}, {Tsou},
  {Toppani}, {Tomioka}, {Tomeoka}, {Teslich}, {Taheri}, {Susini}, {Stroud},
  {Stephan}, {Stadermann}, {Snead}, {Simon}, {Simionovici}, {See}, {Robert},
  {Rietmeijer}, {Rao}, {Perronnet}, {Papanastassiou}, {Okudaira}, {Ohsumi},
  {Ohnishi}, {Nakamura-Messenger}, {Nakamura}, {Mostefaoui}, {Mikouchi},
  {Meibom}, {Matrajt}, {Marcus}, {Leroux}, {Lemelle}, {Le}, {Lanzirotti},
  {Langenhorst}, {Krot}, {Keller}, {Kearsley}, {Joswiak}, {Jacob}, {Ishii},
  {Harvey}, {Hagiya}, {Grossman}, {Grossman}, {Graham}, {Gounelle}, {Gillet},
  {Genge}, {Flynn}, {Ferroir}, {Fallon}, {Ebel}, {Dai}, {Cordier}, {Clark},
  {Chi}, {Butterworth}, {Brownlee}, {Bridges}, {Brennan}, {Brearley},
  {Bradley}, {Bleuet}, {Bland}, \& {Bastien}}]{Zolensky2006}
{Zolensky}, M.~E., {Zega}, T.~J., {Yano}, H., {et~al.} 2006, Science, 314, 1735

\end{thebibliography}
\end{document}